\documentclass[aip,reprint,amsmath,amssymb,amsfonts,floatfix,superscriptaddress,showkeys,showpacs]{revtex4-1}

\usepackage{dcolumn} 
\usepackage{multirow}
\usepackage[T1]{fontenc}
\usepackage{dsfont}
\usepackage{verbatim}
\usepackage{bm}
\usepackage[bookmarks=false]{hyperref} 
\hypersetup{
           breaklinks=true,   % splits links across lines
           colorlinks=true,   % displays links as colored text instead of blocks
           pdfusetitle=true,  % \title and \author values into pdf metadata
        }
\usepackage{cleveref}  
\usepackage{floatrow}  
\usepackage{natbib}
\usepackage{graphicx}
\usepackage{epstopdf}
\usepackage{ifpdf}
\usepackage{epsfig}
\usepackage{color}
\usepackage[normalem]{ulem}
\usepackage[usenames,dvipsnames]{xcolor}
\usepackage{enumitem} 

\usepackage{scalerel}
\newcommand{\mypara}{{\stretchrel*{\parallel}{\perp}}}
\newcommand{\myperp}{{\!\perp}}

\begin{document}

\title{Brownian noise effects on magnetic focusing of prolate and oblate spheroids in channel flow}

\author{Mohammad Reza Shabanniya}
\affiliation{School of Physics, Institute for Research in Fundamental Sciences (IPM), Tehran 19538-33511, Iran}
\author{Ali Naji}
\thanks{Corresponding author. Email: \texttt{a.naji@ipm.ir}}
\affiliation{School of Nano Science, Institute for Research in Fundamental Sciences (IPM), Tehran 19538-33511, Iran}
 
\begin{abstract} 
We investigate Brownian noise effects on magnetic focusing of prolate and oblate spheroids carrying permanent magnetic dipoles in channel (Poiseuille) flow subject to a uniform magnetic field. The focusing is effected by the low-Reynolds-number wall-induced hydrodynamic lift which can be tuned via tilt angle of the field relative to the flow direction. This mechanism is incorporated in a steady-state Smoluchowski equation that we solve numerically to analyze the noise effects through the joint position-orientation probability distribution function of spheroids within the channel. The results feature partial and complete pinning of spheroidal orientation as the field strength is increased and reveal remarkable and even counterintuitive noise-induced phenomena (specifically due to translational particle diffusivity) deep into the strong-field regime. These include field-induced defocusing, or lateral broadening of the focused spheroidal layer,  upon strengthening the field. We map out focusing `phase' diagrams based on the field strength and tilt angle to illustrate different regimes of behavior including centered focusing and defocusing in transverse field, and off-centered focusing in tilted fields. The latter encompasses two subregimes of optimal and shouldered focusing where spheroidal density profiles across the channel width display either an isolated off-centered peak or a skewed peak with a pronounced shoulder stretching toward the channel center. We corroborate our results by analyzing stability of deterministic fixed points and a reduced one-dimensional probabilistic theory which we introduce to semiquantitatively explain noise-induced behavior of pinned spheroids under strong fields. We also elucidate the implications of our results for efficient shape-based sorting of magnetic spheroids.
\end{abstract}

%\pacs{}
%\keywords{}

\maketitle

%%%%%%%%%%%%%%%%%%%%%%%%%%%%%%%%%%%%%
%%%%%%%%%%%%%%%%%%%%%%%%%%%%%%%%%%%%%
\section{Introduction}
  
Magnetic separation of nano/micropar\-ticles in micro\-fluidic setups have emerged as an important technological tool especially in biomedical research  \cite{Pamme2006,Pamme2007,Lenshof2010,Gijs2010,Behdani2018,Xuan2019}. The separation methods typically involve a mixture of magnetic or magnetically labeled biological or synthetic particles in an imposed shear flow within properly designed microchannels. The system is subjected to an external magnetic field to produce controlled alterations in the flow-driven motion of  target particles based on their size, shape and magnetic properties \cite{Liu2009,Suwa_Review,Hejazian2015,Xuan2019}. A host of bioparticles can be manipulated  using such  setups. These include proteins, nucleic acids, bacteria such as {\em Escherichia coli}, and cells such as red blood cells and circulating tumor cells \cite{Xia2006,Zborowski2011,Pamme2012,Chen2014,Surendran_Rev}. Similar methods can be used for detection and sorting of artificial magnetic beads. The latter often consist of a  nonmagnetic  (e.g., spherical or ellipsoidal) core coated or doped by magnetic materials (e.g., ferromagnetic chrome dioxide, superparamagnetic magnetite, etc.) \cite{Pankhurst2003,Gijs2004,Espy2006,Jiang2006,Shen2012,Smits2016,Ruffert2016,Erb2016,Rogowski2021}. 

Magnetic separation techniques often use spatially nonuniform magnetic fields to create net magnetic forces on the target particles. The force magnitude is tuned by adjusting, e.g.,  the field gradient and/or the magnetic susceptibility mismatch between the particles and the background fluid, enabling cross-stream magnetophoretic migration and, hence, separation of the particles   \cite{Zborowski2011book}. 
 
Utilizing uniform magnetic fields has also received mounting attention in recent years as an alternative route to separation of  magnetic (nonmagnetic) particles in nonmagnetic (magnetic) microfluidic flows \cite{Zhu2014,Zhou2017,ZhouPRA2017,Cao2017,GolestanianFocusing, Matsunaga2018,Zhang2018b,Zhang2018,Sobecki2018,Sobecki2020,Zhang2019}.  In a uniform field, the magnetophoretic effect is absent, as magnetic particles experience no net force from the applied field but only a torque that augments the torque experienced from the imposed shear. These torques are both inherently shape dependent. Thus, other than offering practical advantages (e.g., access to large field-exposed areas and high-throughput parallelization of multiple  channels  \cite{Zhou2017}),  uniform fields facilitate a robust shape-based approach to magnetic particle separation \cite{Behdani2018,Xuan2019}. 

Recent experiments have indeed established the above strategy for  separation of prolate  paramagnetic spheroids from  paramagnetic spheres of the same volume \cite{Zhou2017,ZhouPRA2017}. These experiments utilize a pressure-driven channel flow at low Reynolds number and a uniform (static) magnetic field applied either transversally or with a tilt angle relative to the flow direction. In the absence of field, the particles were shown to undergo standard Jeffery rotation \cite{Jeffery,kim_microhydrodynamics} due to the shear-induced torque. In a transverse field  \cite{Zhou2017}, the up-down symmetry pertinent to these rotations breaks down for prolate spheroids but not the spheres (magnetic torque vanishes on paramagnetic spheres). This rotational asymmetry is coupled with  lateral particle translation across the channel width  through the particle-wall hydrodynamic interactions, specifically, the wall-induced hydrodynamic lift, which in turn effects net migration of prolate spheroids to the channel center. This mechanism holds under weak fields, also for nonmagnetic particles in ferrofluid flows \cite{Zhou2017,ZhouPRA2017}. It has been scrutinized by deterministic simulations and theories   
\cite{Zhou2017,ZhouPRA2017,Cao2017,Zhang2018b,Zhang2018,Sobecki2018,Sobecki2020,Zhang2019} which also illuminate the more complex particle migration pattern observed  \cite{ZhouPRA2017} under tilted fields and the orientational pinning of spheroids \cite{Zhou2017,ZhouPRA2017,Cao2017,Zhang2018,Zhang2018b,Sobecki2018,Sobecki2019,Sobecki2020,Kumaran2020,GolestanianFocusing,Matsunaga2018} that follows  from the counterbalance between field and shear-induced torques  at  strong  fields  \cite{Almog1995}.

The strong-field mechanism for separation of  magnetic spheroids with pinned orientation was first  addressed in Refs. \cite{GolestanianFocusing,Matsunaga2018}. These studies considered prolate spheroids of permanent (ferro)magnetic dipole moment in two-dimensional (2D) or plane Poiseuille flow and also in three-dimensional (3D) flows within rectangular and cylindrical  channels.  Using far-field hydrodynamic calculations of the wall-induced  lift, they showed that orientationally pinned  spheroids  laterally translate  within the channel and focus at specific shape-dependent latitudes upon adjusting the field direction. The far-field theory closely reproduces boundary element simulations especially at spheroid-wall separations exceeding a couple particle sizes. When the applied field is strong enough, particle translation and rotation largely occur within the field plane \cite{Zhou2017,ZhouPRA2017,Almog1995,GolestanianFocusing}, and the results in 3D rectangular channels of sufficiently large cross-sectional aspect ratio reproduce those in the 2D flow \cite{GolestanianFocusing,Matsunaga2018}. In this latter setting, the differences between cross-stream migration of prolate ferromagnetic and paramagnetic spheroids under weak and strong magnetic fields have also been studied using direct finite-element numerical simulations \cite{Zhang2018b}. 

In this paper, we also consider  spheroidal particles of permanent magnetic dipole in plane Poiseuille flow under a uniform magnetic field (of arbitrary tilt angle within the flow plane) according to the models in Ref. \cite{GolestanianFocusing,Zhang2018b}. In the theoretical and numerical studies cited above, lateral migration and focusing of (para/ferro)magnetic particles have mainly been analyzed based on deterministic particle trajectories and, to our knowledge, noise effects have been examined only briefly by  Brownian Dynamics (BD) simulations in Ref. \cite{GolestanianFocusing}. Our goal here is to elucidate salient, unexplored aspects of Brownian noise, or particle diffusivity,  which we show to have important qualitative ramifications that go beyond the deterministic picture. 

We use  three complementary approaches and consider not only prolate  but also oblate spheroids that have received much less attention \cite{Mody2005,Einarsson2015,Michelin2016,Michelin2017}. The approaches used here are (i) a continuum formulation based on a steady-state Smoluchowski equation to describe the joint position-orientation probability distribution (PDF) of spheroids from weak to strong fields  (without having to deal with sampling fluctuations of analogous BD simulations); this will be referred to as the {\em full probabilistic approach}; (ii) a systematic phase-space stability analysis to classify fixed points of {\em deterministic} spheroidal dynamics  at strong fields; (iii) a {\em reduced probabilistic theory} to describe the noise-induced behavior of   spheroids on a one-dimensional (1D) pinning curve in the position-orientation coordinate space at strong fields. To achieve our goals, the far-field hydrodynamic contributions (including the hydrodynamic lift) due to the channel confinement need to be derived without imposing the strict-pinning constraint of Ref. \cite{GolestanianFocusing} on the spheroidal orientation. We provide these derivations in the Supplementary Material (SM) and use them within approach (i) and also within our linear stability analysis and determination of the eigenvalues associated with fixed points of the deterministic spheroidal dynamics in approach (ii). The strict-pinning form of our generalized far-field expressions are used within approach (iii) and the nonlinear stability analysis of higher-order fixed points in approach (ii).  

We show that the translational noise (particle diffusivity) across the channel width can specifically impart unexpected effects in the strong-field regime which is of main interest in this work. In particular, under a transverse (longitudinal) field, where prolate (oblate) spheroids are focused at the channel center, we find that the translational noise causes a peculiar flat-top subGaussian density profile for the spheroids across the channel width. It also triggers a counterintuitive {\em field-induced defocusing} upon amplifying the applied field beyond a certain threshold; that is, strengthening the field leads to continual broadening of the spheroidal layer well within the strong-field regime. The defocusing  follows because (1) the dynamical stability of centered focusing turns out to be of {\em nonlinear} order (i.e., the first and second derivatives of the lift  relative to lateral position within the channel vanish at its center); and (2) the {\em higher-order stability} progressively weakens (i.e., the nonvanishing third derivative also tends to zero) as the field is amplified. These features make the deterministic focusing of spheroids at the channel center prone to significant noise-induced alterations. They also stand at odds with the linear stability with a finite first derivative which is reported for the corresponding case in Refs. \cite{GolestanianFocusing,Matsunaga2018}. 

When the applied field is tilted relative to the flow, the spheroids are focused at an off-centered latitude within the channel,  reflected by a localized peak in their steady-state density profile across  the channel width. While our results in this case partly corroborate those in Ref. \cite{GolestanianFocusing}, they unravel the subtle nature of the hydrodynamic lift at the channel center which we find to be a {\em half-stable} fixed   point under a tilted field. The half-stability engenders  a broad noise-induced `shoulder' in the spheroidal density profiles. The shoulder stretches from the focusing  latitude to the channel center. It can be strong enough to undermine the  off-centered focusing peak. Since an optimal focusing peak  (e.g., for particle separation purposes) is desired to be sharp and isolated, we map out comprehensive `phase' diagrams based on the field strength and tilt angle to identify various regimes of magnetic focusing and, in particular, to distinguish the regime of optimal focusing from that with noise-induced density shoulders. 

We show that the reduced probabilistic theory (approach iii) can provide a direct link between the deterministic fixed points and the full probabilistic results by means of a {\em virtual lift potential} defined along the pinning curve. To our knowledge, the full probabilistic approach and also the deterministic analysis as we report here have not been considered in the present context before. A quadratic lift potential has however been shown to exist in the vicinity of the off-centered focusing latitude via a Langevin equation  in Ref. \cite{GolestanianFocusing}. Our derivation of the reduced theory shows that this insightful observation can formally be extended to the whole pinning curve, leading to useful (albeit semiquantitative) analytical predictions.  

We discuss our model and methods in Sections \ref{sec:Model} and \ref{sec:prob}, our results for prolate spheroids in transverse field  in Section \ref{sec:centered_focusing} and for prolate/oblate spheroids in tilted fields in Sections \ref{sec:offcentered_focusing}-\ref{sec:phase_diagram}. The implications of our results for shape-based sorting of spheroids are given in Section \ref{sec:separation}. Details of our generalized far-field calculations and deterministic analysis appear in the SM and details of our reduced probabilistic theory appear in Appendix \ref{app:U_p}.

%%%%%%%%%%%%%%%
\begin{figure}[t!]
\begin{center}
\includegraphics[width=0.8\linewidth]{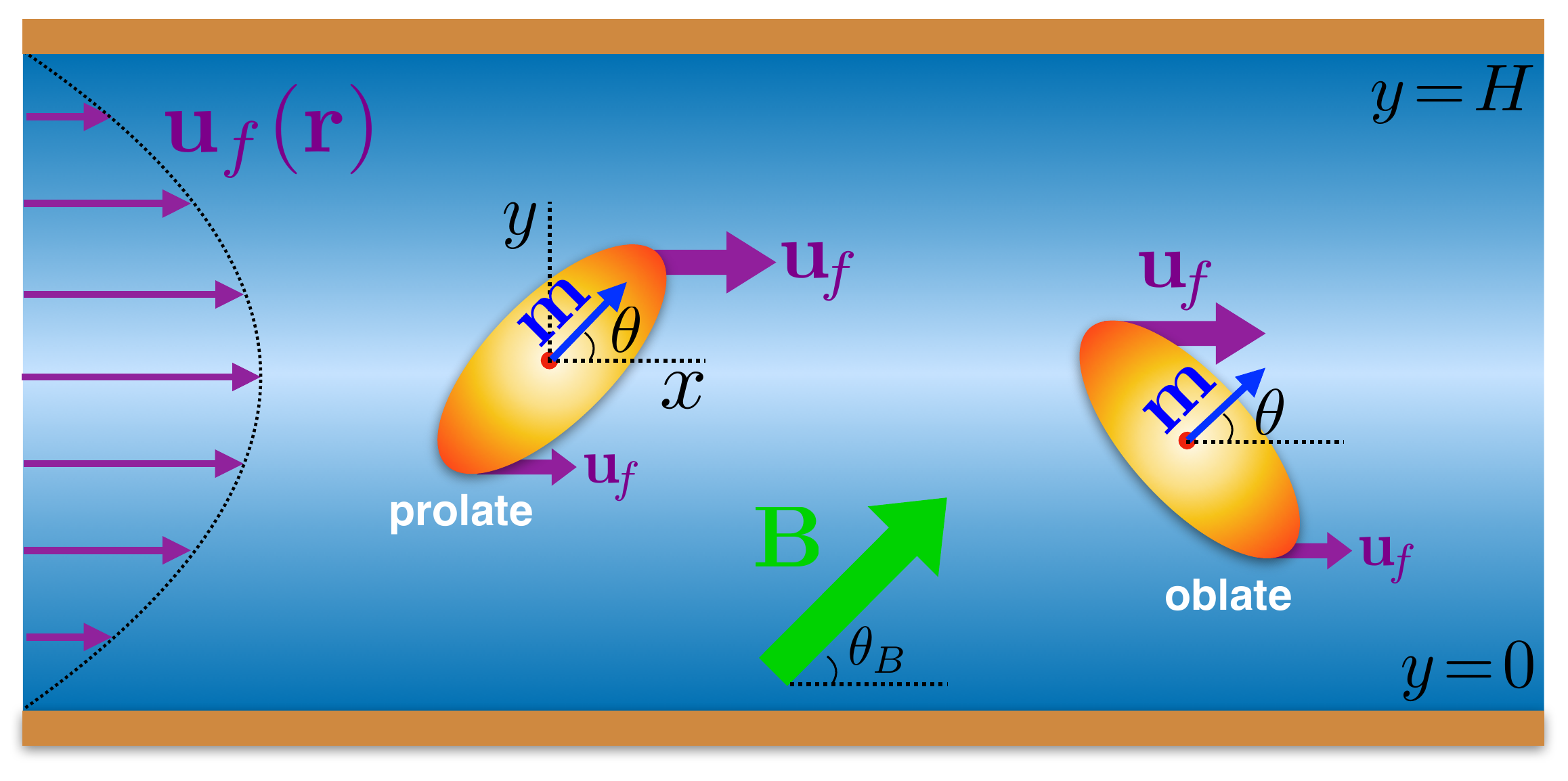} 
\vskip-4mm
\caption{Schematic view of magnetic spheroids (not drawn to scale) with permanent dipole moment $\mathbf{m}$ along their symmetry axis in a channel of width $H$ subject to plane Poiseuille flow $\mathbf{u}_\mathrm{f}({\mathbf r})$ and uniform magnetic field $\mathbf{B}$ of tilt angle $\theta_B$. 
}
\label{fig:Fig1}
\end{center}
\end{figure}

%%%%%%%%%%%%%%%%%%%%%%%%%%%%%%%%%%%%%
%%%%%%%%%%%%%%%%%%%%%%%%%%%%%%%%%%%%%
\section{Model}
\label{sec:Model}

The magnetic particles in our model are taken as rigid, passive and neutrally buoyant  colloidal spheroids within a long planar channel of width $H$ and rigid no-slip walls at $y\!=\!0$ and $H$; see Fig. \ref{fig:Fig1}. The channel is subjected to a stationary plane Poiseuille flow $\mathbf{u}_\mathrm{f}({\mathbf r}) \!=\! u_\mathrm{f}(y) \hat{\mathbf x}$ where  $\mathbf{r}\!=\!(x,y,z)$ is the position vector within the channel, $\hat{\mathbf x}$  is the unit vector in $x$-direction, and  $u_\mathrm{f}(y) \!=\! \dot{\gamma}y \left(1\!-\! {y}/{H}\right)$ is the velocity profile in $y$-direction with maximum shear rate $\dot \gamma\!=\!4 U_{\mathrm{max}}/{H}$ and fluid velocity  $U_{\mathrm{max}}\!>\! 0$. The impermeability of channel walls  is modeled via a harmonic steric  potential which creates repulsive transverse forces when the spheroids  come into contact with them; see Appendix \ref{app:u_st}.  A static uniform magnetic field $\mathbf{B}$ is applied within the $x\!-\!y$ plane; i.e., $\mathbf{B}\cdot \hat{\mathbf z}\!=\!0$ where $\hat{\mathbf z}\!=\!\hat{\mathbf x}\times\hat{\mathbf y}$ in the out-of-plane unit vector. The field strength and tilt angle (relative to the flow direction) are  $B\!=\!|\mathbf{B}|$ and $\theta_B$. 

The spheroids are assumed to be of equal volume ${\mathcal V}(\alpha)$ and aspect ratio $\alpha \!=\! a/b$, where $a$ is half-length of the spheroidal axis of symmetry, i.e., the major (minor) body axis for  prolate (oblate) shapes $\alpha\!>\!1$ ($\alpha\!<\!1$),  and $b$ is half-length of the other two axes. The spheroids carry permanent magnetic dipole moments $\mathbf{m}\!=\! m \hat{\mathbf d}$ ($m\!\geq\!0$) along their symmetry axis identified by the orientation unit vector $\hat{\mathbf d}$. Different shapes will be studied at constant volume ${\mathcal V}(\alpha) \!=\!{\mathcal V}_0$, or {\em effective radius} $R_{\mathrm{eff}}\!=\! \left(3{\mathcal V}_0 /4 \pi\right)^{1/3}$ (radius of {\em reference sphere}, $\alpha\!=\!1$)  and, hence, constant dipole moment   \cite{GolestanianFocusing,Zhou2017,ZhouPRA2017}. Since the  shear and field tend to orient the symmetry axis in the $x\!-\!y$   plane, $\hat{\mathbf d}$ is assumed to vary in-plane at strong enough shear/field strengths  \cite{Zhou2017,ZhouPRA2017,Almog1995,GolestanianFocusing}. We can thus use the 2D parametrization $\hat{\mathbf d}\!=\!(\cos \theta, \sin \theta)$ with   $\theta$  being the orientation angle relative to the $x$-axis.  

%%%%%%%%%%%%%%%%%%%%%%%%%%%%%%%%%%%%%
%%%%%%%%%%%%%%%%%%%%%%%%%%%%%%%%%%%%%
\section{Continuum probabilistic formulation}
\label{sec:prob} 

We assume that the translational and rotational dynamics of spheroids within the channel  are governed by deterministic forces and torques that result from the imposed shear, the external magnetic field, the spheroid-wall  steric and hydrodynamic interactions, and also the Brownian (thermal) ambient noise. Such a description is expected to hold for sufficiently dilute suspensions, where interparticle forces are negligible, or for individual spheroids advecting through the channel. The system properties can then be studied using a continuum probabilistic approach based on a noninteracting Smoluchowski equation that governs the joint position-orientation probability distribution function (PDF) $\Psi({\mathbf{r}}; {\hat{\mathbf d}})$ of spheroids with center position $\mathbf{r}$ and axial orientation ${\hat{\mathbf d}}$. In the steady state, the Smoluchowski equation reads   \cite{doiedwards,Dhont}
\begin{equation}
{\nabla}_{\mathbf{r}} \cdot \left( \mathbf{v}\,\Psi \right) +  {\mathcal R}_{\hat{\mathbf d}}\cdot \left(\boldsymbol{\omega}\,\Psi \right) = \nabla_{\mathbf{r}}\cdot{\mathbb D}_\mathrm{T}\cdot \nabla_{\mathbf{r}} \Psi + D_\mathrm{R}\, {\mathcal R}_{\hat{\mathbf d}}^{\,2}\, \Psi, 
\label{eq:smoluchowski_main}
\end{equation}
where ${\mathcal R}_{\hat{\mathbf d}}\!=\!{\hat{\mathbf d}}\times \nabla_{\hat{\mathbf d}}$ is the rotation operator, $\nabla_{\hat{\mathbf d}}$ is the unconstrained Cartesian gradient relative to  ${\hat{\mathbf d}}$, and $\mathbf{v}\!=\!\mathbf{v}(\mathbf{r}; {\hat{\mathbf d}}) $ and $\boldsymbol{\omega}\!=\!\boldsymbol{\omega}(\mathbf{r}; {\hat{\mathbf d}}) $ are the {\em net} deterministic translational and angular flux velocities of spheroids. 

In Eq. \eqref{eq:smoluchowski_main}, ${\mathbb D}_\mathrm{T}(\alpha)$ is the second-rank translational diffusivity tensor and $D_\mathrm{R}(\alpha)$ is the in-plane rotational diffusivity  due to translational and rotational Brownian noises, respectively. The translational diffusivity tensor is written as 
${\mathbb D}_\mathrm{T}(\alpha) \!=\! D_\mypara(\alpha)\, \hat{\mathbf d}\hat{\mathbf d} +  D_\myperp(\alpha) \big(\mathbb{I} - \hat{\mathbf d}\hat{\mathbf d}\big)$ where $\mathbb{I}$ is the identity tensor and ${D}_{\mypara}(\alpha)$ and ${D}_{\myperp}(\alpha)$ are the {\em longitudinal} and {\em transverse} diffusivities in parallel and perpendicular directions relative to the spheroidal symmetry axis. The diffusivities are assumed to follow the Stokes-Einstein relations  for no-slip spheroids. They are expressed using spheroidal shape functions  \cite{Perrin,koenig,happelbook,MRSh1} $\Delta_{\mypara,\perp} (\alpha)\!=\!{D}_{\mypara,\perp}(\alpha)/D_{0\mathrm{T}}$ and  $ \Delta_\mathrm{R}(\alpha) \!=\!D_\mathrm{R}(\alpha)/ D_{0\mathrm{R}}$ whose explicit forms are reproduced in Section 1, SM. Here, $D_{0\mathrm{T}}\!=\! k_{\mathrm{B}} T / (6\pi\eta R_{\mathrm{eff}})$  and $D_{0\mathrm{R}}\!=\! k_{\mathrm{B}} T / (8\pi\eta R_{\mathrm{eff}}^3)$ are translational and rotational diffusivities  of the reference sphere, $\eta$ is the fluid viscosity and  $k_{\mathrm{B}} T$ is the thermal energy. Also, ${\mathbb D}_\mathrm{T}(\alpha)$ and $D_\mathrm{R}(\alpha)$ express {\em bulk} diffusivities of free spheroids, as  the confinement effects due to spheroid-wall  steric and hydrodynamic interactions  are explicitly included via deterministic fluxes that we discuss next.  

%%%%%%%%%%%%%%%%%%%%%%%%%%%%%%%%%%%%%
\subsection{Translational flux velocities}
\label{subsec:transl_flux}

The net deterministic translational  flux velocity of the spheroids can be written as the sum of three terms 
\begin{equation}
\mathbf{v}(\mathbf{r}; {\hat{\mathbf d}}) =  \mathbf{u}_\mathrm{f}(\mathbf{r})+{\mathbf u}^{(\mathrm{st})}(\mathbf{r}; {\hat{\mathbf d}})+{\mathbf u}^{(\mathrm{im})}(\mathbf{r}; {\hat{\mathbf d}}), 
\label{eq:v}
\end{equation} 
where $\mathbf{u}_\mathrm{f}$ is the advective flux, and 
${\mathbf u}^{(\mathrm{st})}$ and  ${\mathbf u}^{(\mathrm{im})}$  are the translational flux velocities due to spheroid-wall  steric and  hydrodynamic interactions, respectively. The steric term ${\mathbf u}^{(\mathrm{st})}$ follows from the wall potential; see Eqs. \eqref{eq:u_st_main} and \eqref{eq:u_st}, Appendix \ref{app:u_st}. The hydrodynamic term ${\mathbf u}^{(\mathrm{im})}$ is derived using the singularity image method of Blake \cite{Blake1971} by approximating the far-field low-Reynolds-number flow around a no-slip spheroid of center position $\mathbf{r}$ and orientation ${\hat{\mathbf d}}$ with that of a second-order stresslet tensor $\mathbb{S}(\mathbf{r}; {\hat{\mathbf d}})$ \cite{kim_microhydrodynamics}. Relative to a given wall, and using  Einstein summation, the  components of ${\mathbf u}^{(\mathrm{im})}$ at position $\mathbf{r}'$ read
 \begin{equation}
u_i^{(\mathrm{im})}(\mathbf{r}'; {\hat{\mathbf d}}) = - \frac{1}{8\pi\eta} K_{ijk}^{(\mathrm{im})}(\mathbf{r}',\mathbf{r}) S_{jk}(\mathbf{r}; {\hat{\mathbf d}}),   
\label{eq:u_im_main}
\end{equation}
where $S_{jk}$  and $K_{ijk}^{(\mathrm{im})}$ are the components of $\mathbb{S}(\mathbf{r}; {\hat{\mathbf d}})$ and the third-order hydrodynamic Green function tensor $\mathbb{K}^{(\mathrm{im})}(\mathbf{r}',\mathbf{r})$ and $i, j, k$ denote Cartesian coordinates which we show  interchangeably by $\{x, y, z\}$ or $\{1,2,3\}$. The term ${\mathbf u}^{(\mathrm{im})}(\mathbf{r}; {\hat{\mathbf d}})$ thus gives the flux velocity from hydrodynamic interactions of a spheroid with its wall-induced singularity images. It  involves  the equal-point Green function (self-mobility tensor) $\mathbb{K}^{(\mathrm{im})}(\mathbf{r},\mathbf{r})$ which we use on the first-image level from Refs. \cite{smart1991,MatsunagaPRE}. 

The explicit forms of $\mathbb{S}$ and ${\mathbf u}^{(\mathrm{im})}$ are derived in Section 2 of the SM for {\em arbitrary} prolate/oblate aspect ratio, field strength and tilt angle and, thus, extend the expressions given in Ref.  \cite{GolestanianFocusing} for strictly pinned prolate spheroids. Such an extension is essential to our later linearized stability analysis and also our probabilistic analysis where field-induced pinning is not imposed but obtained as a result. 

%%%%%%%%%%%%%%%%%%%%%%%%%%%%%%%%%%%%%
\subsection{Angular flux velocities}
\label{subsec:angular_flux}

The net deterministic angular  flux velocity of the spheroids  can also  be written as the sum of three  terms 
\begin{equation}
\boldsymbol{\omega}(\mathbf{r}; {\hat{\mathbf d}}) =  \boldsymbol{\omega}_\mathrm{f}(\mathbf{r}; {\hat{\mathbf d}})+\boldsymbol{\omega}_\mathrm{ext}({\hat{\mathbf d}})+\boldsymbol{\omega}^{(\mathrm{im})}(\mathbf{r}; {\hat{\mathbf d}}), 
\label{eq:w_tot_main}
\end{equation}
where  $ \boldsymbol{\omega}_\mathrm{f}\!=\!  \omega_\mathrm{f}(y, \theta) \,\hat{\mathbf z}$, $\boldsymbol{\omega}_\mathrm{ext}\!=\!\omega_\mathrm{ext}(\theta) \,\hat{\mathbf z}$ and $\boldsymbol{\omega}^{(\mathrm{im})}\!=\!{\omega}^{(\mathrm{im})}(y, \theta) \,\hat{\mathbf z}$  are the shear, field and wall-induced hydrodynamic angular velocities, respectively. The shear and field terms in the signed magnitude of $\boldsymbol{\omega}$, i.e., ${\omega}\!=\!{\omega}_\mathrm{f}+{\omega}_\mathrm{ext}+{\omega}^{(\mathrm{im})}$,  standardly follow as (Section 2.3, SM)
\begin{eqnarray} 
&&{\omega}_\mathrm{f}(y, \theta)=  \frac{\dot\gamma }{2} \left(1-\frac{2y}{H}\right) \left(\beta(\alpha) \cos2\theta-1 \right), 
\label{eq:w_f} \\
&&{\omega}_\mathrm{ext}(\theta) = \frac{D_\mathrm{R}(\alpha)}{k_\mathrm{B} T} mB \sin\left(\theta_B-\theta\right). 
\label{eq:w_ext}
\end{eqnarray} 
Here, $\beta(\alpha)\!=\! (\alpha^2 - 1)/(\alpha^2+1)$ is the Bretherton number \cite{Bretherton1962}. For prolate/oblate spheroids, one has $0\!<\!\pm\beta(\alpha)\!<\!1$. 

The image term ${\omega}^{(\mathrm{im})}$ is derived in Section 2.5 of the SM and is (due to its  lengthy expression) given in rescaled form in Eq. \eqref{eq:w_im}. While this term has not been considered before and is included in our probabilistic analysis, it turns out to be small and of negligible impact on our results relative to the other angular terms (Appendix \ref{app:w_im}). 
 
%%%%%%%%%%%%%%%%%%%%%%%%%%%%%%%%%%%%%
\subsection{Dimensionless representation}
\label{subsec:nondim}

Due to translational symmetry  in $x$-direction, the solutions of Eq. \eqref{eq:smoluchowski_main} admit the form  $\Psi\!=\!\Psi(y, \theta)$. The latitude-orientation  plane $y-\theta$ is then taken as the reduced coordinate space. Also, due to fluid   incompressibility, %$\nabla \!\cdot\! \mathbf{u}_\mathrm{f}({\mathbf r})\!=\!0$, 
$\mathbf{u}_\mathrm{f}$ drops out of the l.h.s. of Eq. \eqref{eq:smoluchowski_main} and only the $y$-components of ${\mathbf u}^{(\mathrm{st})}$ and  ${\mathbf u}^{(\mathrm{im})}$ prevail,  combining into the net transverse velocity $v_y\!=\!  u_y^{(\mathrm{st})} \!+\! u_y^{(\mathrm{im})}$ where $u^{(\mathrm{im})}_y$ is the {\em hydrodynamic lift}. We rescale the length and time units with  $R_\mathrm{eff}$ and the  inverse rotational diffusivity $D_{0\mathrm{R}}^{-1}$, respectively, leading to the rescaling of lateral coordinate and  flux velocities as $\tilde y \!=\!  y/R_{\mathrm{eff}}$, $\tilde v_y(\tilde y, \theta) \!=\! v_y(R_{\mathrm{eff}}\,\tilde y, \theta) / \!\left(R_{\mathrm{eff}} D_{0\mathrm{R}}\right)$  and $\tilde \omega(\tilde y, \theta) \!=\! \omega(R_{\mathrm{eff}}\,\tilde y, \theta) / D_{0\mathrm{R}}$. The dimensionless parameter space is thus spanned by the aspect ratio $\alpha$, tilt angle of the field $\theta_B$, rescaled channel width $\tilde H$,  flow P\'eclet number (rescaled shear strength) $\mathrm{Pe}_\mathrm{f}$, and the magnetic coupling parameter (rescaled field strength) $\chi$, where 
\begin{equation}
 \tilde H = \frac{H}{R_{\mathrm{eff}}},\,\,\,\, \mathrm{Pe}_\mathrm{f}= \frac{\dot\gamma}{D_{0\mathrm{R}}}\,\,\,\, {\mathrm{and}}\,\,\,\, \chi = \frac{mB }{k_\mathrm{B} T}.  
 \label{eq:dimless_para}
\end{equation}

We obtain the rescaled forms of the net transverse velocity,  the hydrodynamic lift and the net angular velocity (see Appendices \ref{app:u_st} and \ref{app:w_im} and Section 2 of the SM) as  
\begin{widetext}
\begin{eqnarray}
&&\tilde v_y(\tilde y, \theta) =  \tilde u_y^{(\mathrm{st})}(\tilde y, \theta) + \tilde u_y^{(\mathrm{im})}(\tilde y, \theta),  
\label{eq:v_y_tot}\\
&&\tilde u^{(\mathrm{im})}_y(\tilde y, \theta) =
 - \zeta(\alpha)\! \left(\frac{1}{\tilde y^2}\!-\!\frac{1}{(\tilde H-\tilde y)^2}\right)\sin 2\theta 
 \Bigg\{\frac{\!15  \mathrm{Pe}_\mathrm{f}}{64} \!  \left(1\!-\! \frac{2\tilde y}{\tilde H}\right)\!
 \bigg[\!\left(-X^\mathrm{M}+2Y^\mathrm{M}\!-\!Z^\mathrm{M}\right)\cos^2\theta 
 + \left(2X^\mathrm{M}\!-\!2Y^\mathrm{M}\right)\sin^2\theta \bigg]
 \nonumber\\
&&\hskip6.9cm
 -\frac{9\beta(\alpha)}{16}\bigg[\frac{1}{2} Y^\mathrm{H} \mathrm{Pe}_\mathrm{f}  \left(1\!-\! \frac{2\tilde y}{\tilde H}\right)\cos 2\theta  + \frac{\chi}{ \zeta(\alpha)} \sin\left(\theta_B-\theta\right) \bigg]\! \Bigg\}, 
\label{eq:u_im}
\\
&&\tilde\omega(\tilde y, \theta) =\frac{\mathrm{Pe}_\mathrm{f}}{2}\!\left(1\!-\! \frac{2\tilde y}{\tilde H}\right) \left(\beta(\alpha) \cos2\theta\!-\!1 \right) + \chi\Delta_\mathrm{R}(\alpha)  \sin\left(\theta_B-\theta\right) + \tilde{\omega}^{(\mathrm{im})}(\tilde y, \theta), 
 \label{eq:w_tot}
\end{eqnarray}
\end{widetext}
{\noindent}where $X^\mathrm{M}, Y^\mathrm{M}, Z^\mathrm{M}$ and $Y^\mathrm{H}$ are shape-dependent  resistance functions   (see Table 1 of the SM),  and $\zeta(\alpha)\!=\!\alpha^2$  and $\alpha^{-1}$ for prolate and oblate spheroids, respectively   \cite{kim_microhydrodynamics}.   

Our expression in Eq. \eqref{eq:u_im} recovers the standard zero-lift result, $u^{(\mathrm{im})}_y\!=\!0$, for  nonmagnetic spheres ($\chi\!=\!0$; $\alpha\!=\!1$, $X^\mathrm{M} \!=\! Y^\mathrm{M} \!=\! Z^\mathrm{M} \!=\!1$, $Y^\mathrm{H} \!=\! 0$) due to the absence of inertia  in the present setting  \cite{Saffman1956,Bretherton1962,Cox1968,Ho1974,Vasseur1976,Yahiaoui2010,Bureau2022}. It also shows that the lift is consistently zero on  (ferro)magnetic spheres ($\chi\!>\!0$) under a  uniform field of arbitrary tilt angle. For this reason, spheres are  excluded from our later discussion given their lack of  magnetic focusing in the present context. 

%%%%%%%%%%%%%%%%%%%%%%%%%%%%%%%%%%%%%
\subsection{Rescaled Smoluchowski equation}
\label{subsec:nondim_Smol}

The rescaled form of the Smoluchowski equation \eqref{eq:smoluchowski_main} for the joint latitude-orientation PDF of spheroids reads 
\begin{eqnarray}
&&\frac{\partial }{\partial \tilde y} \big[ \tilde v_y(\tilde y, \theta) \tilde \Psi  \big]+ \frac{\partial }{\partial \theta} \big[ \tilde \omega(\tilde y, \theta)  \tilde\Psi \big]=
\label{eq:smoluchowski}\\
&&\qquad\quad \frac{4}{3}\big(\Delta_+(\alpha) - \Delta_-(\alpha) \cos 2\theta \big) \frac{\partial^2 \tilde \Psi }{\partial \tilde y^2} + \Delta_\mathrm{R}(\alpha) \frac{\partial^2 \tilde \Psi }{\partial \theta^2},   
\nonumber
\end{eqnarray}
where $\Delta_\pm \!=\! (\Delta_{\mypara}\! \pm \!\Delta_{\myperp} )/2$ combine the shape functions $\Delta_{\mypara,\perp}$. With no loss of generality, we choose the computational domain to solve Eq. \eqref{eq:smoluchowski} as $ y\!\in\! [0, H]$ and $\theta\!\in\! [{\theta_B\!-\!\pi},{\theta_B\!+\!\pi})$  (we may later depict the PDF over other visually suitable intervals of $\theta$). The rescaled PDF is defined  as $\tilde \Psi(\tilde y, \theta)\!=\!  R_{\mathrm{eff}}\Psi(R_{\mathrm{eff}}\,\tilde y, \theta)/{\bar n}$ where ${\bar n}\!=\!\int_0^{H}\!\int_{\theta_B-\pi}^{\theta_B+\pi}\,\Psi(y, \theta)\,{\mathrm{d}}\theta\,{\mathrm{d}} y$ gives the mean number of particles  per unit $x\!-\!z$ area in a 3D realization of the system; hence, the normalization $\int_0^{\tilde H}\!\int_{\theta_B-\pi}^{\theta_B+\pi}\, \tilde \Psi(\tilde y, \theta)\,{\mathrm{d}}\theta\,{\mathrm{d}}\tilde y\!=\!1$.  

Equation \eqref{eq:smoluchowski} is solved numerically using finite-element methods and by imposing  periodic boundary conditions on $\theta$ and, for numerical convenience, no-flux boundary conditions on the channel walls  \cite{MRSh1,Nili}. We vary the system parameters in ranges that cover realistic values (Appendix \ref{app:parameters}). The tilt angle of applied field is restricted to the first quadrant $0\!\leq\!\theta_B \!\leq\! \pi/2$ to avoid redundant solutions, as solutions of Eq. \eqref{eq:smoluchowski} in other $\theta_B$-quadrants can be recovered using the  symmetry transformation $\{\theta_B \rightarrow \theta_B + \pi, \theta \rightarrow \theta + \pi\}$ \cite{Almog1995,Kumaran2020} and, as our numerical inspections for the key regimes of interest  here indicate, also using $ \{ \theta_B \rightarrow \pi - \theta_B, \theta \rightarrow \pi - \theta, \tilde{y} \rightarrow \tilde H - \tilde{y} \}$.  

%%%%%%%%%%%%%%%%%%%%%%%%%%%%%%%%%%%%%
\subsection{Analytical schemes complementing  numerical solutions}
\label{subsec:analytic_schemes} 

In our numerical implementation of Eq. \eqref{eq:smoluchowski}, i.e., the {\em full probabilistic approach}, we retain all flux velocities in Eqs. \eqref{eq:v_y_tot} and \eqref{eq:w_tot}. In the key regimes of magnetic focusing analyzed in this work   (Sections \ref{sec:centered_focusing} and \ref{sec:offcentered_focusing}), the spheroids are focused sufficiently away from the walls. This  renders the steric term $\tilde u_y^{(\mathrm{st})}$ irrelevant. As noted before, the wall-induced hydrodynamic angular velocity $\tilde{{\omega}}^{(\mathrm{im})}$  also stays negligible (Appendix \ref{app:w_im}). Discarding the above two terms enables approximate analytical schemes that can be used to further illuminate our numerical findings. Since we frequently allude to the predictions of these schemes in the upcoming sections, a summary of their scope and applicability will be in order:  
\begin{itemize}[leftmargin=13pt]  
\item[$\bullet$]{{\em Deterministic dynamical stability analysis}: This approach is based on  noise-free translational and rotational dynamical equations for the spheroidal motion,   
\begin{eqnarray}
&&\tilde{\dot y}= {\tilde v}_y(\tilde y, \theta)\simeq \tilde u^{(\mathrm{im})}_y(\tilde y, \theta),
\label{eq:dtm_dyn_eqs1} \\
&& \tilde{\dot\theta} = {\tilde \omega}(\tilde y, \theta) \simeq \tilde {\omega}_\mathrm{f}(\tilde y, \theta)+\tilde{\omega}_\mathrm{ext}(\theta),   
\label{eq:dtm_dyn_eqs2}
\end{eqnarray}
 where $\tilde{\dot\theta}\!=\!\hat{\boldsymbol{\theta}}\!\cdot\!(\tilde{\boldsymbol{\omega}}\!\times\!\hat{\mathbf{d}})$ and  $\hat{\boldsymbol{\theta}}$ is the counterclockwise polar unit vector. This approach is useful in predicting the boundaries of most, albeit {\em not} all, regimes of magnetic focusing that we later identify in this work. The said boundaries are determined based on the dynamical stability of  fixed points $(\tilde y_\ast, \theta_\ast )$ of Eqs. \eqref{eq:dtm_dyn_eqs1} and \eqref{eq:dtm_dyn_eqs2} which follow by setting $\tilde{\dot y}\!=\!\tilde{\dot\theta}\!=\!0$. The fixed points can be obtained from the intersections of {\em nullclines} that give the loci of coordinate-space points where either $\tilde {\dot y}$ or $\tilde {\dot\theta}$ vanishes \cite{Rasband1990,Strogatz2000,Nayfeh2008}.  According  to Eq. \eqref{eq:dtm_dyn_eqs1},  $\tilde{\dot y}\!=\!0$ yields the {\em zero-lift nullclines}. From Eq. \eqref{eq:u_im}, these follow as  the channel centerline  $\tilde y\!=\!\tilde H/2$ and the lines $\theta\!=\! n\pi/2$ for integer $n$. On the other hand, according to Eq. \eqref{eq:dtm_dyn_eqs2}, $\tilde{\dot \theta}\!=\!0$ yields the loci of coordinate-space points where the shear and field-induced torques counterbalance and produce zero net angular velocity. The ensuing nullclines include the orientationally stable (deterministic) {\em pinning curve} $\theta\!=\!\theta_\mathrm{p}(\tilde y)$ along which $\partial \tilde{\dot\theta}/\partial \theta\!<\!0$ and, hence, the spheroidal orientation is pinned    \cite{Almog1995,GolestanianFocusing,Matsunaga2018}. We shall elaborate on this later. 
 
\hskip10pt{}While deterministic equations have previously been used to study noise-free trajectories of prolate spheroids \cite{GolestanianFocusing,Matsunaga2018}, a comprehensive study to classify the phase-space dynamical stability of their fixed points has been missing. We present such an analysis for both prolate and  oblate spheroids in Section 3 of the SM and use its predictions in what follows to interpret the probabilistic solutions obtained from Eq. \eqref{eq:smoluchowski}. While the deterministic results are valuable, they exhibit important departures from the probabilistic ones especially at strong fields, contrasting the intuitive assumption that noise (particle diffusivity) can be neglected in the strong-field regime  \cite{GolestanianFocusing,Matsunaga2018}. 

\hskip10pt{}For future reference, the fixed point stability is standardly determined on the linearization level by the eigenvalues   $\lambda_y$ and $\lambda_\theta$ of the Jacobian matrix
\begin{equation}
{\mathbb J}  = \frac{\partial (\tilde{\dot y}, \tilde{\dot\theta})}{\partial (\tilde y, \theta)} \simeq  
\begin{bmatrix}
 \frac{\partial \tilde u^{(\mathrm{im})}_y}{\partial \tilde y} & \frac{\partial \tilde u^{(\mathrm{im})}_y}{\partial \theta} \vspace{1mm}
 \\ 
\frac{\partial \tilde\omega}{\partial \tilde y } & \frac{\partial \tilde\omega}{\partial \theta }
\end{bmatrix}_{\tilde y_\ast,\theta_\ast}\!\!. 
 \label{eq:Jacobian}
\end{equation}
When $\lambda_y$ and/or $\lambda_\theta$ vanish, we use a higher-order (nonlinear) analysis \cite{Rasband1990,Strogatz2000,Nayfeh2008} to determine the true nature of the fixed point. For the most part, $\lambda_y$ and $\lambda_\theta$ turn out to be real-valued, and $\lambda_\theta<0$ due to the aforesaid orientational  pinning. We emphasize that our stability analysis based on Eq. \eqref{eq:Jacobian} relies on the hydrodynamic lift in its generalized form that we give in  Eq. \eqref{eq:u_im} rather than its constrained strict-pinning form    \cite{GolestanianFocusing}.}

\item[$\bullet$]{{\em Reduced probabilistic theory}: This approach is systematically derived in Appendix \ref{app:U_p} as a limiting approximation to Eq. \eqref{eq:smoluchowski} under an orientational pinning constraint at strong fields. It leads to a {\em reduced}  (1D) Smoluchowski equation that analytically predicts the PDF of pinned spheroids based on a {\em virtual lift potential} over the pinning curve. Despite its  semiquantitative nature, the reduced probabilistic theory gives  useful  insights into subtleties of magnetic focusing that can otherwise be masked by the intricacies of the full (2D) probabilistic solutions as we shall see later. Further details on this reduced approach, along with a brief comparison with a previously introduced quadratic lift potential  \cite{GolestanianFocusing} are given in Appendix \ref{app:U_p}.}
\end{itemize}
  
%%%%%%%%%%%%%%%%%%%%%%%%%%%%%%%%%%%%%
%%%%%%%%%%%%%%%%%%%%%%%%%%%%%%%%%%%%%
\section{Centered focusing}
\label{sec:centered_focusing}

We begin the discussion of our results by considering the cases in which sheared magnetic spheroids are symmetrically focused by the applied field around the channel centerline $\tilde y\!=\!\tilde H/2$. We refer to this scenario as {\em centered focusing} and elucidate its characteristics   in this section. For prolate (oblate) spheroids, $\alpha\!>\!1$ ($\alpha\!<\!1$), centered focusing  arises under a transverse  (longitudinal)  field; e.g., when the tilt angle $\theta_B \!=\! \pi/2$ ($\theta_B \!=\! 0$). Given the qualitative similarities we find for centered focusing of prolate and oblate spheroids, we shall only report the probabilistic results for  prolate shapes  ($\alpha\!>\!1$, $\theta_B \!=\! \pi/2$) and return to oblate shapes in the case of tilted fields in Section \ref{sec:offcentered_focusing}. 

%%%%%%%%%%%%%%%%%%%%%%%%%%%%%%%%%%%%%
\subsection{Baseline behavior in the absence of field ($\chi=0$)}
\label{subsec:prolate_nofield}

\begin{figure}[t!]
\begin{center}
	\includegraphics[width=0.9\linewidth]{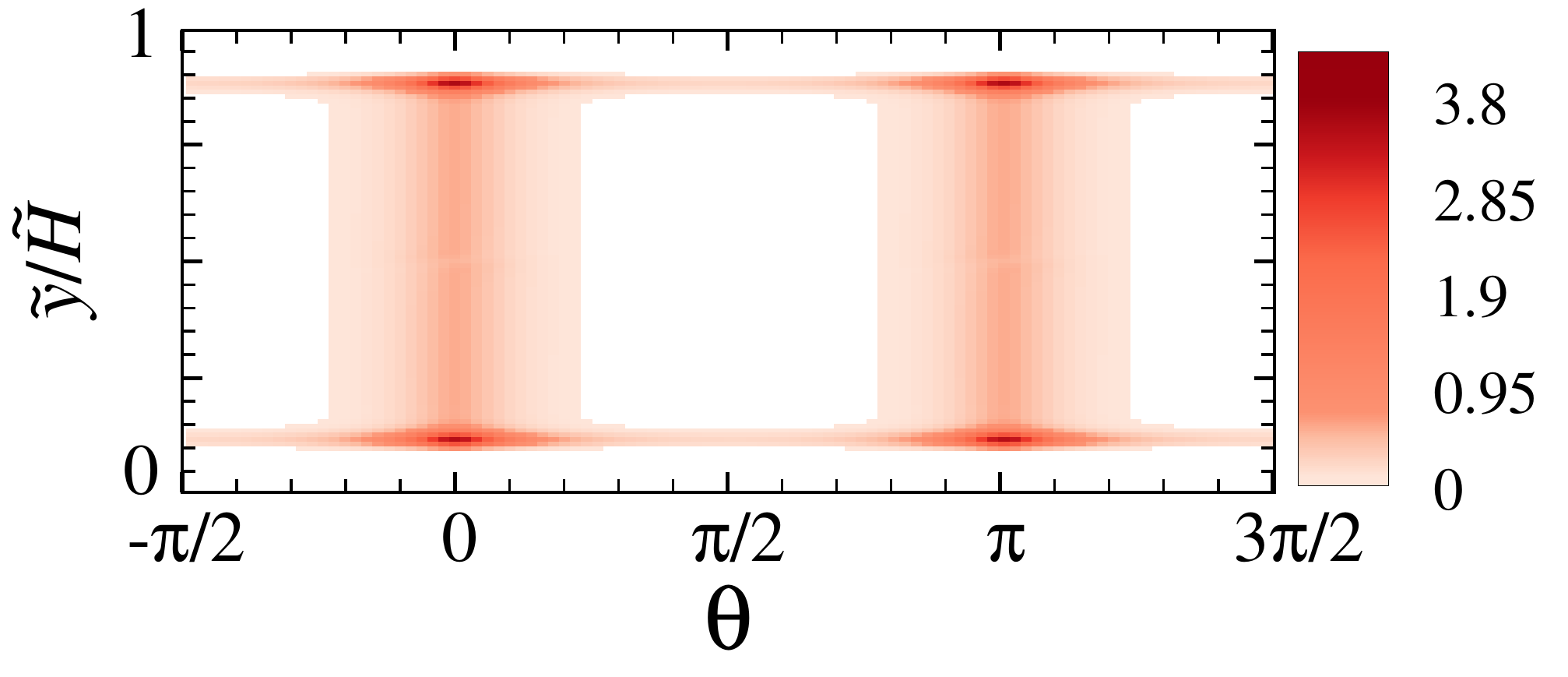} 
	\vskip-2mm
	\caption{Density map of the numerically obtained PDF, $\tilde{\Psi}(\tilde y, \theta)$, for prolate spheroids of aspect ratio $\alpha\!=\!4$ in a planar channel of rescaled width $\tilde H \!=\!20$ subjected to Poiseuille flow of P\'eclet number $\mathrm{Pe}_\mathrm{f} \!=\! 10^4$ and zero external field  ($\chi\!=\!0$). For clearer illustration, regions with  $\Psi (\tilde y,\theta) \!<\! 10^{-3} \Psi_{\mathrm{max}}$, where $\Psi_{\mathrm{max}}$ is the maximum value obtained within the shown domain, are cut out from the plot and appear as white areas. 
	}
\label{fig:Fig2}
\end{center}
\vskip-4mm
\end{figure}

Figure \ref{fig:Fig2} shows the typical form of the computed  PDF for prolate spheroids in the absence of a magnetic field (here with particle aspect ratio $\alpha\!=\!4$, rescaled  channel width $\tilde H \!=\!20$ and  flow P\'eclet number $\mathrm{Pe}_\mathrm{f} \!=\! 10^4$). For the sake of illustration, we shall hereafter depict the PDF in the left-handed $\theta-\tilde y$ plane. As seen, the zero-field PDF has two relatively broad (light orange) columns extending across much of the channel width at orientation angles  $\theta\!=\!0$ and $\pi$. These columns manifest the shear-induced Jeffery oscillation of the spheroidal orientation \cite{Jeffery} in a probabilistic setting. Maximal steady-state probability densities are thus produced expectedly at $\theta\!=\!0$ and $\pi$ where the spheroidal symmetry axis is aligned with the flow and the magnitude of  shear-induced angular velocity  \eqref{eq:w_f} is minimized  \cite{MRSh1,Nili}. For oblate spheroids, the symmetry axis will preferentially be orientated normal to the flow, creating similar columns at  $\theta\!=\! \mp\pi/2$  \cite{MRSh1}. Figure \ref{fig:Fig2} also shows that the PDF is  diminished in  narrow near-wall regions of width about  the closest-approach distance set by the steric wall potential (Appendix \ref{app:u_st}). Both the steric repulsion and the hydrodynamic lift in fact contribute to these near-wall depletion layers.  

%%%%%%%%%%%%%%%
%

%%%%%%%%%%%%%%%%%%%%%%%%%%%%%%%%%%%%%
\subsection{Partial pinning and modified Jeffery oscillations}
\label{subsec:partial_pinning}

\begin{figure}[t!]
\begin{center}
	\includegraphics[width=0.9\linewidth]{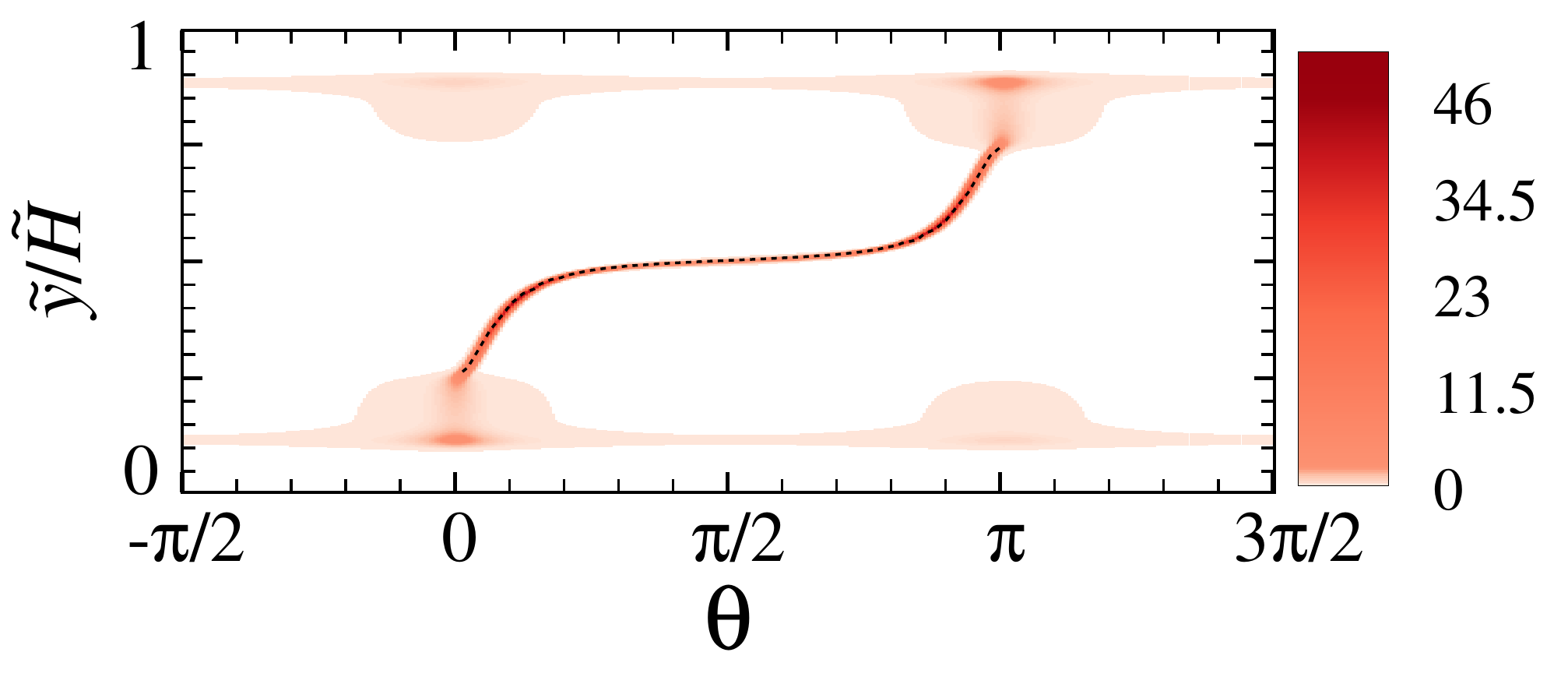}
	\vskip-2mm	
	\caption{Same as Fig. \ref{fig:Fig2} but here the system is subjected to  a transverse magnetic field  ($\theta_B\!=\!\pi/2$) of rescaled strength $\chi \!=\! 10^3$. The dotted black curve shows the pinning curve. 
	}
\label{fig:Fig3}
\end{center}
\vskip-4mm
\end{figure}

Under an applied field, the PDF displays significant morphological changes relative to the zero-field case. Figure \ref{fig:Fig3} depicts the typical form of the PDF at finite field strengths (up to a certain threshold as specified below). The PDF is heavily concentrated over a narrow band across the channel width, connecting two relatively diffuse regions around $\theta \!=\! 0$ ($\pi$) near the bottom (top) wall. 

The diffuse regions represent remnant subpopulations of spheroids undergoing field-modified Jeffery oscillations. Near the bottom wall, around the orientation angle $\theta\!=\!0$ ($\pi$), the field-induced angular velocity $\tilde \omega_\mathrm{ext}$ suppresses   (enhances) the oscillation rate of  spheroids from its zero-field shear-induced value set by $\tilde \omega_\mathrm{f}$  \cite{MRSh1}, leading to higher (lower) steady-state  probability densities as seen in the figure. The situation reverses near the top wall.  

The narrow band in Fig. \ref{fig:Fig3} indicates a distinct subpopulation of spheroids whose orientation is stabilized around a well-defined curve. As noted in Section \ref{subsec:analytic_schemes}, a pinning curve $\theta\!=\! \theta_\mathrm{p}(\tilde y)$ can be predicted by seeking the stable solutions of Eq. \eqref{eq:dtm_dyn_eqs2} for zero net deterministic angular velocity, $\tilde {\omega}_\mathrm{f}+\tilde{\omega}_\mathrm{ext}\!=\!0$. Since the deterministic pinning of sheared prolate (para/ferro)magnetic spheroids has been explored before  \cite{Zhou2017,ZhouPRA2017,Cao2017,Zhang2018,Zhang2018b,Sobecki2018,Sobecki2019,Sobecki2020,Kumaran2020,GolestanianFocusing,Matsunaga2018}, we proceed by giving only a brief summary of the results for the current model. More details can be found in the SM (Section 3.2). 

%%%%%%%%%%%%%%%

Using Eqs. \eqref{eq:w_f}, \eqref{eq:w_ext} and \eqref{eq:dtm_dyn_eqs2},  $\theta_\mathrm{p}(\tilde y)$ is found  to satisfy 
\begin{equation}
 \cos^2\! \theta_\mathrm{p}(\tilde y)\! +\!\frac{ \chi \Delta_\mathrm{R}(\alpha) / \beta(\alpha)}{\mathrm{Pe}_\mathrm{f}(1\!-\! 2\tilde y/\tilde H) }\sin\!\left(\theta_B \!-\! \theta_\mathrm{p}(\tilde y)\right) \!-\! \frac{1\!+\! \beta(\alpha)}{2\beta(\alpha)}\!=\!0.      
\label{eq:pinning}
\end{equation}
For the case under study in this section ($\alpha\!>\!1$, $\theta_B \!=\! \pi/2$), the stable or pinning solution of Eq. \eqref{eq:pinning} at a given latitude $\tilde y$ within the channel follows from a saddle-node bifurcation when $\chi$ is increased above a certain $\tilde y$-dependent threshold. The solutions are found only within a finite interval of width $\tilde w_\mathrm{p}$ around the centerline  $\tilde y\!=\! \tilde H/2$ and are shown by the (dotted black) pinning curve in Fig.  \ref{fig:Fig3}. As seen, the pinning curve closely overlays the computed narrow  (orange) probability-density band in the figure. This confirms that the pinning is consistently reproduced within the probabilistic approach.  

The width of pinning region follows from Eq. \eqref{eq:pinning} by noting that  the endpoints of pinning curve always stay on $\theta\!=\!0,\pi$ when $\tilde w_\mathrm{p}\!<\!\tilde H$ (see Fig. 1, SM). We thus find  
\begin{equation}
\frac{\tilde w_\mathrm{p}}{\tilde H} = \frac{\chi}{\mathrm{Pe}_\mathrm{f}} \frac{2\Delta_\mathrm{R}(\alpha)}{1-\beta(\alpha)}. 
\label{eq:w_p}
\end{equation}

As described, Fig. \ref{fig:Fig3} typifies a mixed regime of {\em partial pinning} where both unpinned and pinned spheroids are present within the channel. In fact, a pinning solution for Eq. \eqref{eq:pinning}  {\em always} exist for $\chi\!>\!0$ at $\tilde y\!=\!\tilde H/2$ where the shear rate from the Poiseuille flow identically vanishes; i.e., an arbitrary field can always   pin the spheroidal orientation at $\theta\!=\!\theta_B$ at  the centerline. This implies that the weak-field scenario (ideally with field-modified Jeffery solutions being present across the whole channel width  \cite{Zhou2017,ZhouPRA2017}) is not fully realized in the present model except for regions sufficiently close  to the channel walls. 
  
Before proceeding further, it is useful to note  that the pinning in probabilistic solutions is realized only up to a small angular width (thickness of the probability-density band in Fig. \ref{fig:Fig3}). Such an angular broadening is caused by the rotational noise. That is, while the field-induced angular velocity $\tilde{\omega}_\mathrm{ext}$  masks the angular Brownian flux  at all nonpinning angles,  its cancellation by $\tilde \omega_\mathrm{f}$ brings out the residual effect of noise  in causing the said broadening around the deterministic pinning curve. 
  
%%%%%%%%%%%%%%%%%%%%%%%%%%%%%%%%%%%%%
\subsection{Whole-channel pinning at strong fields ($\chi>\chi_\ast$)}
\label{subsec:whole_pinning}

\begin{figure}[t!]
\begin{center}
	\includegraphics[width=0.9\linewidth]{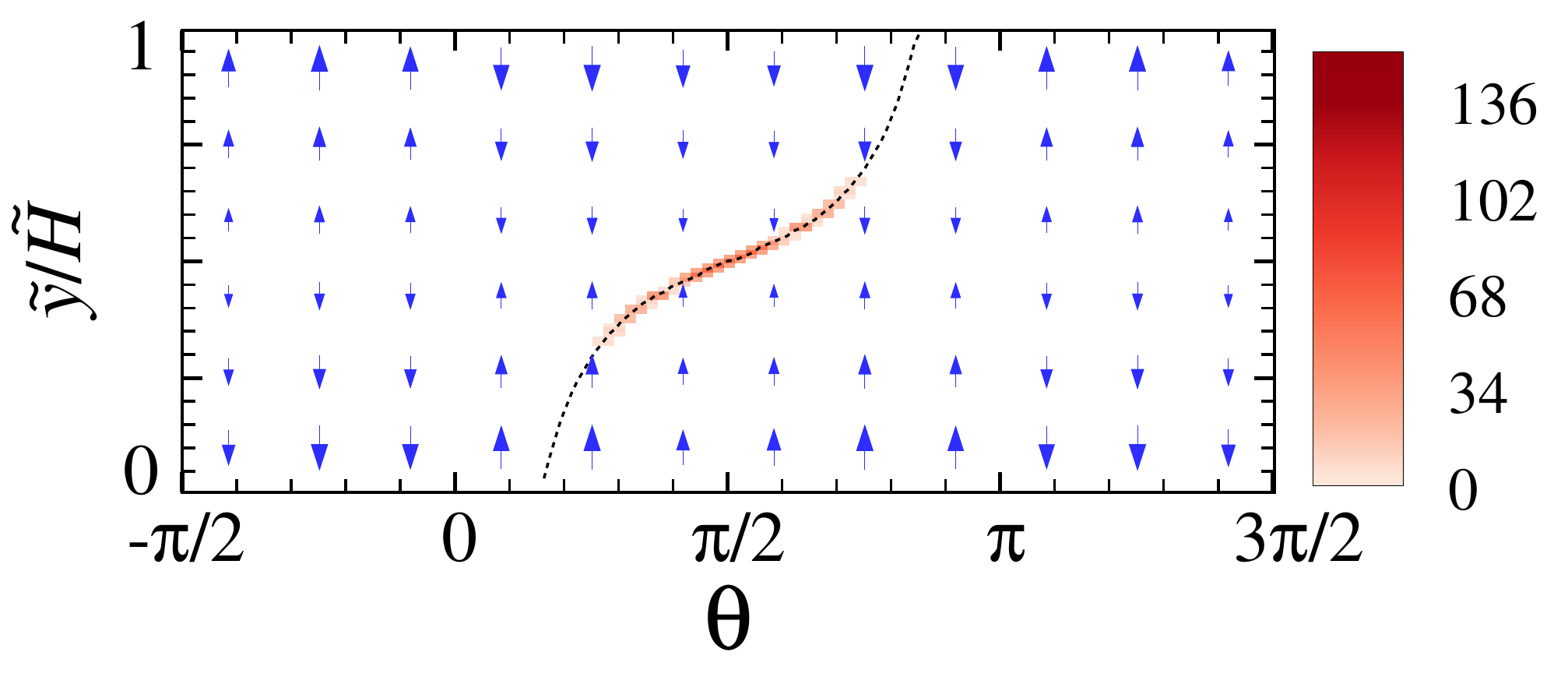}
	\vskip-2mm	
	\caption{Same as Fig. \ref{fig:Fig3} but under a strong field of rescaled strength $\chi \!=\! 10^4$.  Blue arrows show the hydrodynamic lift $\tilde u^{(\mathrm{im})}_y\left(\tilde y, \theta\right)$ whose magnitude  is  rescaled by   $ \lvert\tilde u^{(\mathrm{im})}_y\rvert^{7/8}$. 
	}
\label{fig:Fig4}
\end{center}
\vskip-4mm
\end{figure}

For sufficiently strong fields, the pinning curve can extend beyond  the channel width, $\tilde w_\mathrm{p}\!>\!\tilde H$, permitting no Jeffery orbits but only a single population of orientationally pinned spheroids anywhere within the channel  (compare  dotted black curves in Figs. \ref{fig:Fig3} and \ref{fig:Fig4}). For $\tilde w_\mathrm{p}=\tilde H$, Eq. \eqref{eq:w_p} gives the onset of this {\em strong-field regime} or, interchangeably, the regime of {\em whole-channel pinning}, as
\begin{equation}
\frac{\chi_\ast}{\mathrm{Pe}_\mathrm{f}} = \frac{1-\beta(\alpha)}{2\Delta_\mathrm{R}(\alpha)}     \qquad (\alpha\!>\!1,\, \theta_B\!=\!\pi/2).
\label{eq:chi_ast_app}
\end{equation}
This deterministic relation  accurately describes the onset of whole-channel pinning within the probabilistic approach as well. It is because, as $\chi$ is increased toward $\chi_\ast$ (data not shown), the pinning curve remains overlaid with the narrow PDF band as in Fig. \ref{fig:Fig3}. 

By generalizing it to tilted applied fields ($0\!<\!\theta_B\!<\!\pi/2$), the whole-channel pinning regime can be shown  (Section \ref{sec:phase_diagram}) to encompass, as its different subregimes, all regimes of magnetic focusing that we shall discuss later. 

%%%%%%%%%%%%%%%
%

%%%%%%%%%%%%%%%%%%%%%%%%%%%%%%%%%%%%%
\subsection{Centered focusing: Noise-induced aspects}
\label{subsec:centered_prob}

Figure \ref{fig:Fig4} also displays how the centered focusing in the strong-field regime is realized within the probabilistic setting. As expected, the PDF (orange band) is strongly accumulated on the (dotted black) pinning curve but, due to the wall-induced hydrodynamic lift (blue arrows in Fig. \ref{fig:Fig4}), it is squeezed or focused into a finite  interval of latitudes around the centerline rather than spreading over the entire curve. While the lift-induced origin of focusing in this regime is well established, our probabilistic results reveal salient effects with significant departures from the deterministic picture as we discuss next.

%%%%%%%%%%%%%%%%%%%%%%%%%%%%%%%%%%%%%
\subsubsection{NonGaussian flat-top density profiles}
\label{subsubsec:centered_vs_point}

According to the 2D analysis of Ref. \cite{GolestanianFocusing}, centered focusing of noise-free, prolate ferromagnetic spheroids in a transverse field is linked with a {\em stable} central fixed point in the lateral dynamics of spheroids. The fixed point results from a vanishing lift velocity with a finite negative {\em first} derivative in $\tilde y$-direction at the centerline $\tilde y\!=\!\tilde H/2$, where the pinning angle is set by the field as $\theta\!=\!\pi/2$. Even as a systematic analysis with the corresponding eigenvalues have not been provided \cite{GolestanianFocusing}, the reported negative first derivative directly implies an eigenvalue $\lambda_y\!<\!0$ using Eq. \eqref{eq:Jacobian}. Before examining the nature of the fixed point more carefully (Section \ref{subsubsec:higherFP_subG}), it is insightful to consider implications of a stable fixed point for the density profiles of spheroids across the channel width and make a direct comparison with our probabilistic results. 

In the deterministic setting, the density profile of spheroids is given by a central Dirac $\delta$-peak $\sim\! \delta(\tilde y\! - \!\tilde H/2)$ (Appendix \ref{app:deterministic_RT}). When the {\em translational} noise is introduced (note that the rotational noise is redundant due to orientational pinning), this idealized peak is  broadened in $\tilde y$-direction but the aforesaid negative first derivative (eigenvalue) creates a restoring linear `stiffness', making the peak to emerge only as a {\em Gaussian}, preserving the centerline as the most-probable latitude in the channel.   

%%%%%%%%%%%%%%%%
\begin{figure}[t!]
\begin{center}
	\includegraphics[width=0.75\linewidth]{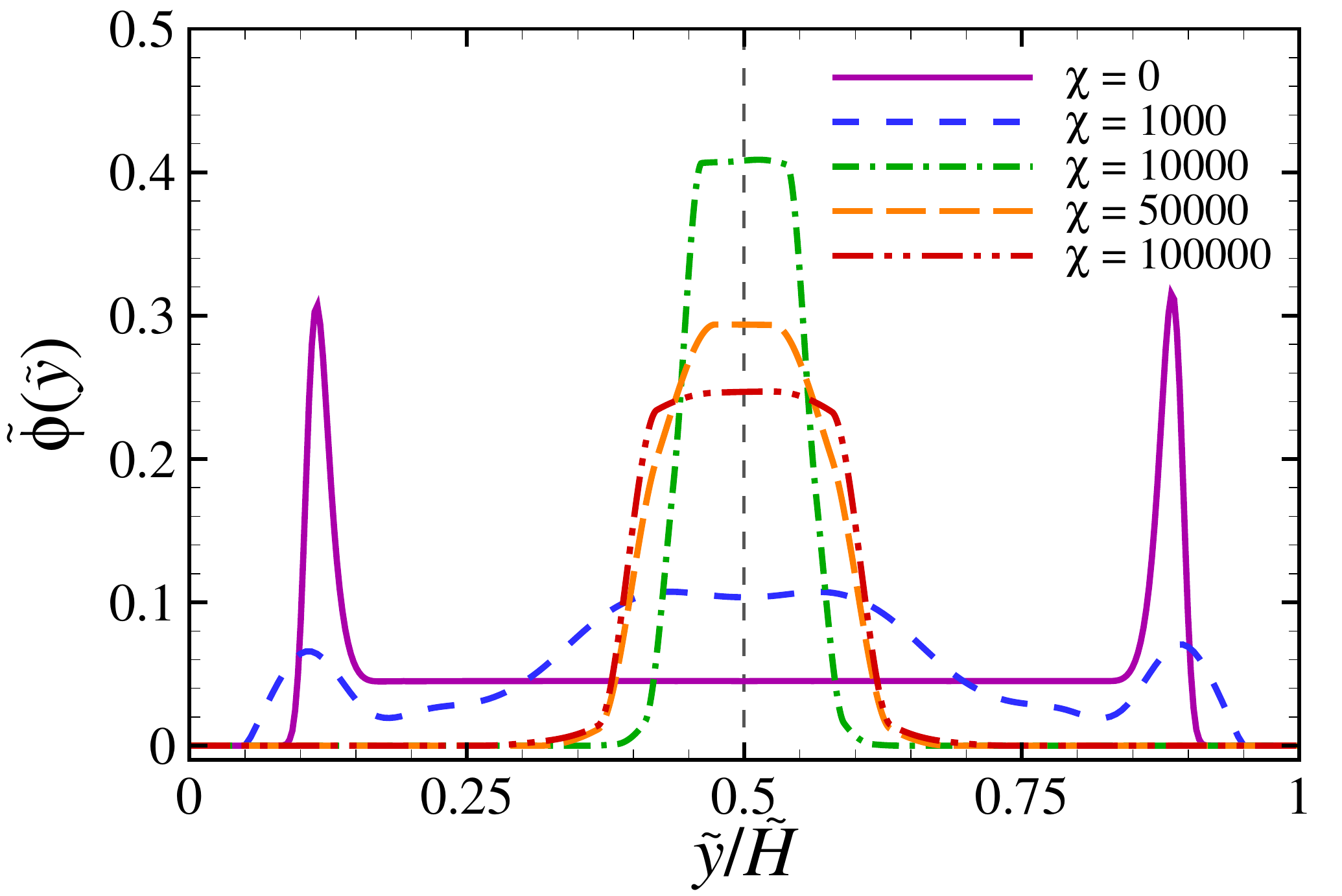}
	\vskip-2mm		
	\caption{Density profiles of prolate magnetic spheroids in a transverse  field as functions of the rescaled lateral  position $\tilde y$ for fixed $\alpha\!=\!4$, $\tilde H \!=\!20$, $\mathrm{Pe}_\mathrm{f} \!=\! 10^4$ and different rescaled field strengths as indicated on the graph.
	}
\label{fig:Fig5}
\end{center}
\vskip-4mm
\end{figure} 

Our probabilistic results, however, point to a qualitatively different scenario. The computed PDFs (see, e.g., Fig. \ref{fig:Fig4}) indicate a relatively uniform noise-induced broadening over a $\tilde y$-interval that extends up to several effective particle radii (in units of $R_\mathrm{eff}$), with no particular  preference given to the centerline. To show this more clearly, we compute the rescaled (lateral) number density profile   
\begin{equation}
\tilde{\phi} (\tilde y) = \!\int_{\theta_B-\pi}^{\theta_B+\pi} \tilde\Psi \left(\tilde y, \theta \right){\mathrm{d}} \theta,  
\label{eq:dens_def}
\end{equation}
where $\int_0^{\tilde H}\!\tilde{\phi} (\tilde y)\,{\mathrm{d}} \tilde y\!=\!1$ by definition (Section \ref{subsec:nondim_Smol}).  The typical behavior of $\tilde{\phi} (\tilde y) $ is shown in Fig. \ref{fig:Fig5} (here, $\alpha\!=\!4$, $\mathrm{Pe}_\mathrm{f} \!=\! 10^4$ and $\tilde H\!=\!20$, and the strong-field threshold $\chi_\ast/\mathrm{Pe}_\mathrm{f}\!\simeq\! 0.2$). In the zero-field limit  (purple curve in Fig. \ref{fig:Fig5}), $\tilde{\phi} (\tilde y) $ reflects the main aspects of the  PDF in Fig. \ref{fig:Fig2}; i.e., sharp depletion-induced near-wall peaks, joining through the midchannel by a low-density plateau. For $\chi \!=\! 10^3$ (dashed blue curve), $\tilde{\phi} (\tilde y) $ shows suppressed near-wall peaks and a broad midchannel bump, portraying unpinned  and pinned subpopulations of the partial-pinning PDF (Fig. \ref{fig:Fig3}), respectively.  Finally, the  pronounced central peaks of the strong-field density  profiles (green, orange and red curves) indicate centered focusing of the spheroids, with PDFs (not shown) resembling the typical one in Fig. \ref{fig:Fig4}. These relatively wide {\em flat-top} peaks clearly signify the {\em nonGaussian} nature of noise-induced effects in centered focusing which we scrutinize further below.

%%%%%%%%%%%%%%%%%%%%%%%%%%%%%%%%%%%%%
\subsubsection{Higher-order stability and subGaussianity}
\label{subsubsec:higherFP_subG}

Our analysis of deterministic phase-space flows using Eqs. \eqref{eq:dtm_dyn_eqs1} and   \eqref{eq:dtm_dyn_eqs2} reveals that the coordinate-space center $(\tilde y_\ast, \theta_\ast) \!=\!(\tilde H/2,\pi/2)$ and $(\tilde H/2,0)$ are indeed solitary fixed points for prolate and oblate magnetic spheroids under transverse and longitudinal fields, respectively. However, in contrast to Ref.  \cite{GolestanianFocusing}, not only the {\em first} $\tilde y$-derivative of the lift velocity but also its {\em second} $\tilde y$-derivative vanishes at this fixed point; see Section 3.5 of the SM. The zero first derivative produces a zero Jacobian eigenvalue as Eq. \eqref{eq:Jacobian} yields $\lambda_y \!=\! 0$ and $\lambda_\theta\!=\! - \chi\Delta_\mathrm{R}(\alpha)\!<\!0$. This means that the central fixed point is a {\em nonhyperbolic, neutrally stable} one in the linearization and its true stability has  to be determined by nonlinear analysis \cite{Rasband1990,Strogatz2000,Nayfeh2008}. This is done in the SM (see also Appendix \ref{app:centered_off_RT} for a recap) where higher-order stability of the fixed point is shown to emerge from a finite negative {\em third} $\tilde y$-derivative of the lift velocity. 
 
The reduced probabilistic theory succinctly links the higher-order nature of the fixed point to the flatness of central density peaks found within the full probabilistic approach (Fig. \ref{fig:Fig5}).  As derived in Appendix \ref{app:U_p}, the reduced theory semiquantitatively describes the  lateral broadening of central (or any other) fixed point  along the pinning curve as a diffusive process driven by the translational noise against a deterministic background. The noise enters via the effective diffusivity $ \Delta_\mathrm{p}(\tilde y) \!=\! 4[\Delta_+(\alpha) \!-\! \Delta_-(\alpha) \cos 2\theta_\mathrm{p}(\tilde y)]/3$ and the  background via the {\em virtual lift potential} $\tilde U_\mathrm{p} (\tilde y) \!=\! - \int^{\tilde y}_{0} \mathrm{d} \tilde y_1\,\tilde u^{(\mathrm{im})}_\mathrm{p} (\tilde y_1)/\Delta_\mathrm{p}(\tilde y_1)$ where $\tilde u^{(\mathrm{im})}_\mathrm{p}(\tilde y) \!=\!  \tilde u^{(\mathrm{im})}_y(\tilde y, \theta_\mathrm{p}(\tilde y))$ is the hydrodynamic lift  along the pinning curve; see Eqs.  \eqref{eq:u_Delta_p1}-\eqref{eq:Up2b}. The reduced density profile of pinned spheroids across the channel width is then predicted as  $\tilde{\phi} (\tilde y) \!\sim\! \Delta_\mathrm{p}^{-1}(\tilde y) \exp [- \tilde U_\mathrm{p}(\tilde y)]$.  

In centered focusing, the virtual lift potential admits a {\em quartic} shape on the leading order in lateral distance from the  central fixed point as $\tilde U_\mathrm{p} (\tilde y) \!\simeq\! k (\tilde y \!-\! \tilde H/2)^4$ where $k\!>\!0$ is a parameter-dependent factor that scales inversely with the translational noise strength and directly with  the aforementioned  third derivative, $k\!\sim\! |\partial^3 \tilde u^{(\mathrm{im})}_\mathrm{p}/\partial\tilde y^3|_{\tilde H/2}$ (Appendix \ref{app:centered_off_RT}). The quartic $\tilde U_\mathrm{p}$ thus reveals the peculiar flat-top form of centered-focusing peaks, Fig. \ref{fig:Fig4}, to be of {\em subGaussian} type  $\sim\! \exp [- k(\tilde y \!-\! \tilde H/2)^4]$.  

%%%%%%%%%%%%%%%%%%%%%%%%%%%%%%%%%%%%%
\subsubsection{Field-induced defocusing in the strong-field regime}
\label{subsec:defocusing}

The density profiles of Fig.  \ref{fig:Fig5} also reveal a striking {\em nonmonotonic} dependence  on the rescaled field strength $\chi$ (or $\chi/\mathrm{Pe}_\mathrm{f}$) at fixed flow P\'eclet number $\mathrm{Pe}_\mathrm{f}$:  While the central peak is amplified initially and the spheroids are expectedly more strongly focused by increasing $\chi$ (green  curve), the trend is reversed and the peak is broadened and its amplitude is suppressed at larger $\chi$ (orange and red curves). This amounts to a counterintuitive case of {\em field-induced defocusing}; i.e., at fixed imposed shear, centered focusing cannot be enhanced  continually by excessive amplification of the  field which can only produce a {\em transient maximal focusing} (largest possible peak) before the defocusing sets in well within the strong-field regime. The defocusing thus transpires in a manner that stands  at odds with the deterministic picture. The latter predicts the pinned spheroids  to remain ideally point-focused (i.e., with a $\delta$-distributed PDF; Appendix \ref{app:deterministic_RT}) at the central fixed point regardless of the field strength. 

The defocusing can be viewed as progressive weakening of the {\em higher-order stability} at the central fixed point (i.e., the nonvanishing third derivative of the lift also tends to zero) as $\chi$ is increased. It can also be understood based on the subGaussian distribution of the reduced probabilistic theory (Section \ref{subsubsec:higherFP_subG}) by noting that $k^{-1}$ gives a measure for the width of central density peak.  As we discuss in Appendix \ref{app:centered_off_RT} and Section 3.5 of the SM, the prefactor  $k$ in the quartic virtual potential  is also proportional to the slope of pinning curve at the channel center; i.e.,  $k\!\sim\! |\partial\theta_\mathrm{p}/\partial \tilde y|_{\tilde H/2}$. This slope decreases with $\chi/\mathrm{Pe}_\mathrm{f}$ and, hence, the width set by $k^{-1}$ increases with $\chi/\mathrm{Pe}_\mathrm{f}$, explaining the defocusing effect. For clarity, we  emphasize that the pinning curve in Fig. \ref{fig:Fig4} is shown by its inverse function $\tilde y \!=\! \tilde y_\mathrm{p}(\theta)$ and that the slope of this curve increases on increasing $\chi/\mathrm{Pe}_\mathrm{f}$ (as it eventually tends to the zero-lift nullcline $\theta\!=\!\pi/2$); see Fig. 1 of the SM. 
 
%%%%%%%%%%%%%%%%
\begin{figure}[t!]
\begin{center}
\includegraphics[width=0.75\linewidth]{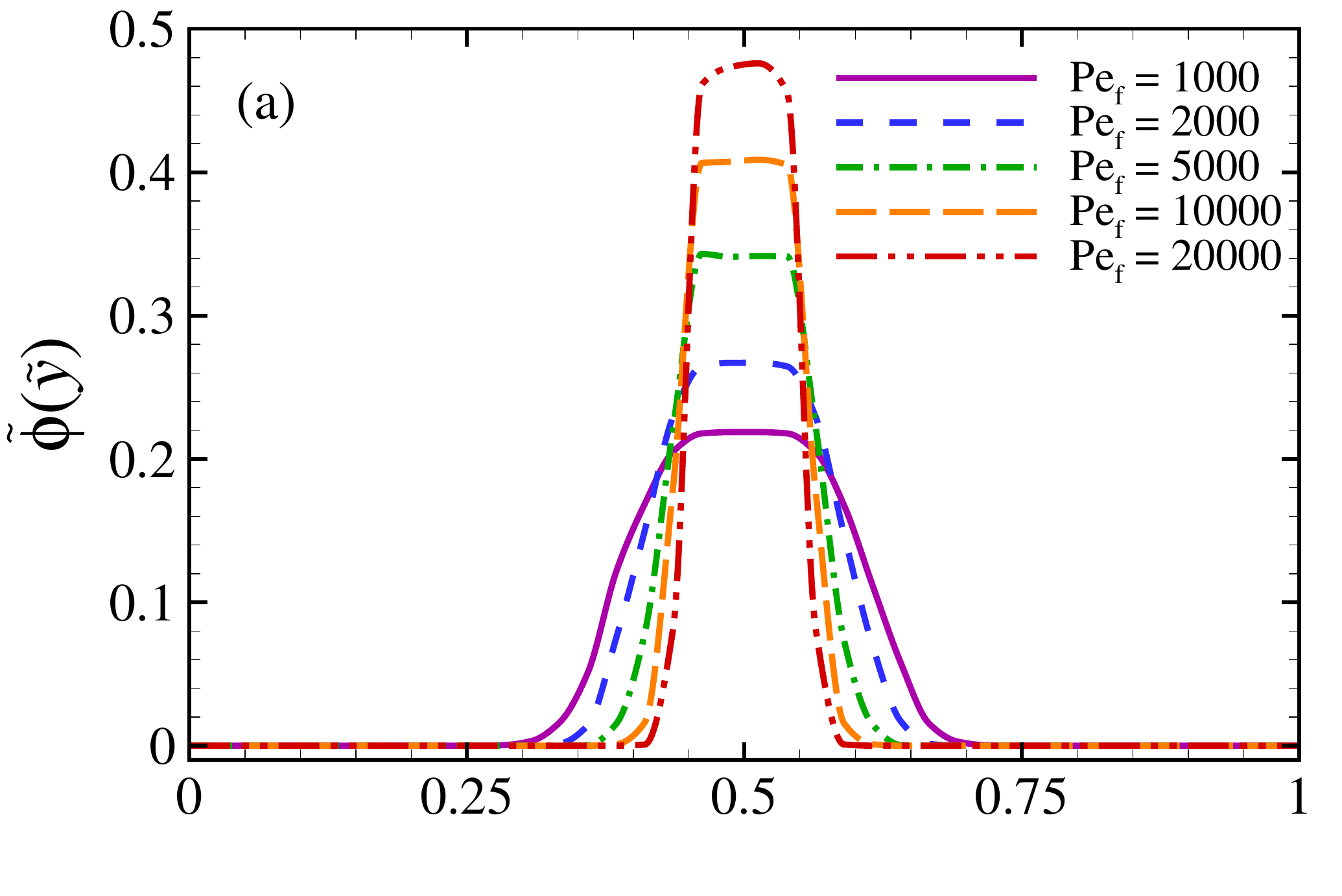}\\
\vskip-4mm
\includegraphics[width=0.75\linewidth]{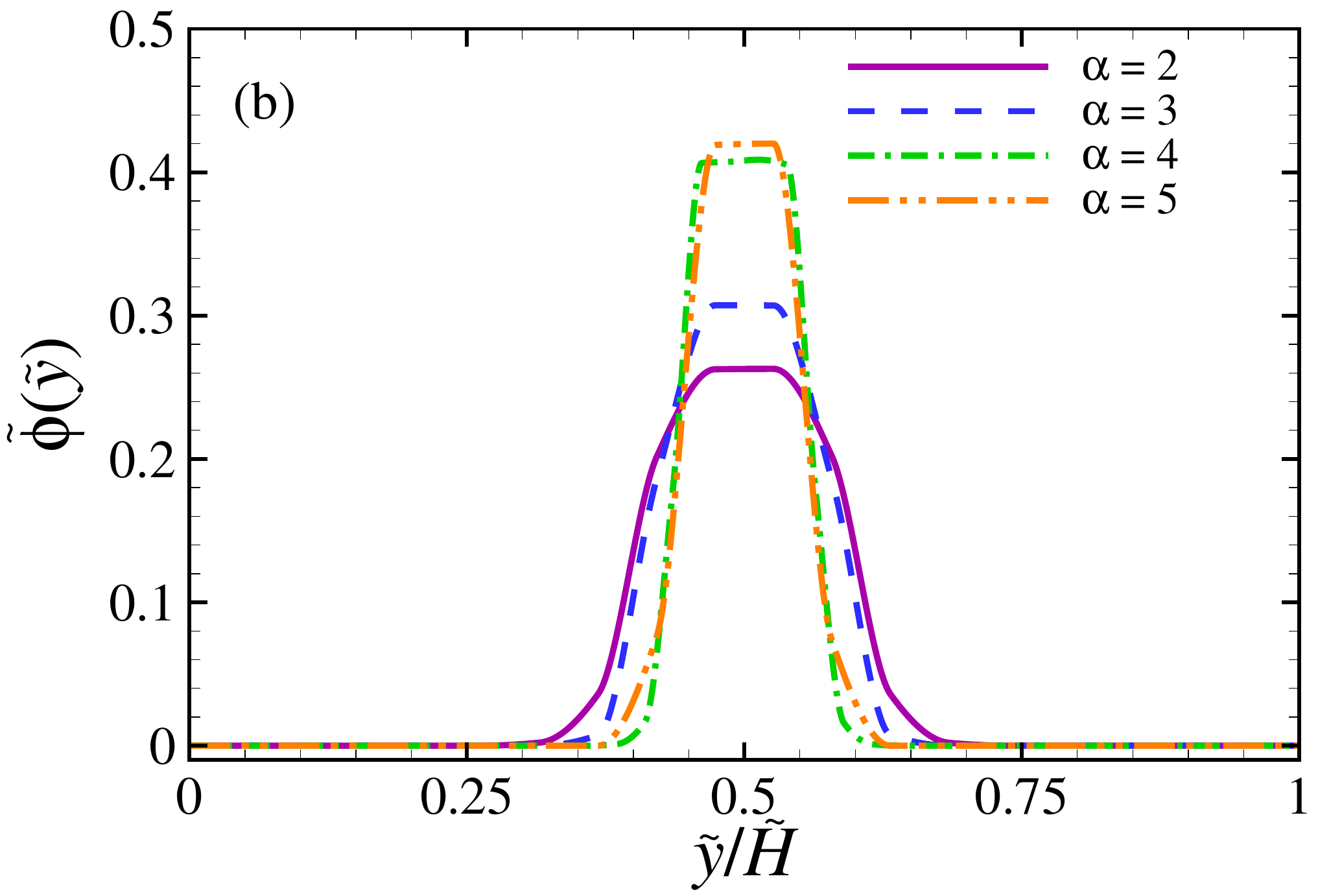}
	\vskip-2mm	
	\caption{Same as Fig. \ref{fig:Fig5} but here we plot the density profiles for (a) $\alpha\!=\!4$ and different  $\mathrm{Pe}_\mathrm{f}$ at fixed $\chi/\mathrm{Pe}_\mathrm{f}\!=\! 1$ and (b) $\chi \!=\! \mathrm{Pe}_\mathrm{f} \!=\! 10^4$ and different particle aspect ratios. 
	}
	\vskip-4mm
\label{fig:Fig5plus2}
\end{center}
\end{figure} 
%

%%%%%%%%%%%%%%%
\begin{figure*}[t!]
\begin{center}
	\begin{minipage}[t]{0.3425\linewidth}\begin{center}
		\includegraphics[width=\linewidth]{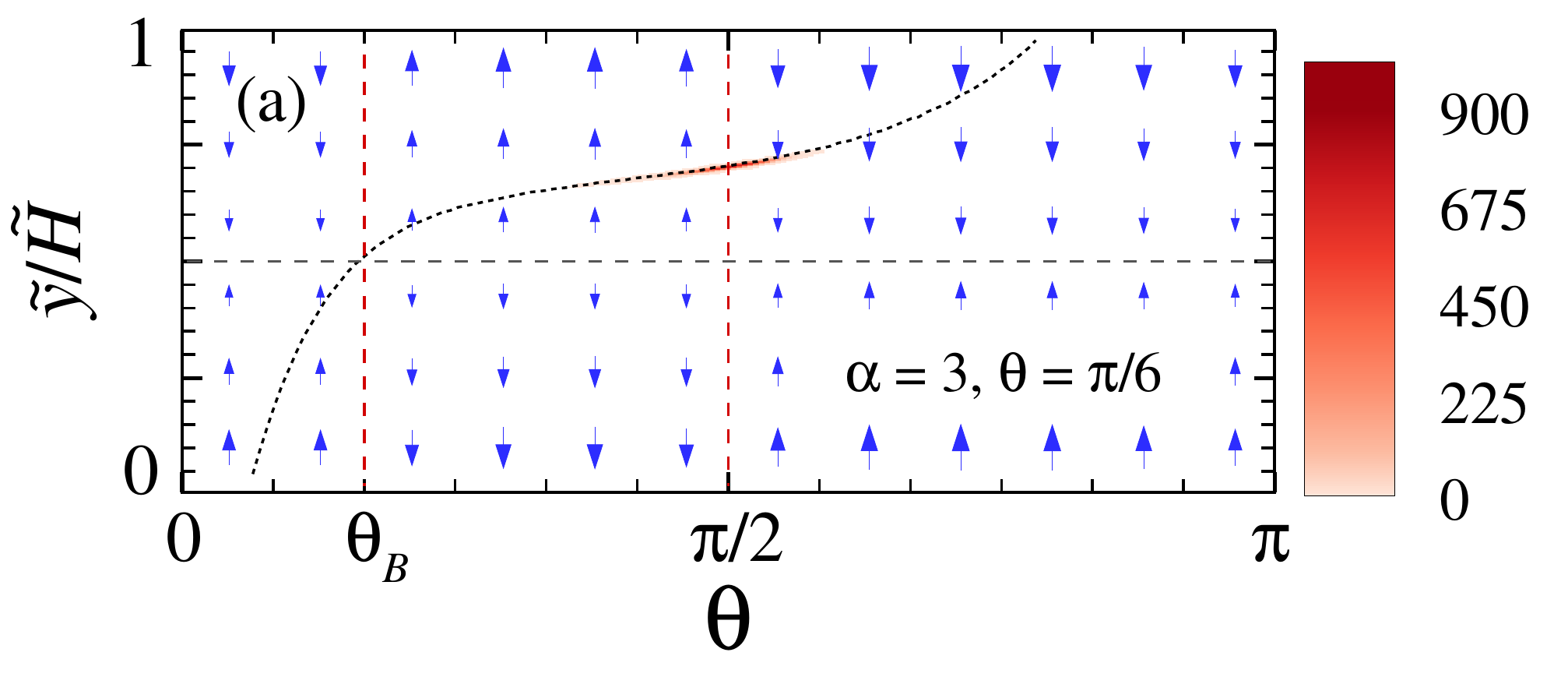} \vskip-1mm %(a) $\alpha \!=\! 3, \theta_B \!=\! \pi/6$
	\end{center}\end{minipage}\hskip0mm
	\begin{minipage}[t]{0.3425\linewidth}\begin{center}
		\includegraphics[width=\linewidth]{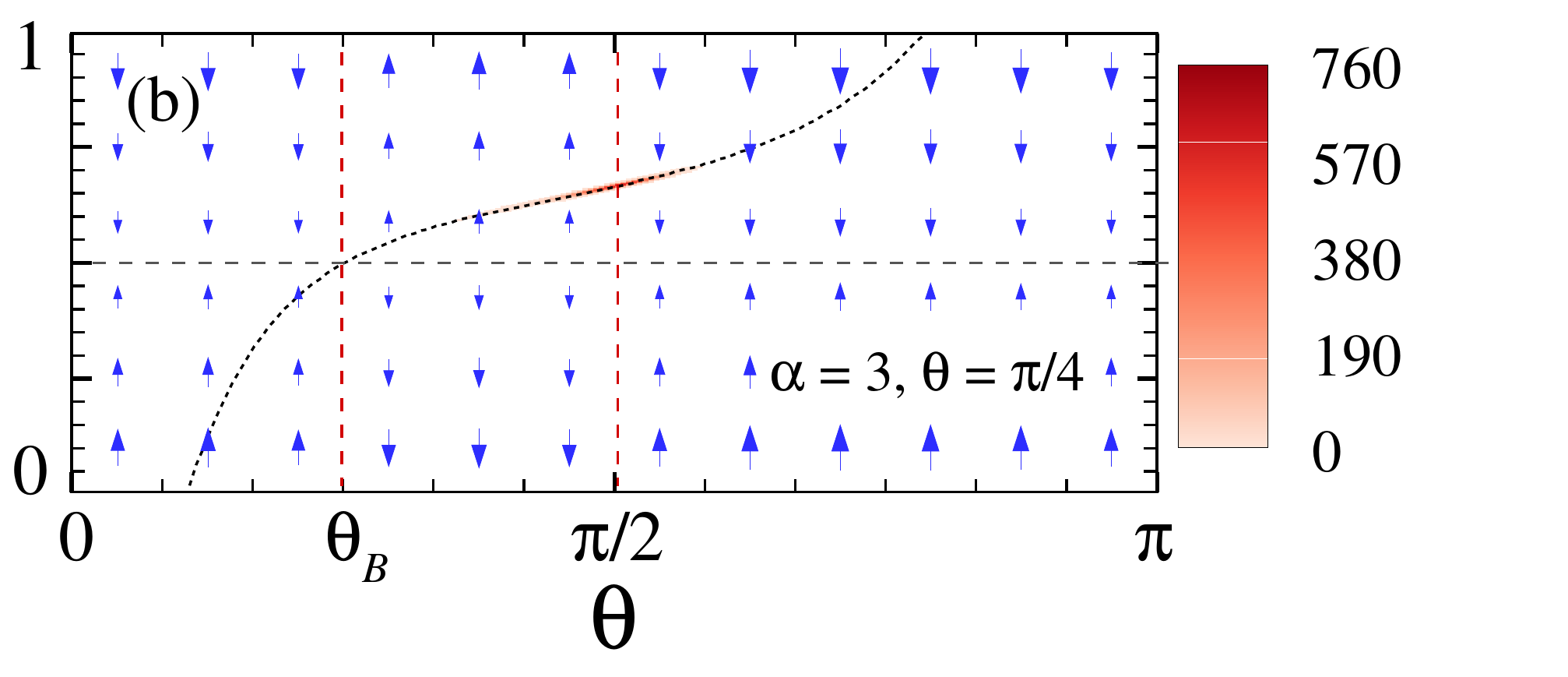} \vskip-1mm %(b) $\alpha \!=\! 3, \theta_B \!=\! \pi/4$
	\end{center}\end{minipage}\hskip-5mm	
	\begin{minipage}[t]{0.3425\linewidth}\begin{center}
		\includegraphics[width=\linewidth]{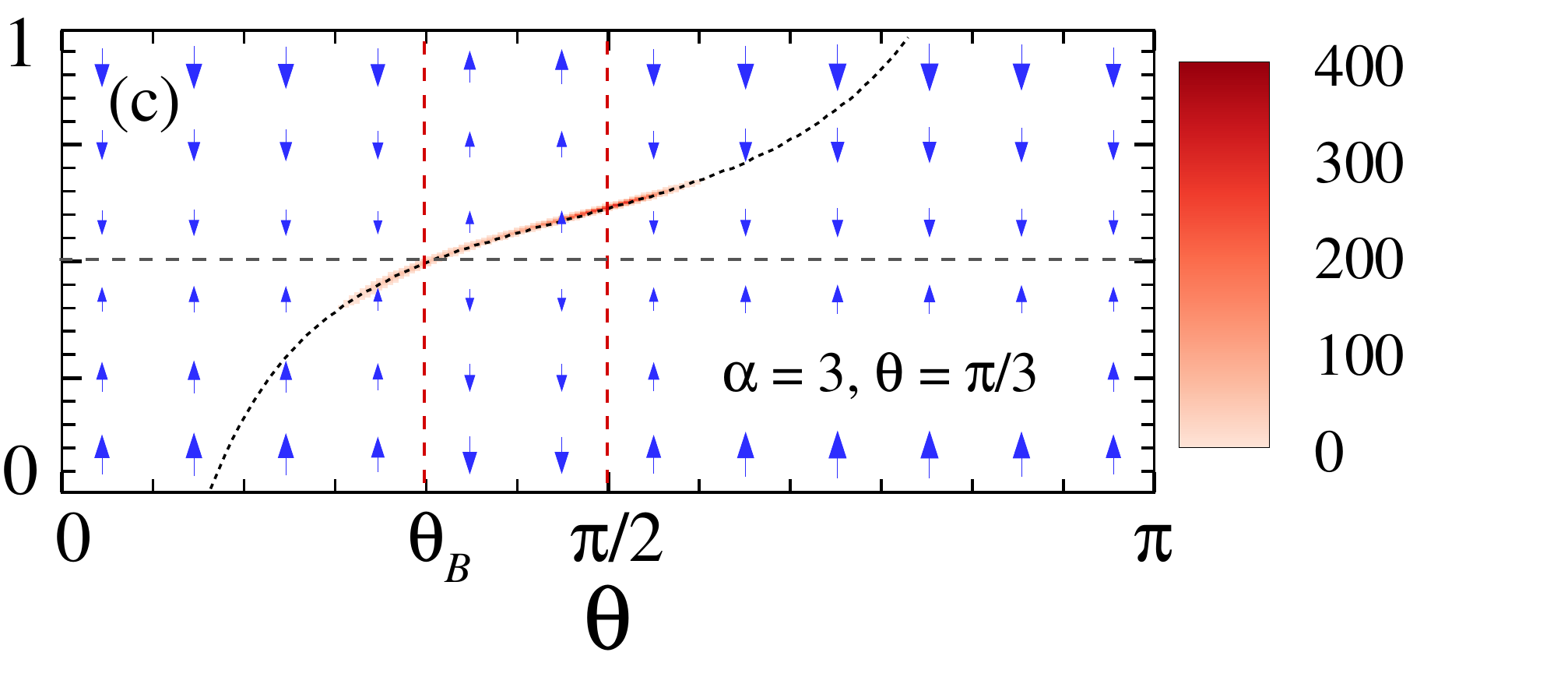} \vskip-1mm %(c) $\alpha \!=\! 3, \theta_B \!=\! \pi/3$
	\end{center}\end{minipage}\vskip-1mm
	\begin{minipage}[t]{0.3425\linewidth}\begin{center}
		\includegraphics[width=\linewidth]{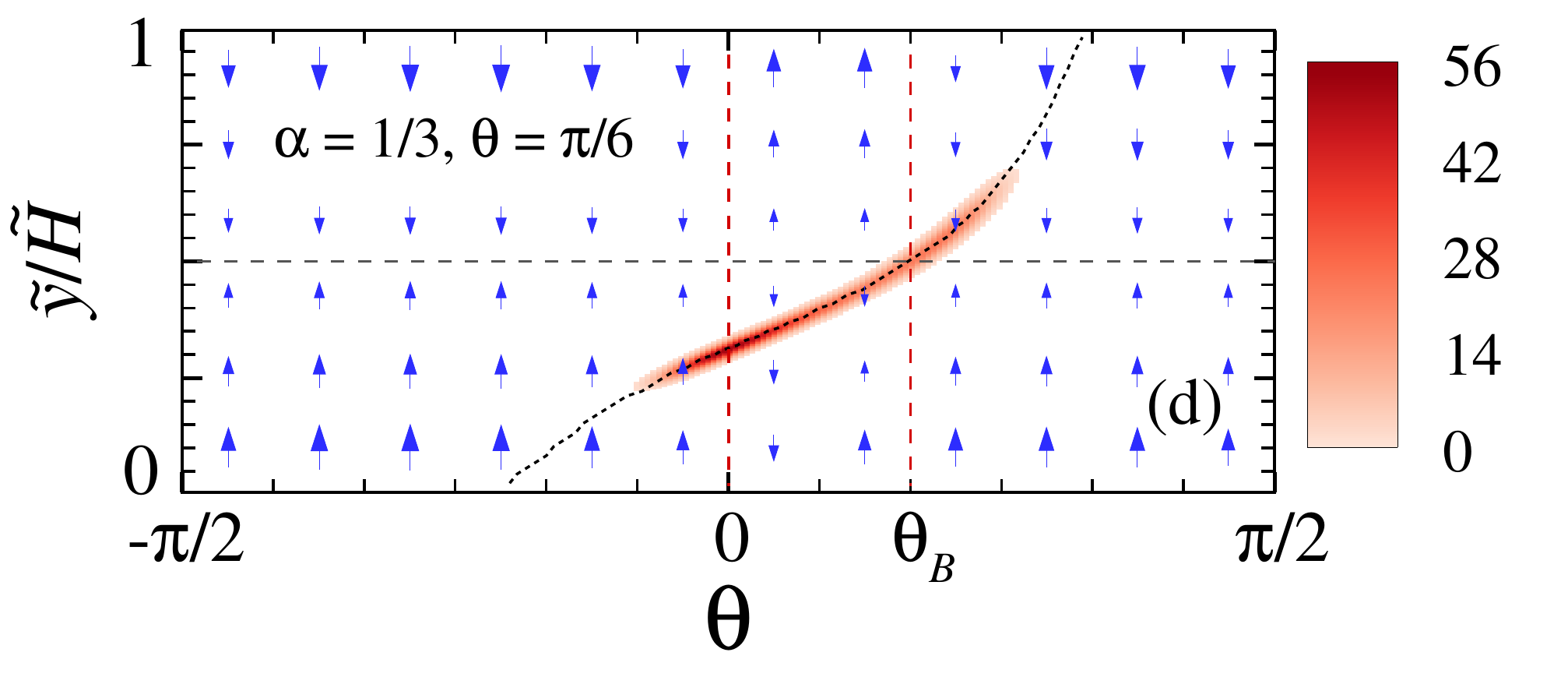} \vskip-1mm %(d) $\alpha \!=\! 1/3, \theta_B \!=\! \pi/6$
	\end{center}\end{minipage}\hskip0mm
	\begin{minipage}[t]{0.3425\linewidth}\begin{center}
		\includegraphics[width=\linewidth]{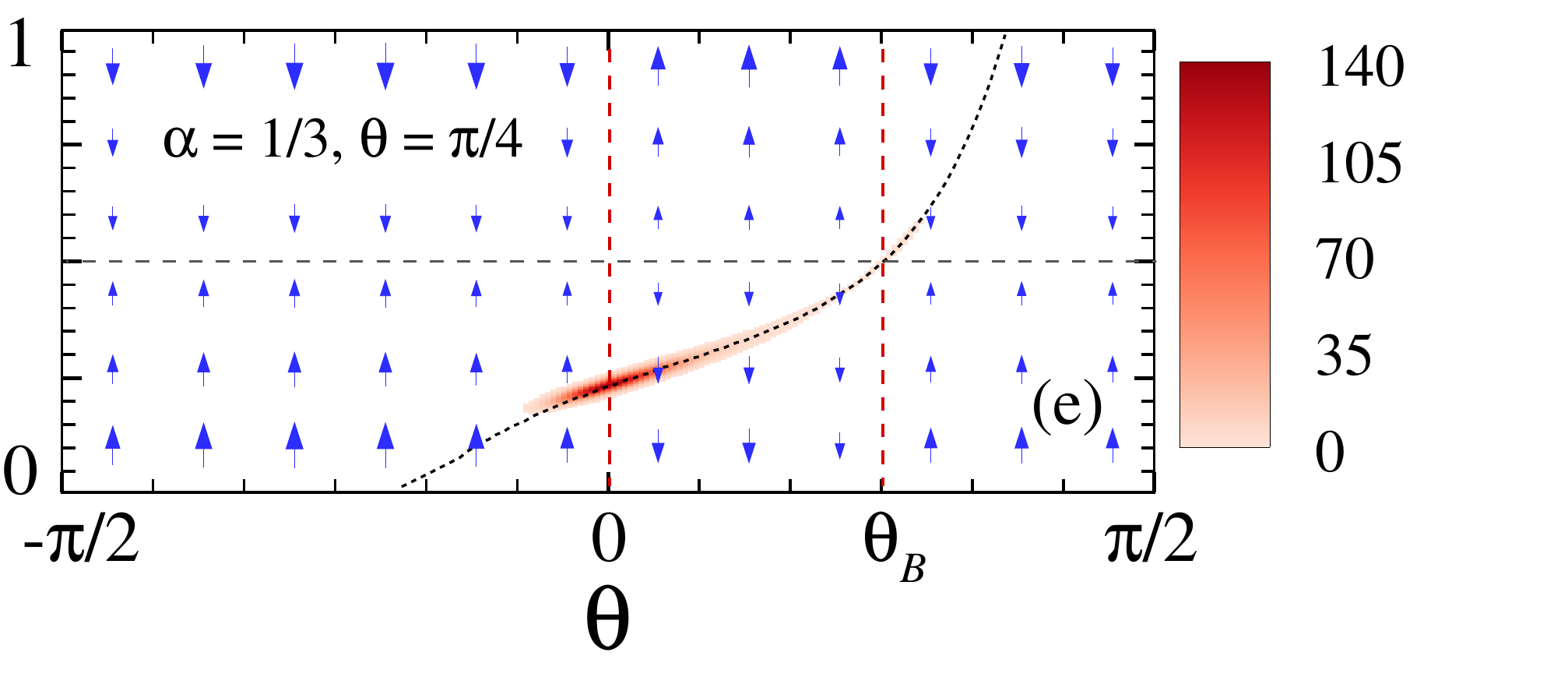} \vskip-1mm %(e) $\alpha \!=\! 1/3, \theta_B \!=\! \pi/4$
	\end{center}\end{minipage}\hskip-5mm		
	\begin{minipage}[t]{0.3425\linewidth}\begin{center}
		\includegraphics[width=\linewidth]{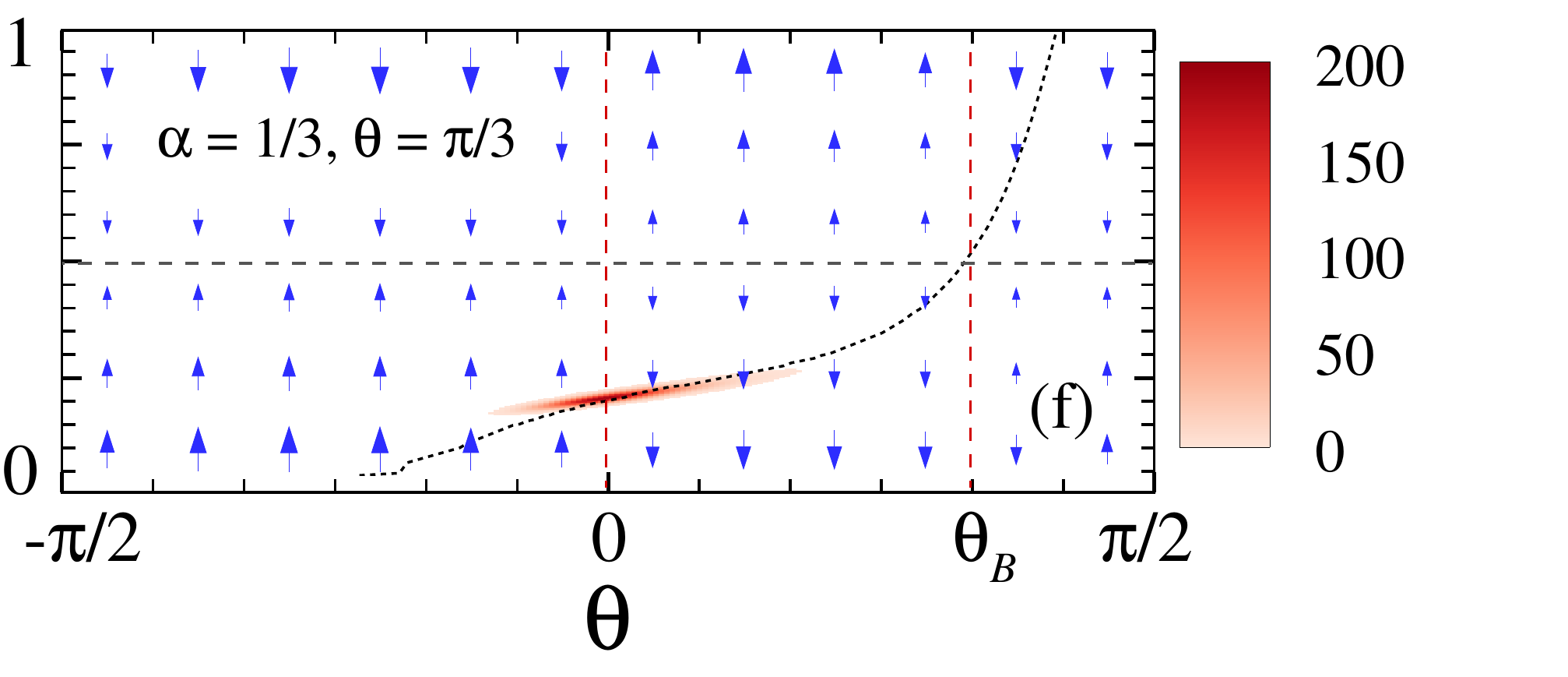} \vskip-1mm %(f) $\alpha \!=\! 1/3, \theta_B \!=\! \pi/3$
	\end{center}\end{minipage}
	\vskip-6mm \caption{Density maps of the numerically obtained PDF, $\tilde{\Psi}(\tilde y, \theta)$, for prolate spheroids of aspect ratio $\alpha\!=\!3$ (panels a to c) and oblate spheroids of aspect ratio $1/3$ (panels d to f) under uniform fields of tilt angle  $\theta_B\!=\!\pi/6, \pi/4, \pi/3$ (left to right). In (a) to (c), $\mathrm{Pe}_\mathrm{f} \!=\!\chi\!=\! 10^4$ and, in (d) to (f), $\mathrm{Pe}_\mathrm{f} \!=\!\chi\!=\! 10^3$ and, in all panels, $\tilde H \!=\! 20$. Orange shadings show the PDFs visualized using the same cutoff as in Fig. \ref{fig:Fig2}. Blue arrows are the hydrodynamic lift with the same rescaling as in Fig. \ref{fig:Fig4}. The dotted black curve is the pinning curve, the horizontal dashed line is the  centerline, and the vertical dashed lines are the angular loci of the coordinate-space center $\theta_B$ and the nullclines $\pi/2$ (prolate, a-c) and $0$ (oblate, d-f) intersecting the pinning curve.  
	}
\label{fig:Fig6}
\vskip-4mm
\end{center}
\end{figure*}

%%%%%%%%%%%%%%%%%%%%%%%%%%%%%%%%%%%%%
\subsubsection{Circumventing the strong-field defocusing}
\label{subsec:defocusing_circumvented}

The field-induced defocusing can be of significant ramification to experiments. It entails   that intensifying the field would be a {\em self-limiting} factor in achieving precision centered focusing of ferromagnetic particles under transverse (longitudinal) field for prolate (oblate) spheroids. As a strategy to circumvent the defocusing, one can increase the rescaled shear strength, $\mathrm{Pe}_\mathrm{f}$, at fixed field-to-shear strength ratio $\chi/\mathrm{Pe}_\mathrm{f}$ (chosen in a way that the whole-channel pinning is still fulfilled). In this case, the central peak is amplified continually as shown in  Fig. \ref{fig:Fig5plus2}a.  

The tunability of central peak with $\mathrm{Pe}_\mathrm{f}$  signifies another departure from the deterministic picture where the governing equations scale with $\chi/\mathrm{Pe}_\mathrm{f}$ rather than depending on $\chi$ and $\mathrm{Pe}_\mathrm{f}$ separately (see, e.g., Eqs. \eqref{eq:pinning}-\eqref{eq:chi_ast_app} and Section 3, SM). This scaling is  broken  by the noise terms in Eq. \eqref{eq:smoluchowski} and its breakdown is unmasked when noise  effects become relevant as in the cases noted above. 

It is worth mentioning that the central density peak can also be enhanced at fixed  $\mathrm{Pe}_\mathrm{f}$ and $\chi$ by increasing the particle aspect ratio $\alpha$ as shown in Fig. \ref{fig:Fig5plus2}b.

%%%%%%%%%%%%%%%%%%%%%%%%%%%%%%%%%%%%%
\section{Off-centered focusing}
\label{sec:offcentered_focusing} 

We now turn to the more general situation in which prolate and oblate spheroids are  focused at noncentral latitudes within the channel. This {\em off-centered focusing}  is realized for a wide range of applied field strengths  but only when the field is tilted relative to the flow direction. In what follows, the tilt angle  $\theta_B$ of the applied field is restricted to the first quadrant, $0\!<\!\theta_B\!<\!\pi/2$ (Section \ref{subsec:nondim_Smol}) and its rescaled strength $\chi$ is kept above the onset of whole-channel pinning $\chi_\ast(\theta_B)$ which, for arbitrary $\theta_B$, is obtained numerically from Eq. \eqref{eq:pinning}. 

%%%%%%%%%%%%%%%%%%%%%%%%%%%%%%%%%%%%%
\subsection{Off-centered PDFs: Overview of main features}
\label{subsec:offcentered_overview}

Under a tilted field, the point-reflection symmetry of the pinning curve relative to the coordinate-space center $(\tilde H/2, \theta_B)$ is broken. As seen in Figs.  \ref{fig:Fig6}a-f, the (dotted black) pinning curves are shifted to the top/bottom channel half for prolate/oblate spheroids as compared with the symmetric case in Section \ref{sec:centered_focusing}. Here, the numerically obtained PDFs are again accumulated as narrow (orange) bands on the pinning curves. They are also focused by the wall-induced hydrodynamic lift (blue arrows) within finite, primarily off-centered $\tilde y$-intervals.  There are however remarkable variations in the shape of PDF with the tilt angle of field $\theta_B$ and the aspect ratio of particles $\alpha$. These can be best seen in the number density profiles,  $\tilde{\phi} (\tilde y) $,   Eq. \eqref{eq:dens_def}, that are shown in  Figs. \ref{fig:Fig7}a-c. These panels show the results for different prolate/oblate $\alpha$ at fixed $\theta_B\!=\!\pi/6$ (a), $\pi/4$ (b) and $\pi/3$ (c). The dot-dashed green curves correspond to the PDFs in Fig. \ref{fig:Fig6}. 

%%%%%%%%%%%%%%%
\subsubsection{Optimal focusing subregime}
\label{subsubsec:optimal_subregime}

\begin{figure*}[t!]
\begin{center}
	\begin{minipage}[t]{0.335\linewidth}\begin{center}
		\includegraphics[width=\linewidth]{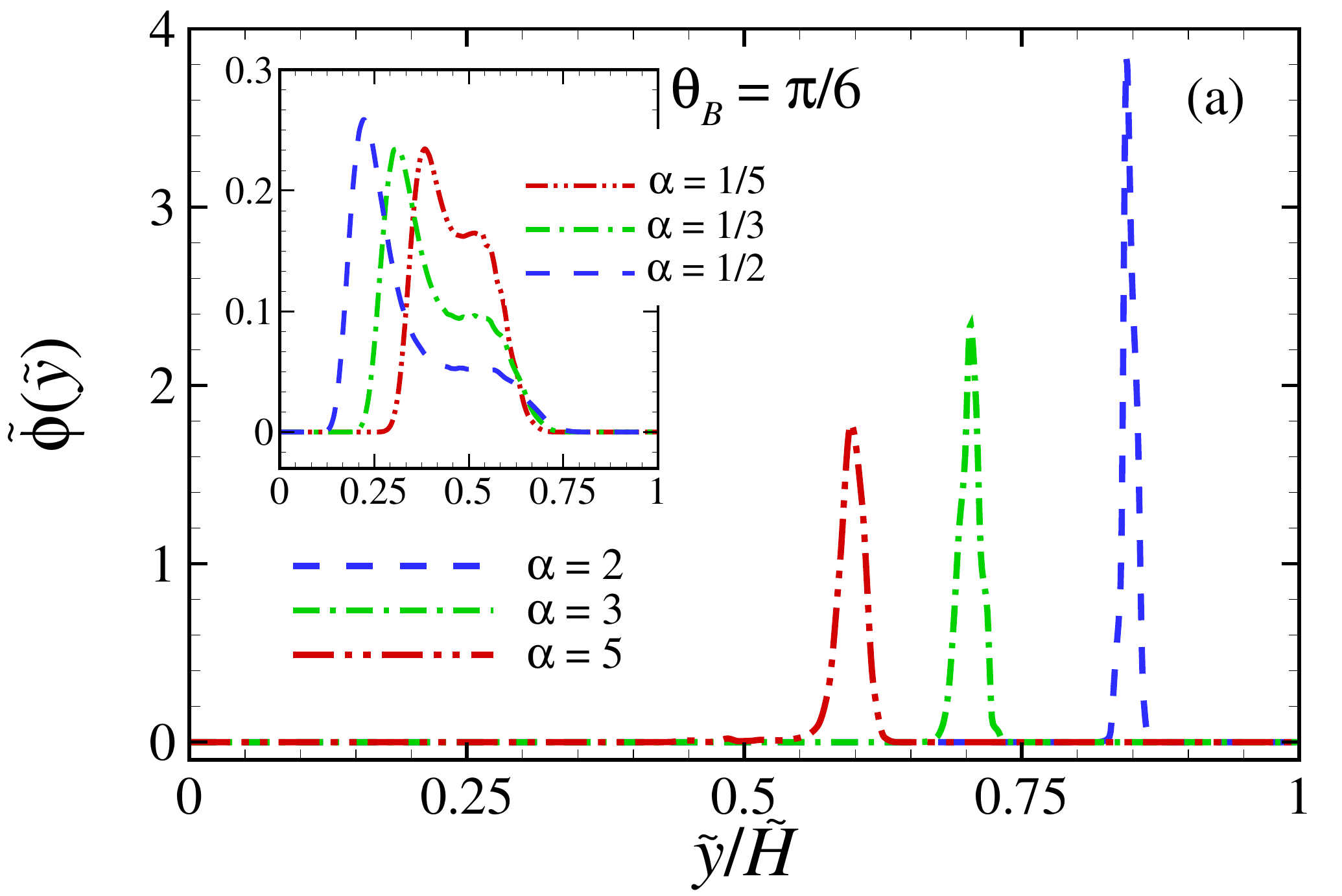} \vskip-1mm %(a) $\theta_B \!=\! \pi/6$
	\end{center}\end{minipage}\hskip-1mm
	\begin{minipage}[t]{0.335\linewidth}\begin{center}
		\includegraphics[width=\linewidth]{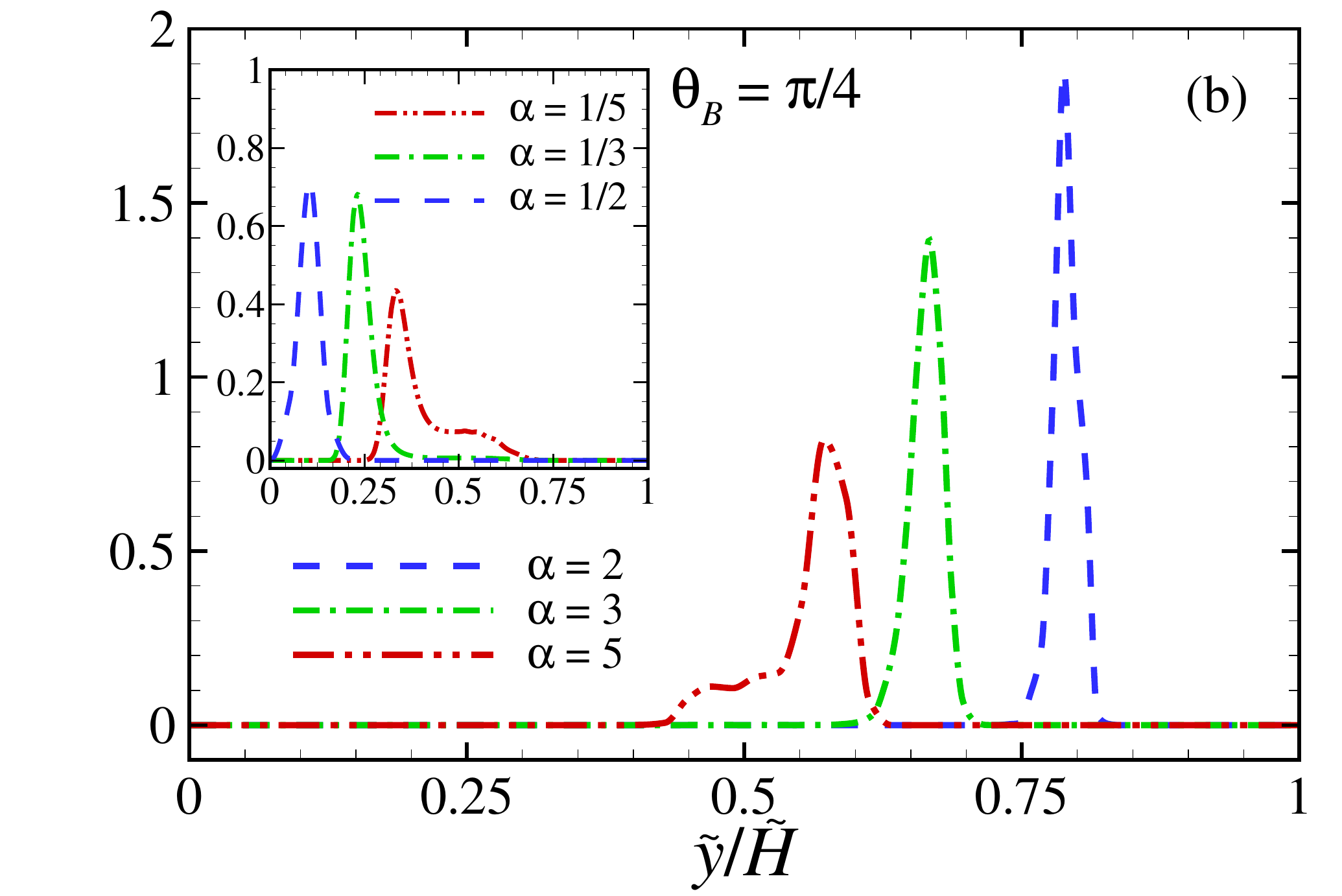} \vskip-1mm %(b)  $\theta_B \!=\! \pi/4$
	\end{center}\end{minipage}\hskip-1mm		
	\begin{minipage}[t]{0.335\linewidth}\begin{center}
		\includegraphics[width=\linewidth]{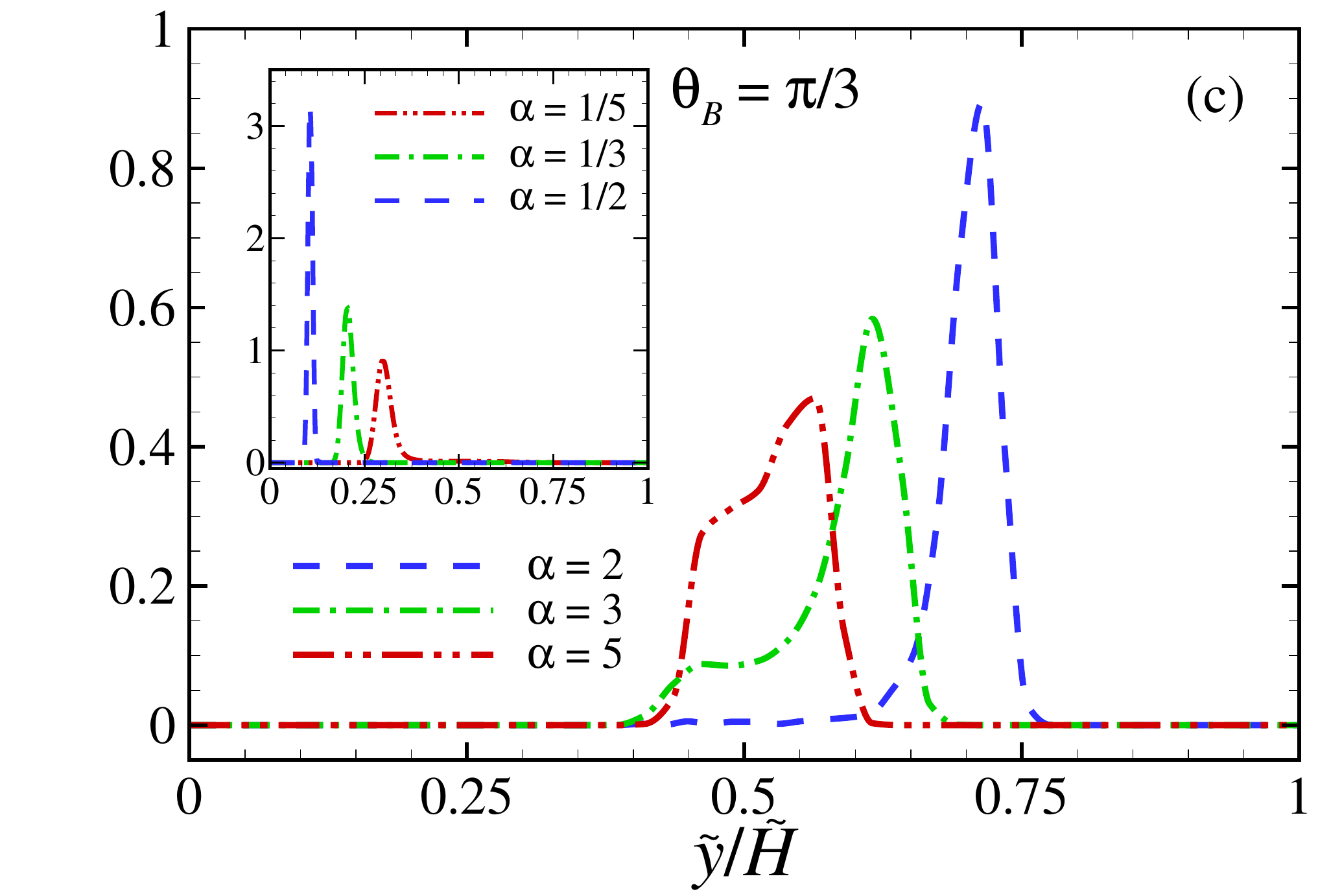} \vskip-1mm %(c)  $\theta_B \!=\! \pi/3$
	\end{center}\end{minipage}
		\caption{Numerically obtained spheroidal density profiles, $\tilde{\phi} (\tilde y) $,   Eq. \eqref{eq:dens_def}, plotted  in rescaled units as functions of the rescaled lateral  position $\tilde y$ for $\tilde H \!=\!20$ and different particle aspect ratios as indicated on the plots for (a) $\theta_B \!=\! \pi/6$, (b) $\theta_B \!=\! \pi/4$, and (c) $\theta_B \!=\! \pi/3$. The main sets (insets) correspond to prolate (oblate) spheroids with  $\mathrm{Pe}_\mathrm{f} \!=\!  \chi\!=\!10^4$ ($10^3$).
}
\label{fig:Fig7}
\end{center}
\vskip-4mm
\end{figure*}

To explore the results in Figs. \ref{fig:Fig6} and \ref{fig:Fig7}, we begin with the case of prolate spheroids in an applied field of relatively {\em small} tilt angle; here $\theta_B\!=\!\pi/6$. The corresponding density profiles in  Fig. \ref{fig:Fig7}a (main set) give an unequivocal demonstration of off-centered focusing with a prominent peak in the top channel half. The locus of the peak, or the focusing latitude, varies markedly with the aspect ratio $\alpha$: Decreasing $\alpha$ (from $\alpha\!=\!5$ to 2 in Fig. \ref{fig:Fig7}a, main set) shifts the peak by several particle sizes away from the centerline toward the top channel wall (from $\tilde y\simeq 11.9$ to $\simeq 16.9$). This trend is accompanied by a sizable (up to 50\%)  decrease  in peak width and inversely its amplitude. The density peaks nevertheless maintain two key properties: First, they are highly localized, having relatively small widths (comparable to unity, or the effective particle radius $R_\mathrm{eff}$ in rescaled units)  along the $\tilde y$-axis. Second, the peaks are sharp and symmetric at the top (even as their tails are slightly skewed toward the center). These properties starkly differ  from those found under centered focusing in Section \ref{sec:centered_focusing}, indicating an inherently different focusing mechanism that we clarify later in Section \ref{subsec:fixedpoints_off}.

We designate the parametric region where off-centered  peaks appear with the above two properties as the {\em optimal focusing subregime}. This subregime can be delineated quantitatively (Section \ref{subsec:optimal_criterion}) and its properties can  be utilized for efficient sorting of spheroids by tuning the system parameters (Section  \ref{sec:separation}).  

The optimal focusing  for oblate spheroids is achieved when the tilt angle $\theta_B$ is chosen at the other end of the angular interval; see Fig. \ref{fig:Fig7}c, inset, for $\theta_B\!=\!\pi/3$.  Even as the  qualitative features here are similar to those of prolate spheroids  (i.e., Fig. \ref{fig:Fig7}a), some of the trends are expectedly reversed: The  oblate focusing peaks arise  in the bottom half of the channel and, on decreasing $\alpha$ (from $\alpha\!=\!1/2$ to $1/5$ in Fig. \ref{fig:Fig7}c, inset), their loci  shift toward the channel center. Despite these reverse similarities, a duality prolate-oblate correspondence for spheroids of aspect ratios $\alpha$ and $1/\alpha$ does not exist (Section 3.2, SM). 

%%%%%%%%%%%%%%%

%%%%%%%%%%%%%%
\subsubsection{Shouldered focusing subregime}
\label{subsubsec:shouldered_subregime}

For prolate spheroids, increasing the tilt angle (compare main sets of Figs. \ref{fig:Fig7}a-c) leads to a peculiar change in the density profiles. The density peaks become skewed and eventually develop a left {\em shoulder}, stretching from the off-centered peak to the channel center. For a  given particle aspect ratio  (e.g., red curve for $\alpha\!=\!5$), the amplitude of  off-centered  peak is suppressed and its location shifts (rather weakly) toward the channel center as $\theta_B$ is increased; meanwhile, the left shoulder is amplified, even as its  location persists on the centerline. As seen, the shoulder is enhanced also as $\alpha$ is increased at fixed $\theta_B$. 

Since the off-centered peaks emerge either as localized (optimal) peaks or as skewed ones with shoulders, we designate the latter as a subregime complementary to optimal focusing, referring to it as the  {\em shouldered focusing subregime}.  This points to an underlying  bimodality and  pronounced noise-induced effects as we discuss below.  

For oblate spheroids, shouldered focusing occurs similarly to prolate spheroids except that the  tilt angle of the field in their case needs to be decreased (see insets of Fig. \ref{fig:Fig7}c to a). The shoulders here are enhanced by lowering the aspect ratio $\alpha$. 

%%%%%%%%%%%%%%%%%%%%%%%%%%%%%%%%%%%%%
\subsection{Deterministic fixed points}
\label{subsec:fixedpoints_off}

Under a tilted field, our analysis of deterministic phase-space flows based on Eqs. \eqref{eq:dtm_dyn_eqs1} and   \eqref{eq:dtm_dyn_eqs2} reveals {\em two} distinct fixed points as detailed in Section 3.6 of the SM. The essential aspects of these fixed points are as follows.  
 
The said fixed points result from two distinct nullcline intersections; one  between the pinning curve and the zero-lift nullcline $\tilde y\!=\! \tilde H/2$ and the other between the pinning curve and the zero-lift nullcline  $\theta\!=\!\pi/2$ and $0$ for prolate and oblate spheroids, respectively. The former intersection produces a {\em central} fixed point at $(\tilde y_\ast, \theta_\ast) \!=\!(\tilde H/2, \theta_B)$ and the latter intersection produces an {\em off-centered} fixed point at $(\tilde y_\ast, \theta_\ast) \!=\!(\tilde y_\mathrm{F}, \pi/2)$ and  $(\tilde y_\mathrm{F}, 0)$ for prolate and oblate spheroids, respectively. 

The central fixed point is found to be a nonhyperbolic, neutrally stable one in the linearization with a zero first $\tilde y$-derivative for the lift velocity and the same eigenvalues as in Section \ref{subsubsec:higherFP_subG}. These make it similar to the solitary fixed point in centered focusing. However, our  higher-order analysis in the SM shows that, for the present case  ($0\!<\!\theta_B\!<\!\pi/2$), the {\em second} $\tilde y$-derivative of the lift velocity is finite and positive (negative) for prolate (oblate) spheroids; i.e., the central fixed point here is attractive (repulsive) from the bottom and  repulsive  (attractive)  from the top. It is thus a {\em half-stable higher-order} fixed point and can produce qualitatively different effects as compared to the centered focusing (Section \ref{subsec:optimal_vs_shoulder}). 

The off-centered fixed point is, on the other hand,  found to be located in the top (bottom) channel half for prolate (oblate) spheroids at the latitude 
\begin{align}
\!\!\frac{\tilde y_\mathrm{F}}{\tilde H}  = 
\left\{ \begin{array}{ll}
\!\!\dfrac{1}{2} \!+\! \dfrac{\chi}{\mathrm{Pe}_\mathrm{f}} \dfrac{\Delta_\mathrm{R}(\alpha)}{1\!+\! \beta(\alpha)} \cos\theta_B & \!:\,  \alpha\!>\!1\,\,\, (\theta_\ast\!=\!\pi/2),  \\ 
 \\
\!\!\dfrac{1}{2} \!-\!  \dfrac{\chi}{\mathrm{Pe}_\mathrm{f}} \dfrac{\Delta_\mathrm{R}(\alpha)}{1\!-\! \beta(\alpha) } \sin\theta_B & \!:\,  \alpha\!<\!1 \,\,\, (\theta_\ast\!=\!0). 
\end{array}\right.
\label{eq:yF}
\end{align}
Given the negative eigenvalues $\lambda_y$ and $\lambda_\theta$ obtained in the SM, this fixed-point turns out to be a {\em stable node} in the linearization and its basin of attraction is found to be the entire channel half where $\tilde y_\mathrm{F}$ is located; i.e., the top (bottom) half $ \tilde H /2 \!<\! \tilde y \!\leq\! \tilde H $ ($ 0 \!\leq\! \tilde y  \!<\!\tilde H/2 $) for prolate (oblate) spheroids. It is thus evident from the above discussion that particles trajectories repelled from the repulsive side of the half-stable central fixed point will be attracted by the stable off-centered fixed point. 

The first relation in Eq. \eqref{eq:yF} and the presence of a finite negative first derivative for the lift velocity, as a stability condition for magnetic focusing of prolate spheroids at $\tilde y_\mathrm{F}$, were previously obtained in  Ref. \cite{GolestanianFocusing}. The eigenvalue analysis, the association of off-centered focusing with two distinct fixed points, the implications of fixed-point stability for the boundaries of this regime and the spheroidal density profiles, which we shall discuss next, have not however been addressed before. 

%%%%%%%%%%%%%%%%%%%%%%%%%%%%%%%%%%%%%
\subsection{Delineation of the off-centered focusing regime}
\label{subsec:boundaries_off}

The  boundaries for the regime of off-centered focusing follow by requiring that, first, the off-centered fixed point remains inside the channel, i.e., $\tilde y_\mathrm{F}\! <\! \tilde H$ ($\tilde y_\mathrm{F} \!>\! 0$) for prolate (oblate) spheroids, and, second, no new nullcline intersection, other than the two noted above,  occurs in this regime. These lead to the {\em upper} and {\em lower} thresholds for $\chi$ (see Section 3.6 of the SM), respectively, as     
\begin{align}
\frac{\chi_{\ast\ast}(\theta_B)}{\mathrm{Pe}_\mathrm{f}}  = 
\left\{ \begin{array}{ll}
\!\dfrac{1+\beta(\alpha)}{2 \Delta_\mathrm{R}(\alpha)} \dfrac{1}{\cos\theta_B} &\,\,\,\, : \,\,\,  \alpha>1, \\ \\
\!\dfrac{ 1-\beta(\alpha)}{2 \Delta_\mathrm{R}(\alpha)} \dfrac{1}{\sin\theta_B} &\,\,\,\, : \,\,\,  \alpha<1. 
\end{array}\right.
\label{eq:chi_2ast}
\end{align}
\vskip-2mm
\begin{align}
\frac{\chi^{\prime}_{\ast\ast}(\theta_B)}{\mathrm{Pe}_\mathrm{f}}  = 
\left\{ \begin{array}{ll}
\!\dfrac{1-\beta(\alpha)}{2\Delta_\mathrm{R}(\alpha)}\dfrac{1}{\sin\theta_B} &\,\,\,\, : \,\,\,  \alpha>1, \\ \\
\!\dfrac{1+\beta(\alpha)}{2\Delta_\mathrm{R}(\alpha)}\dfrac{1}{\cos\theta_B} &\,\,\,\, : \,\,\,  \alpha<1.
\end{array}\right.
\label{eq:chi_2astPrime}
\end{align}  
The self-consistency condition $\chi_{\ast\ast}^\prime(\theta_B)\!<\chi_{\ast\ast}(\theta_B)$ then restricts  the tilt angle as  $\theta_B^{\ast} \!<\!\theta_B \!<\! \pi/2$  ($0\!<\!\theta_B \!<\! \theta_B^{\ast} $) for prolate (oblate) spheroids. Here,  $\theta_B^{\ast}\!=\! \tan^{-1} [(1-\beta)/(1+\beta)]$ is the tilt angle where the boundary curves $\chi_{\ast\ast}^\prime(\theta_B)$ and $\chi_{\ast\ast}(\theta_B)$ meet in the $\theta_B\!-\!\chi$ plane. The parametric span of off-centered focusing can thus be expressed as 
\begin{equation}
\!\!\chi_{\ast\ast}^\prime(\theta_B)\!<\!\chi\!<\!\chi_{\ast\ast}(\theta_B)\,\, \&\, 
 \left\{ \begin{array}{ll}
\!\theta_B^{\ast} \!<\!\theta_B \!<\! \pi/2 & : \,  \alpha\!>\!1, \\  \\
\!0\!<\!\theta_B \!<\! \theta_B^{\ast}  & : \,  \alpha\!<\!1. 
\end{array}\right.
 \label{eq:offcentered_condition}
 \end{equation}
 
%%%%%%%%%%%%%%%%%%%%%%%%%%%%%%%%%%%%%
\subsection{Implications for spheroidal distributions}
\label{subsec:optimal_vs_shoulder}

The attractive off-centered fixed point ensures focusing of prolate (oblate) spheroids  in the top (bottom) channel half  at the latitude  $\tilde y_\mathrm{F}$, Eq. \eqref{eq:yF}. The predicted focusing latitude in fact perfectly agrees with  peak locations of the numerically obtained  density profiles in Fig. \ref{fig:Fig7}.  An illustration of this agreement is provided in Section \ref{sec:separation} (compare symbols and dashed curves in Fig. \ref{fig:Fig9}) where shape-based sorting of spheroids is discussed. 

Despite its accurate prediction of focusing latitude \cite{GolestanianFocusing}, the deterministic approach falls short in capturing other qualitative aspects of off-centered focusing as unraveled by the probabilistic approach. In the deterministic limit, the spheroids are predicted to accumulate (albeit with different pinned orientations, see Section \ref{subsec:fixedpoints_off}) on both fixed-point latitudes $\tilde y \!=\! \tilde H/2$ and $\tilde y_\mathrm{F}$ where the phase-space flow in $\tilde y$-direction vanishes. This leads to a density profile with two $\delta$-function peaks (Appendix \ref{app:deterministic_RT}). It is worth noting that particle-size effects beyond the far-field theory used here can facilitate particle crossing through the centerline \cite{GolestanianFocusing}, in which case a single $\delta$-peak only at $\tilde y_\mathrm{F}$ would emerge. Therefore, the predicted deterministic distribution of spheroids either  drastically differs (with two peaks) from the probabilistic profiles or  (with its one peak) entirely excludes the shoulder formation. This holds regardless of the tilt angle $\theta_B$ or the aspect ratio $\alpha$, while our results in Fig. \ref{fig:Fig7} indicates that these parameters are indeed crucial to the regulation of density shoulders. 

The significance of density shoulders can be underscored by noting that they tie the translational noise effects with the half-stable (higher-order) nature of the central fixed point. That is, rather being absorbed by the attractive side of this fixed point, the spheroids are triggered by the noise to `funnel' through the channel centerline from the attractive side of the fixed point to its repulsive side while they remain orientationally pinned. The steady-state `funneling' flow of spheroids (from the bottom to the top channel half for prolate spheroids and vice versa for oblate spheroids) creates the   plateau-like  shoulders between $\tilde y \!=\! \tilde H/2$ and  $\tilde y_\mathrm{F}$ (the effect can visually be discerned from the 2D PDFs in  Fig. \ref{fig:Fig6} as well). 

We note in passing that the BD simulations of Ref.  \cite{GolestanianFocusing} for prolate spheroids  also appear  to indicate a low-amplitude density tail, extending from the off-centered focusing latitude toward the centerline, which seemingly resembles the density shoulders predicted by us. This is seen from the data in Fig. 2e therein where the tilt angle $\theta_B \!=\! -0.4\pi$. Possible origins or significance of this effect have however not been discussed in Ref. \cite{GolestanianFocusing}. For the mentioned tilt angle, our analysis predicts an optimal peak with no shoulders.  Therefore, the nature of the tail inferred from the said data and its possible connection with the  shoulders reported here remain to be clarified. 

The qualitative picture discussed above can be placed on a semiquantitative basis using the reduced probabilistic theory (Appendix \ref{app:centered_off_RT}) where the virtual lift potential in the regime of off-centered focusing takes an intuitive shape: It features the stable off-centered fixed point as its global minimum and the half-stable central fixed point as its stationary inflection point (for a plot, see Fig. \ref{fig:Up_RT}, main set). The depth of the global minimum varies depending on $\theta_B$ and $\alpha$. Hence, both optimal and shouldered density profiles can be realized  when the minimum depth is sufficiently large and small relative to the noise strength, respectively  (see Fig. \ref{fig:Up_RT}, right inset). 

%%%%%%%%%%%%%%%%%%%%%%%%%%%%%%%%%%%%
\subsection{Optimal focusing criterion}
\label{subsec:optimal_criterion}

Because shouldered focusing may be less advantageous to particle separation purposes, it is desirable to have a measure to differentiate it from optimal focusing. A heuristic  optimal focusing criterion follows by requiring that the ratio between the width of off-centered density peaks (Fig. \ref{fig:Fig7}), which we denote by $2\sigma$, and the effective particle radius $R_{\mathrm{eff}}$ remains smaller than or comparable to a prescribed `resolution' factor $\epsilon$; i.e., in rescaled units,
\begin{equation}
2\tilde \sigma \leq \epsilon.  
\label{eq:opt_criterion}
\end{equation}
The width $\tilde \sigma=\sigma/R_{\mathrm{eff}}$ can be taken as half width at half maximum of the numerically obtained density peaks (Fig. \ref{fig:Fig7}) or as  standard deviation of the analytically obtained peaks within the reduced probabilistic theory (Appendix \ref{app:U_p}). When implemented across  the  $\theta_B\!-\!\chi$ plane, the above criterion gives a boundary curve,  $\chi_\mathrm{o}(\theta_B)$, separating  the two mentioned subregimes. Given the lower and upper boundaries of off-centered focusing, $\chi_{\ast\ast}^\prime(\theta_B)$ and $\chi_{\ast\ast}(\theta_B)$, respectively (Section \ref{subsec:boundaries_off}), the shouldered and optimal focusing subregimes will thus be  bracketed as $\chi^\prime_{\ast\ast}(\theta_B)\!<\! \chi\! < \!\chi_\mathrm{o}(\theta_B)$ and $\chi_\mathrm{o}(\theta_B) \!\leq\! \chi \!<\!\chi_{\ast\ast}(\theta_B)$, respectively, with the range of $\theta_B$ given in Eq. \eqref{eq:offcentered_condition}. This procedure will be used in Section \ref{sec:phase_diagram} to discriminate these subregimes  within a focusing `phase' diagram using the resolution factor $\epsilon\!=\!1$ (a rather stringent choice) and the reduced probabilistic theory to calculate $\sigma$. 

We emphasize that the boundary $\chi_\mathrm{o}(\theta_B)$ can vary depending on how $\epsilon$ and $\sigma$ are prescribed. This is especially true when the density shoulder is highly pronounced and masks the off-centered peak (see, e.g., Fig. \ref{fig:Fig7}c). For this reason, in its potential applications, the criterion \eqref{eq:opt_criterion} can be amended for practical convenience. 

%%%%%%%%%%%%%%%%%%%%%%%%%%%%%%%%%%%%%%
\section{Beyond off-centered focusing}
\label{sec:NearWallFocusing}

The deterministic fixed points  change as  the  field strength and its tilt angle are tuned {\em outside} the regime of off-centered focusing, Eq.  \eqref{eq:offcentered_condition}. These changes  occur under two different scenarios which we detail in Section 3.7 of the SM and summarize  briefly in this section. 

In the first scenario, a {\em third} fixed point emerges due to an intersection between the pinning curve and the nullcline $\theta \!=\! 0$ ($\theta \!=\!\pi/2$) for prolate (oblate) spheroids. This is  additional to the stable off-centered and the half-stable central fixed points of Section \ref{subsec:fixedpoints_off}. It turns out to be unstable and, specifically, a {\em saddle point} \cite{Rasband1990,Nayfeh2008}  at a latitude $\tilde y_\mathrm{S}$ with eigenvalues $\lambda_y \!>\! 0 $ and $\lambda_\theta \!<\!0$; see Eqs. (66) and (67) in the SM. In this case, the off-centered fixed point may continue to exist (as a stable focus) at the latitude $\tilde y_\mathrm{F}$, Eq. \eqref{eq:yF}, or it may completely disappear, while the saddle point and the central fixed point remain intact. When the three fixed points coexist, $\tilde y_\mathrm{S}$ and $\tilde y_\mathrm{F}$  are found in the {\em opposite} halves of the channel.  As a result of the saddle point, deterministic particle trajectories are guided toward the channel wall adjacent to the saddle point, causing partial accumulation of spheroids on that wall. There will be no accumulation on the other wall (opposite channel half) as the stable off-centered fixed point attracts all particle trajectories in its respective channel half. When the off-centered fixed point disappears and the saddle point and the half-stable central fixed point coexist, particle trajectories can also be guided toward the opposite wall upon funneling via the latter fixed point; hence, giving partial spheroid accumulation on {\em both} walls.   

In the second scenario, all other fixed points disappear except for the half-stable central fixed point. Being governed by this solitary fixed point, deterministic particle trajectories are  pushed toward only {\em one} of the walls; the one facing the repulsive side of the half-stable fixed point. 

In the above scenarios, the spheroids are driven toward the wall(s) by the lift velocity along the pinning curve and the wall accumulation is itself effected by the assumed impermeability of the walls. The results in these regimes can thus depend on the specific and detailed modeling of near-wall phenomena, taking them outside the scope of our model. They are included here only for the sake of completeness; see further remarks in Section \ref{sec:Conclusion}. 

%%%%%%%%%%%%%%%
\begin{figure*}[t!]
\begin{center}
	\begin{minipage}[t]{0.4\linewidth}\begin{center}
		\includegraphics[width=\linewidth]{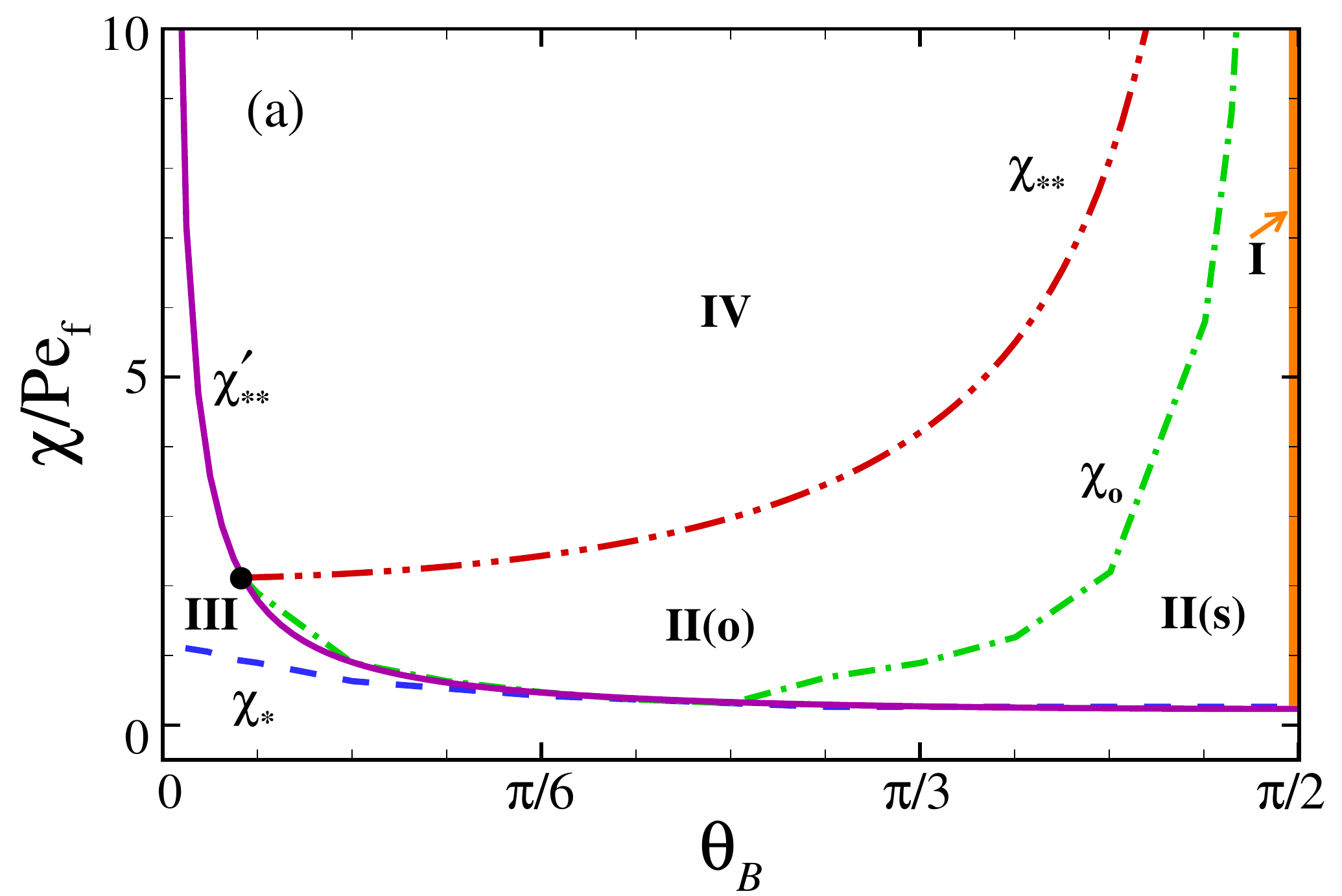} \vskip-2mm %(a) %$\theta_B = \pi/6$
	\end{center}\end{minipage}
	\hskip5mm
	\begin{minipage}[t]{0.4\linewidth}\begin{center}
		\includegraphics[width=\linewidth]{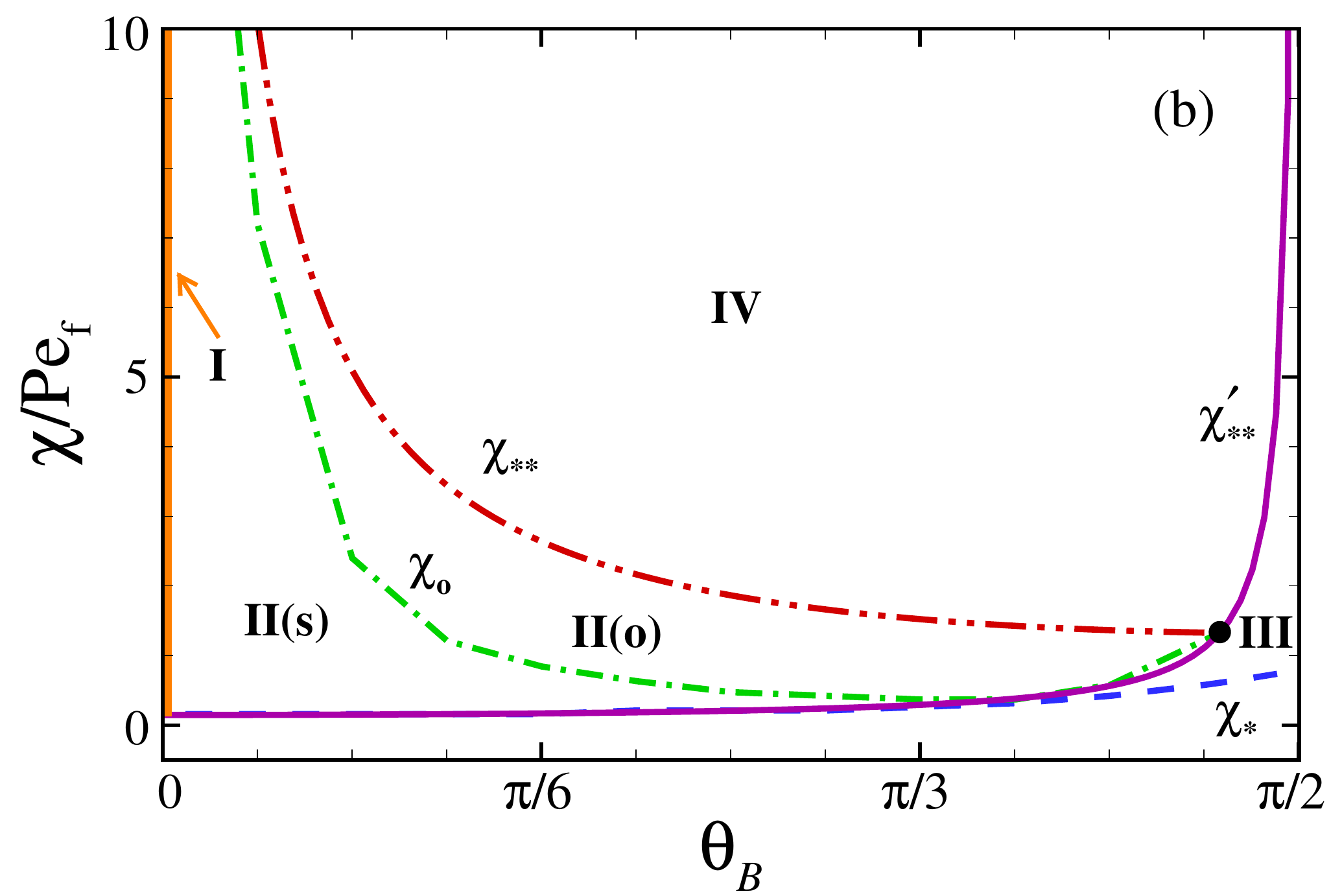} \vskip-2mm %(b)  %$\theta_B = \pi/4$
	\end{center}\end{minipage}
	\vskip-1mm\caption{`Phase' diagrams portraying different (sub)regimes of magnetic focusing for (a)  prolate and (b) oblate spheroids over the $\theta_B\!-\!\chi$ plane, where $\chi$ is the rescaled strength of the applied and $\theta_B$ its tilt angle. We have fixed the rescaled channel width as $\tilde H \!=\! 20$, the flow P\'eclet number as $\mathrm{Pe}_\mathrm{f} \!=\!10^4$, and the particle aspect ratio as $\alpha\!=\!3$ (a) and $1/3$ (b). The focusing (sub)regimes are indicated by Roman numerals according to their labeling in Eq. \eqref{eq:summary}. $\chi_{\mathrm{o}}$ is determined by acquiring data at intervals of $\pi/24$ over the $\theta$-axis using the probabilistic criterion \eqref{eq:opt_criterion} with $\epsilon\!=\!1$ and the reduced probabilistic theory to calculate $\sigma$. 
}
\label{fig:Fig8}
\end{center}
\vskip-4mm
\end{figure*}
%

%%%%%%%%%%%%%%%%%%%%%%%%%%%%%%%%%%%%%
%%%%%%%%%%%%%%%%%%%%%%%%%%%%%%%%%%%%%
\section{Focusing `phase' diagrams}
\label{sec:phase_diagram}

To visualize in $\theta_B\!-\!\chi$ `phase' diagrams the different  (sub)regimes of magnetic focusing that we have identified, we first summarize and label them as follows:  
\begin{align}
\left\{\begin{array}{ll}
\!\textrm{- Partial pinning (no focusing)}&\\
\,\,\,\,0\!<\!\chi\!\leq\! \chi_{\ast}(\theta_B) &
\vspace{3mm}\\
\!\textrm{- Whole-channel pinning (strong fields):} &
\vspace{2mm}\\
\,\,\,\,\textrm{{\bf I}: Centered focusing:} &\\
\,\,\,\,\quad\,\,\chi\!>\!\chi_{\ast}(\theta_B),\,\,\theta_B \!=\! \pi/2\, \mathrm{(prolate)}, 0\, \mathrm{(oblate)} &
\vspace{1mm}\\
\,\,\,\,\textrm{{\bf II}: Off-centered focusing:}&
\vspace{1mm}\\
\,\,\,\,\quad\quad\!\!\textrm{{\bf II(s)}: Shouldered $\chi^\prime_{\ast\ast}(\theta_B)\!<\! \chi \!<\! \chi_\mathrm{o}(\theta_B)$} &
\vspace{1mm}\\
\,\,\,\,\quad\quad\!\!\textrm{{\bf II(o)}: Optimal $\chi_{\mathrm{o}}(\theta_B) \!\leq\! \chi\!<\!\chi_{\ast\ast}(\theta_B)$} &
\vspace{1mm}\\
\,\,\,\,\quad\quad\!\!\theta_B^{\ast} \!<\! \theta_B \!<\! \pi/2\,\, \mathrm{(prolate)} &\\
\,\,\,\,\quad\quad\!\!0\!<\!\theta_B \!<\! \theta_B^{\ast}\,\, \mathrm{(oblate)} &
\vspace{1mm}\\ 
\,\,\,\,\textrm{{\bf III}: Near-wall accumulation (saddle point):}&
\vspace{1mm}\\
\,\,\,\,\quad\quad\!\!\chi_\ast(\theta_B) < \chi \leq \chi_{\ast\ast}^\prime(\theta_B) &
\vspace{1mm}\\ 
\,\,\,\,\textrm{{\bf IV}: Near-wall accumulation (half-stable center):} &
\vspace{1mm}\\
\,\,\,\,\quad\quad\!\!\chi \! \geq\! \chi_{\ast\ast}(\theta_B)\,\, \&\,\, \chi \!>\! \chi^\prime_{\ast\ast}(\theta_B) &\\ 
\end{array}\right.
\label{eq:summary}
\end{align}
where $\theta_B$ is restricted as  $0\!\leq\! \theta_B\!\leq\!\pi/2$ (Section \ref{subsec:nondim_Smol}) and the following points are to be borne in mind: 
\begin{itemize}[leftmargin=13pt] 
\item[$\bullet$]{$\chi_\ast(\theta_B)$ in the onset of whole-channel pinning obtained numerically from Eq. \eqref{eq:pinning}. For prolate spheroids in transverse field ($\theta_B\!=\!\pi/2$), one can use Eq. \eqref{eq:chi_ast_app} which can also be used for oblate spheroids in a longitudinal field ($\theta_B\!=\!0$) by replacing $\beta\! \rightarrow\! -\beta$ (Section 3.2, SM).}

\item[$\bullet$]{$\chi_{\ast\ast}(\theta_B)$ \& $\chi_{\ast\ast}^\prime(\theta_B)$ are the upper/lower boundaries of off-centered focusing from Eqs.  \eqref{eq:chi_2ast} and \eqref{eq:chi_2astPrime}. They meet at $\theta_B^{\ast}\!=\! \tan^{-1} [(1\!-\!\beta)/(1\!+\!\beta)]$ (Section \ref{subsec:boundaries_off}).}

\item[$\bullet$]{$\chi_{\mathrm{o}}(\theta_B)$ is the lower boundary of optimal focusing subregime determined using the procedure in Section \ref{subsec:optimal_criterion}.}

\item[$\bullet$]{All separating boundaries depend on spheroidal shape and aspect ratio $\alpha$. $\chi_{\ast\ast}(\theta_B)/\mathrm{Pe}_\mathrm{f}$ and $\chi^\prime_{\ast\ast}(\theta_B)/\mathrm{Pe}_\mathrm{f}$ are independent of $\mathrm{Pe}_\mathrm{f}$ but $\chi_\mathrm{o}(\theta_B)/\mathrm{Pe}_\mathrm{f}$ can vary with $\mathrm{Pe}_\mathrm{f}$ as it is affected by the noise-induced funneling (Section \ref{subsec:optimal_vs_shoulder}). The noise effects can break the $\chi/\mathrm{Pe}_\mathrm{f}$-scaling in line with our remarks in Sections \ref{subsec:defocusing_circumvented} and 3.3, SM.}

\item[$\bullet$]{Strong-field regimes {\bf I} and {\bf II} constitute the main part of our analysis in this work, while regimes of partial pinning (Section \ref{subsec:partial_pinning}) and wall accumulation {\bf III} and {\bf IV} (Sections \ref{sec:NearWallFocusing} and 3.7, SM) are included for  insights into what may transpire in peripheral regions of  $\theta_B\!-\!\chi$  plane  beyond centered and off-centered focusing.}
\end{itemize}

The representative form of focusing `phase' diagrams summarizing the above (sub)regimes  are shown in  Figs.  \ref{fig:Fig8}a and b  for prolate and oblate spheroids with  $\alpha\!=\! 3$ and $1/3$, respectively,  at fixed $\mathrm{Pe}_\mathrm{f}\!=\!10^4$.  The different (sub)regimes are indicated by their corresponding labels {\bf I}, {\bf II(s)}, {\bf II(o)},  {\bf III} and  {\bf IV}. The boundary curves $\chi_\ast(\theta_B)$,  $\chi_{\ast\ast}^\prime(\theta_B)$, $\chi_\mathrm{o}(\theta_B)$ and $\chi_{\ast\ast}(\theta_B)$  are shown by the dashed blue,  solid purple, dot-dashed green and double-dot-dashed red curves, respectively.   Note that the  centered focusing regime ({\bf I}) coincides  with the vertical (thick orange) line on $\theta_B\!=\!\pi/2$ (0)  in panel a (b).

As seen, the  onset of whole-channel pinning $\chi_\ast(\theta_B)$ increases rather weakly and remains bounded nearly as $\chi_\ast(\theta_B)/\mathrm{Pe}_\mathrm{f}\lesssim 1$ as the tilt angle is decreased (increased) from $\theta_B \!=\! \pi/2$ ($0$) to $\theta_B \!=\! 0$ ($\pi/2$) for prolate (oblate) spheroids.  The diverging trends displayed by $\chi^\prime_{\ast\ast}(\theta_B)$ and $\chi_{\ast\ast}(\theta_B)$ at the limiting ends of the plots are in accord with Eqs. \eqref{eq:chi_2astPrime} and \eqref{eq:chi_2ast}.  The point where $\chi_{\ast\ast}^\prime(\theta_B)$ and $\chi_{\ast\ast}(\theta_B)$ meet is indicated by a bullet, corresponding to $\theta_B^{\ast}\simeq\pi/14$ ($29\pi/30$) in panel  a (b). It is thus clear from the plots that $\theta_B^{\ast}$ represents the {\em minimum} (\emph{maximum}) tilt angle of the field for which off-centered focusing can still be achieved for prolate (oblate) spheroids. For the specific choice of $\alpha$ in the plots, $\chi_\ast(\theta_B)$, $\chi^\prime_{\ast\ast}(\theta_B)$ and $ \chi_{\mathrm{o}}(\theta_B)$ overlap at intermediate values of $\theta_B$ and display a {\em minimum} field-to-shear strength ratio $\chi/\mathrm{Pe}_\mathrm{f}$ (at around $\theta_B\!\simeq\!\pi/4$ and $3\pi/8$ in panels a and b, respectively) for which optimal focusing can still be achieved. 
%{Furthermore, $\chi_\ast(\theta_B)$ and $\chi_{\ast\ast}^\prime(\theta_B)$ merge together at $\theta_B^{\ast\ast}$ (see Section 3.7 of the SM).}  
While these latter specificities of the `phase' diagrams may vary with $\alpha$ or,  in the case of $\chi_{\mathrm{o}}(\theta_B)$, with $\mathrm{Pe}_\mathrm{f}$ or other choices of optimal focusing criterion,  the overall aspects of the diagrams remain the same. Another point to be noted is that, by increasing $\mathrm{Pe}_\mathrm{f}$, the noise effects can be reduced (similar to what we discussed in Section \ref{subsec:defocusing_circumvented}); as a result, the optimal focusing subregime expands due to a downward shift in $ \chi_{\mathrm{o}}(\theta_B)$ (not shown).  

Finally, it is worth mentioning that the right (left) margin of the shouldered focusing subregime {\bf II(s)} in Fig. \ref{fig:Fig8}a (Fig. \ref{fig:Fig8}b) is marked by a smooth crossover from off-centered to centered focusing. That is,  by increasing (decreasing) the tilt angle $\theta_B$ toward $\pi/2$ (0) for prolate (oblate) spheroids at fixed $\chi/\mathrm{Pe}_\mathrm{f}$, the off-centered focusing peak becomes strongly suppressed and merges into the shoulder as it shifts toward the centerline (not shown). This makes the density profiles hardly discernible from those of centered focusing (Fig. \ref{fig:Fig5}) in these marginal crossover regions, where the field-induced defocusing (Section \ref{subsec:defocusing}) will also be present.

%%%%%%%%%%%%%%%%%%%%%%%%%%%%%%%%%%%%%%%%%%
%%%%%%%%%%%%%%%%%%%%%%%%%%%%%%%%%%%%%%%%%%
\section{Application to shape-based sorting}
\label{sec:separation}

Our results can be used to predict efficient strategies for shape-based sorting and separation of sheared magnetic spheroids (when they are subjected to sufficiently strong magnetic fields to ensure their whole-channel pinning).  As noted before, the characteristically sharp and isolated peaks of spheroidal density profiles in the subregime of  optimal focusing (Section \ref{subsubsec:optimal_subregime}, Fig. \ref{fig:Fig7}) provide clear advantages for the purpose. The peaks can be adjusted by tuning the particle aspect ratio $\alpha$, the tilt angle of applied field $\theta_B$, and the field-to-shear strength ratio $\chi/\mathrm{Pe}_\mathrm{f}$ to fulfill the criterion \eqref{eq:opt_criterion}. This should enable access to focused lanes of same-shape spheroids (from a mixture of nonidentical ones) with sufficient lateral separation as also proposed within BD simulations of Ref. \cite{GolestanianFocusing}. This strategy can be substantiated using our  results to identify the range of tilt angles that yield such an efficient shape-based lane separation within the channel. 

For two spheroidal species of aspect ratios $\alpha_1$ and $\alpha_2$ and focusing latitudes $ y_\mathrm{F} (\alpha_1)$ and $ y_\mathrm{F} (\alpha_2)$, an efficient lane separation can thus envisaged when $\Delta y_\mathrm{F} \!\equiv\! | y_\mathrm{F} (\alpha_1) \!-\!  y_\mathrm{F} (\alpha_2)|$  is larger than a set minimal interlane distance (e.g., sum of lane widths, etc). Since we assume optimal focusing via  the criterion \eqref{eq:opt_criterion} (with the choice of $\epsilon\!=\!1$), we merely take the said distance to be twice the effective particle radius; i.e., $\Delta\tilde y_\mathrm{F} \!>\! 2$ in rescaled units. For clarity, this latter criterion is slightly different from that used in Ref. \cite{GolestanianFocusing}, requiring the width of the density peak from BD simulations to be smaller than $\Delta y_\mathrm{F}$. Using Eq. \eqref{eq:yF}, one can express the aforementioned criterion for prolate-prolate and oblate-oblate lane separation as  
\begin{align}
\left\{\begin{array}{ll}
&\!\!\!\!\! \tilde H\! \cos\theta_B \dfrac{\chi}{\mathrm{Pe}_\mathrm{f}}\! \left\lvert \dfrac{\Delta_\mathrm{R}(\alpha_2) }{\beta(\alpha_2) + 1}\! - \!\dfrac{\Delta_\mathrm{R}(\alpha_1) }{\beta(\alpha_1) + 1} \right\rvert \! >\!2  \,\,:\,  \alpha_1, \alpha_2>1, \\
\\
&\!\!\!\!\! \tilde H\! \sin\theta_B \dfrac{\chi}{\mathrm{Pe}_\mathrm{f}}\! \left\lvert \dfrac{\Delta_\mathrm{R}(\alpha_2) }{\beta(\alpha_2) - 1}\! - \!\dfrac{\Delta_\mathrm{R}(\alpha_1) }{\beta(\alpha_1) - 1} \right\rvert \!  >\!2  \,\,:\, \alpha_1, \alpha_2<1,  
\end{array}\right.
\label{eq:sep_criterion_3}
\end{align}
 (note that prolate and oblate shapes are readily separated  as their  focusing latitudes fall in different channel halves). 

\begin{figure}[t!]
\begin{center}
\includegraphics[width=0.8\linewidth]{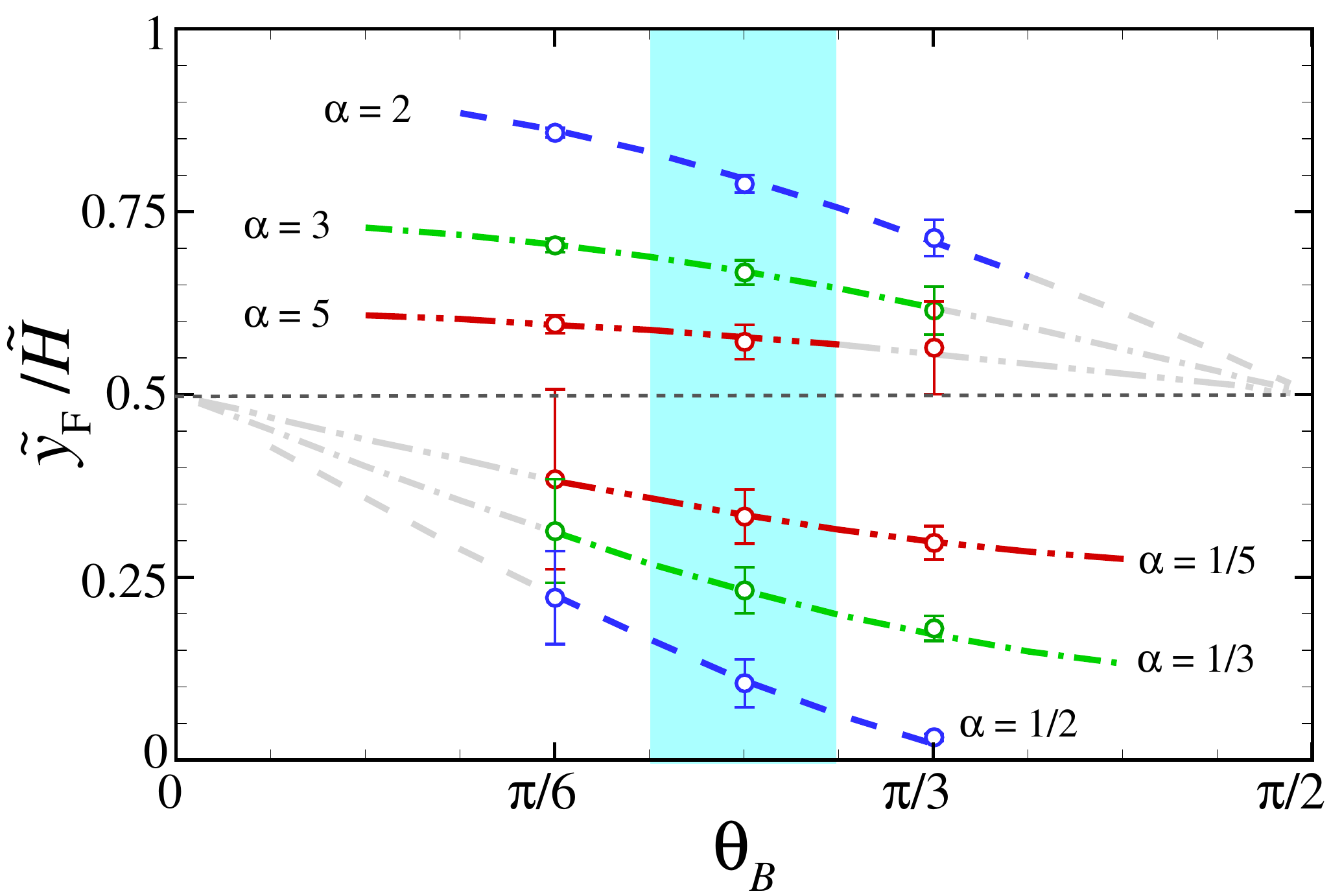}
	\vskip-2mm	
\caption{Focusing latitudes for prolate and oblate spheroids with aspect ratios indicated on the graph  as functions of the tilt angle $\theta_B$ of applied field at fixed $\chi/\mathrm{Pe}_\mathrm{f}\!=\! 1$ and $\tilde H \!=\!20$. Colored (gray) curves show $\tilde y_\mathrm{F}$ from Eq. \eqref{eq:yF} depicted over intervals of $\theta_B$ that fall within the optimal focusing subregime where criterion \eqref{eq:opt_criterion} is satisfied for $\mathrm{Pe}_\mathrm{f}\!=\! 10^4$ ($\mathrm{Pe}_\mathrm{f}\!=\! 10^6$). Note that the gray curves are overlaid with  the colored ones. Open circles show the results  from numerically computed density profiles (Fig. \ref{fig:Fig7}). The error bars for the circles show the width of the corresponding density peaks (see the text).
}
\label{fig:Fig9}
\end{center}
\vskip-4mm
\end{figure}

The utility of inequalities \eqref{eq:sep_criterion_3} is illustrated through Fig. \ref{fig:Fig9} where the focusing latitudes of prolate ($\alpha\!=\!2,3,5$) and oblate spheroids ($\alpha\!=\!1/2,1/3,1/5$) are shown as functions of the tilt angle $\theta_B$ at  fixed $\chi/\mathrm{Pe}_\mathrm{f} \!=\! 1$ and $\tilde H \!=\! 20$. The colored curves show $ y_\mathrm{F}$, Eq. \eqref{eq:yF},  over intervals of $\theta_B$ that fall within the optimal focusing subregime where criterion \eqref{eq:opt_criterion} is satisfied for the chosen set of parameters and $\mathrm{Pe}_\mathrm{f} \!=\! 10^4$. The open circles show the focusing latitudes directly extracted from the full numerical solutions; i.e., from the peak locations of the density profiles in Fig. \ref{fig:Fig7}. The error bars show the full width at half maximum of the PDFs in latitude-orientation space and, thus, the width of focused spheroidal lanes in each case. The inequalities \eqref{eq:sep_criterion_3} for efficient prolate-prolate and oblate-oblate sorting are simultaneously satisfied in the $\theta_B$-interval  (here,  $5\pi/24\!\lesssim\!\theta_B\!\lesssim\!7\pi/24$) shaded in cyan. The lane widths (error bars) can significantly exceed the interlane separations outside the shaded interval, making them less suitable for particle sorting. 

The gray curves in Fig. \ref{fig:Fig9} are also obtained from Eq. \eqref{eq:yF} but with the larger  $\mathrm{Pe}_\mathrm{f}\!=\! 10^6$ at fixed $\chi/\mathrm{Pe}_\mathrm{f} \!=\! 1$. For these cases,  the optimal focusing criterion \eqref{eq:opt_criterion} can be satisfied for tilt angles up to $\pi/2$ (down to $0$) for prolate (oblate) spheroids. Thus, while increasing $\chi$ and $\mathrm{Pe}_\mathrm{f}$  at fixed  $\chi/\mathrm{Pe}_\mathrm{f}$ suppresses  the noise effects and expands the subregime of optimal focusing, it does not create a more advantageous situation for particle separation purposes; i.e., the criterion \eqref{eq:sep_criterion_3} is still not satisfied outside the shaded interval in Fig. \ref{fig:Fig9} due to insufficient interlane separations between particles of different $\alpha$.

%%%%%%%%%%%%%%%

%%%%%%%%%%%%%%%%%%%%%%%%%%%%%%%%%%%%%%%%%%%%%%
%%%%%%%%%%%%%%%%%%%%%%%%%%%%%%%%%%%%%%%%%%%%%%
\section{Summary and concluding remarks} 
\label{sec:Conclusion}

We investigate  Brownian noise (particle diffusivity) effects on magnetic focusing of prolate and oblate spheroids subject to a uniform magnetic field and a plane Poiseuille flow in a microfluidic channel at low Reynolds numbers. The spheroids carry permanent magnetic dipoles, the external field is applied at an arbitrary tilt angle relative to the flow direction, and the field strength is varied across a wide range of values to cover weak to strong fields. The noise effects are  studied via a Smoluchowski equation that governs the joint latitude-orientation PDF of noninteracting or individual spheroids advecting through the channel. Brownian noise is represented by (Stokes) translational and rotational diffusivities for no-slip spheroids and shear and field effects  through  deterministic forces and torques (translational and rotational flux velocities) that they impart on the spheroids. In this approach, pinning of spheroidal orientation (along a specific pinning curve within the latitude-orientation coordinate space) follows self-consistently as a result of the interplay between deterministic shear and field-induced torques without imposing a pinning constraint \cite{GolestanianFocusing,Matsunaga2018}. The magnetic focusing follows as a result of the wall-induced hydrodynamic lift \cite{Zhou2017,ZhouPRA2017,Cao2017,GolestanianFocusing, Matsunaga2018,Zhang2018b,Zhang2018,Sobecki2018,Sobecki2020,Zhang2019}, here bearing signatures of the noise.  Since  the rotational noise is strongly diminished  by orientational pinning of spheroids, the effects reported herein primarily arise  from the translational noise. 

The foregoing full probabilistic approach is supplemented with two other complementary schemes of analysis in the regime of strong fields; i.e., linear and nonlinear stability analysis of fixed points (and their corresponding linearization eigenvalues) for the deterministic spheroidal dynamics, and a semiquantitative (analytical) reduced probabilistic theory describing the lateral broadening of focused spheroidal layer using a virtual lift potential. %These are provided  to elucidate quantitative and qualitative effects that we find at elevated field strengths from the presence of noise. 

We provide novel results and insights into magnetic focusing of spheroids especially within the strong-field regime where noise-induced effects may inaccurately be assumed to be negligible due to the dominant role of the applied field. Our findings are to the contrary. For prolate (oblate) spheroids under transverse (longitudinal) field, where the spheroids are focused at the channel center, the noise effects are most clearly mirrored by  the subGaussian density profile and  field-induced defocusing of spheroids. We show  these to result from a higher-order (nonlinear) stability  of the central fixed point (rather than a linear stability inaccurately noted in the case of prolate particles in Refs. \cite{GolestanianFocusing,Matsunaga2018}). This higher-order stability weakens upon strengthening the field, exacerbating the impact of translational  noise in broadening the width of focused spheroidal layer in a counterintuitive  way amounting to defocusing. Hence, we find that amplifying the applied field to enhance centered focusing of spheroids is a self-limiting factor and becomes counterproductive at sufficiently strong fields.  

Our results also demonstrate that, under a tilted field, where a stable fixed point arises and focuses prolate (oblate) spheroids at an off-centered latitude in the top (bottom) channel half, the central fixed point persists and maintains its  higher-order nature but only as a half-stable one. This enables partial funneling of the spheroids from the bottom to top (top to bottom) channel half in the presence of translational noise, producing pronounced density shoulders that extend from the off-centered focusing latitude to the channel center. We thoroughly discuss the consequences of such noise-induced phenomena on optimal focusing and shape-based sorting of magnetic spheroids as may be desirable in practical applications. Finally, we introduce appropriate `phase' diagrams based on the field strength and tilt angle to illustrate the distinct regimes of magnetic focusing that we identify by the different methods of analysis in this work.

Since our predictions arise in experimentally amenable ranges of parameters (Appendix \ref{app:parameters}), it is useful  to emphasize the limitations of our model and extensions that may be undertaken to improve its extent of applicability. 

First, our analysis is done by assuming that the spheroidal symmetry axis can orient itself only in the 2D plane of applied field. Boundary element simulations and far-field calculations \cite{GolestanianFocusing,Matsunaga2018} confirm the relative accuracy of 2D results by comparing them with those  in 3D rectangular channels of large cross-sectional aspect ratio. It is nonetheless interesting to scrutinize the (translational) noise effects in 3D channels by repeating  our analysis  in the presence of hydrodynamic lift that ensues from the channel confinement in the third dimension \cite{Matsunaga2018}. 

Second, we make use of unconstrained far-field calculations (i.e., without assuming strict  pinning of spheroidal orientation by the applied field) to determine the hydrodynamic lift. Indeed, far-field calculations under strict pinning are shown  by  boundary element simulations to be accurate at spheroid-wall separations beyond a couple particle sizes ($R_{\mathrm{eff}}$) when the channel width is only moderately large (e.g., $H/R_{\mathrm{eff}}\!=\!20$) \cite{GolestanianFocusing,Matsunaga2018}. The finite size of spheroids and near-field hydrodynamics (e.g., lubrication effects, higher-order reflections to improve the first-image self-mobility tensor  \cite{smart1991,MatsunagaPRE} used in the present context, variations of the shear along the spheroids, etc. \cite{kim_microhydrodynamics,Liron1976,ONeill1967,Goldman1967,Goldman1967b,DeMestre1975,lighthill1975,Cichocki1998,Moses2001,Ekiel2008,Cichocki1998,Spagnolie2012,Mathijssen:2016b}) are among other factors whose impact on the deterministic fixed points and their stability against noise-induced broadening can be explored in the future. 

Third, our analysis points to the existence of near-wall accumulation regimes in peripheral regions of  the focusing `phase' diagrams  in Figs. \ref{fig:Fig8}a and b (i.e., regimes {\bf III} and {\bf IV}  which themselves involve finer subregimes that are discussed in Section 3.7 of the SM but not in the main text). These regimes appear when the applied field is very strong and/or its direction is nearly parallel  (perpendicular) to the flow for prolate (oblate) spheroids. Aside from particle-size and near-field effects, the results in these regimes can be influenced by the precise form of the particle-wall steric potential, wall-induced steric torques (not included here) and any additional model-specific boundary features. These would take the near-wall accumulation regimes beyond the scope of our model, which is best suited for field-dominant and far-field  phenomena occurring sufficiently away from the  walls. A consistent approach to these aforementioned regimes should first include the necessary amendments in the model itself.  

Other extensions of the current model include study of paramagnetic particles \cite{Zhou2017,ZhouPRA2017,Cao2017,Zhang2018b,Zhang2018,Sobecki2018,Sobecki2019,Sobecki2020,Zhang2019}, diamagnetic particles in ferrofluid and multiphase flows (see, e.g., Refs. \cite{Zhu2014,Zhou2016,Zhou2016b,Banerjee2019}), and  alternating or rotating fields (see, e.g., Refs. \cite{Abbas2017,Huang2017}). Exploring these venues can shed further light on noise effects in the wider context of magnetic focusing.

%%%%%%%%%%%%%%%%%%%%%%%%%%%%%%%%%%%%%%%%%%%%%
%%%%%%%%%%%%%%%%%%%%%%%%%%%%%%%%%%%%%%%%%%%%% 
\section*{supplementary material}
The Supplementary Material includes expressions for free-spheroid diffusivities, self-mobility tensor, derivations for hydrodynamic stresslet tensor and hydrodynamic image velocities, and stability analyses for deterministic pinning and fixed points of spheroidal dynamics. 

%\section*{Author Declarations}
%\subsection*{Conflict of interest}
\section*{Conflict of interest}
The authors have no conflicts to disclose.

%\subsection*{Author contributions}
\section*{Author contributions}
{\bf Mohammad Reza Shabanniya:} Conceptualization (equal); data curation; formal analysis (lead); investigation (lead); methodology (lead); software; validation (lead); visualization; writing -- original draft preparation (supporting); writing -- review and editing (supporting). {\bf Ali Naji:} Conceptualization (equal); formal analysis (supporting); project administration; investigation (supporting); methodology (supporting); supervision; validation (supporting); writing -- original draft preparation (lead); writing -- review and editing (lead).

\section*{Data Availability}
The data that support the findings of this study are available within the article and its supplementary material.

%%%%%%%%%%%%%%%%%%%%%%%%%%%%%%%%%%%%%
%%%%%%%%%%%%%%%%%%%%%%%%%%%%%%%%%%%%%

\appendix

%%%%%%%%%%%%%%%%%%%%%%%%%%%%%%%%%%%%%
\section{Steric wall potential $V^{(\mathrm{st})}$ and flux velocity $\tilde u_y^{(\mathrm{st})}$}
\label{app:u_st}
 
We model the physical impermeability of channel walls via a repulsive harmonic steric potential as 
\begin{align}
\!\! V^{(\mathrm{st})}(y, \theta)\!=\!\frac{\kappa}{2}
\left\{ \begin{array}{ll}
\!\!\big(y\!-\!y_0(\theta)\big)^2
&: \, y \!<\! y_0(\theta),
\\ 
\!\!\big(y\!-\!H\!+\!y_0(\theta)\big)^2
\!\!\!&: \, y \!>\! H - y_0(\theta), 
\\
\!0
\!\!\!&: \, \mathrm{otherwise},
\end{array}\right.
\label{eq:steric_potential}
\end{align}
where $\kappa$ is the rigidity of the potential and $y_0(\theta)/R_{\mathrm{eff}}\!=\!\alpha^{2/3}(\sin^2\! \theta+\alpha^{-2}\cos^2\!\theta)^{1/2}$ is the closest-approach distance of spheroidal center from the channel walls.  The steric flux velocity, $\mathbf{u}^{(\mathrm{st})}$, is obtained from Eq. \eqref{eq:steric_potential}  as     
\begin{equation}
\mathbf{u}^{(\mathrm{st})}(\mathbf{r}; {\hat{\mathbf d}})\! =\! -  \frac{\partial V^{(\mathrm{st})}}{\partial y} \!\!\left[ \frac{D_{\mypara}(\alpha)}{k_{\mathrm{B}} T} {\hat{\mathbf d}}{\hat{\mathbf d}}  + \! \frac{D_{\myperp}(\alpha)}{k_{\mathrm{B}} T} \!\left(\mathbb{I}- {\hat{\mathbf d}}{\hat{\mathbf d}}\right) \!  \right] \cdot \hat{\mathbf{y}},  
\label{eq:u_st_main}
\end{equation} 
reflecting localized repulsive forces on the spheroids along the outward wall-normals ($\pm\hat{\mathbf{y}}$) when  their centers come closer to the walls than $y_0(\theta)$. Using Eq.  \eqref{eq:u_st_main} and  the rescaling $\tilde V^{(\mathrm{st})}(\tilde y, \theta) \!=\! V^{(\mathrm{st})}(R_{\mathrm{eff}}\,\tilde y, \theta)/(k_{\mathrm{B}} T)$, the rescaled steric flux velocity is found as
\begin{equation}
\!\!\! \tilde u^{(\mathrm{st})}_y(\tilde y, \theta) \!=\!  - \frac{4}{3}\!\left(\! \frac{\partial \tilde V^{(\mathrm{st})}}{\partial \tilde y}\! \right)\!  \big(\Delta_+(\alpha) \!-\! \Delta_-(\alpha) \cos{2\theta}\big),  
\label{eq:u_st}   
\end{equation} 
where  $\Delta_\pm \!=\! (\Delta_{\mypara} \pm \Delta_{\myperp} )/2$ and the shape functions $\Delta_{\mypara,\perp}$ are defined in Section \ref{sec:Model}. In Eq. \eqref{eq:steric_potential}, $\kappa$  can be fixed at a sufficiently large value with little impact on the outcomes; see further notes in  Sections \ref{subsec:analytic_schemes} and  \ref{sec:NearWallFocusing}.

%%%%%%%%%%%%%%%%%%%%%%%%%%%%%%%%%%%%%
\section{Hydrodynamic angular velocity $\tilde{{\omega}}^{(\mathrm{im})}$}
\label{app:w_im}

The derivation of hydrodynamic (image) angular velocity, ${\omega}^{(\mathrm{im})}( y, \theta)$,  requires both components  of the (image) translational velocity, $  u_x^{(\mathrm{im})}( y, \theta)$ and $  u_y^{(\mathrm{im})}( y, \theta)$, as derived in Section 2 of the SM. In rescaled units, we find
\begin{widetext}
\begin{eqnarray}
&&\tilde{{\omega}}^{(\mathrm{im})}(\tilde y, \theta)
= - \frac{\mathrm{Pe}_\mathrm{f}}{64}\frac{\zeta(\alpha)}{\tilde{H}}\left(\!\frac{\tilde{H} - \tilde{y}}{\tilde{y}^3}-\frac{\tilde{y}}{\big(\tilde{H}-\tilde{y}\big)^3}\!\right) \beta(\alpha)  \bigg\{\frac{ 15 X^\mathrm{M}+5 Z^\mathrm{M}}{\beta(\alpha)}+\cos^3 2\theta \left(15X^\mathrm{M}-20Y^\mathrm{M}+5Z^\mathrm{M}-6\beta(\alpha) Y^\mathrm{H}\right) 
\nonumber\\
&& \hskip0cm
 - \cos^2 2\theta \left(15X^\mathrm{M} -20Y^\mathrm{M} + 5 Z^\mathrm{M} +12 \beta(\alpha) Y^\mathrm{H}\right) 
  -  \cos 2\theta \left(15X^\mathrm{M} + 5 Z^\mathrm{M} - 18 Y^\mathrm{H} \right)
 +  \frac{\sin 4\theta}{4} \left(45X^\mathrm{M}-30Y^\mathrm{M}-15Z^\mathrm{M}\right)
\nonumber\\
&& \hskip0cm
  - \frac{\sin 2\theta}{2}   \left( 15 X^\mathrm{M} + 30 Y^\mathrm{M} - 45 Z^\mathrm{M} \right)
\!\!\bigg\}  
+  \frac{3\chi  }{16}\!\left(\frac{1}{\tilde{y}^3}\!+\!\frac{1}{(\tilde{H}\!-\!\tilde{y})^3}\right)\beta(\alpha)
\sin\left(\theta_B\!-\!\theta\right)
\left(  \beta(\alpha) \cos^2 2\theta  +2  \cos 2\theta - 3 \beta(\alpha) \right).
\label{eq:w_im}
\end{eqnarray}
\end{widetext}
{\noindent}Because of its relatively rapid (inverse cube) decay with the distance from the channel walls, $\tilde{\omega}^{(\mathrm{im})}$ remains small relative to the other terms in the net  angular velocity \eqref{eq:w_tot_main} in sufficiently wide channels, including the rescaled channel width $\tilde H\!=\!20$ which is used as a representative value in our work and also in Ref. \cite{GolestanianFocusing} where the validity of far-field hydrodynamic calculations in this case is explicitly established. We have numerically ascertained the negligible impact of $\tilde{\omega}^{(\mathrm{im})}$ especially in the key regimes of centered and off-centered focusing  where the spheroids are sufficiently distanced from the walls.   

%%%%%%%%%%%%%%%%%%%%%%%%%%%%%%%%%%%%%
\section{Choice of parameter values}
\label{app:parameters}

The range of dimensionless parameter values used in our study can be mapped to a wide range of experimentally relevant actual parameter values  for the shear rate $\dot\gamma$, applied field strength $B$, particle dipole moment $m$, channel width $H$, etc. For instance, to explore weak to strong-field effects, we have increased the magnetic coupling parameter (rescaled field strength) up to  $\chi\!=\!10^5$.  Our approach can be used to study even larger values of $\chi$ as desired. For real-life examples, one can consider magnetotactic bacteria whose internal magnetosome chain behaves as a permanent  dipole \cite{Dunin1998}. As representative values for the magnetic moment and the applied field strength, one can use  $m \!\simeq \! 6\!\times\! 10^{-16}\,\mathrm{A}\cdot\mathrm{m}^2$  \cite{Nadkarni2013,Klumpp2019} and $B\! \simeq\! 0.03\,\mathrm{T}$  \cite{Kornig2014}.  These give magnetic couplings up to $\chi \!=\! m B / (k_\mathrm{B}T) \!\simeq\!  4 \!\times\! 10^3$ at room temperature. In synthetic ferromagnetic microparticles (e.g., made by polystyrene microparticles coated with chrome dioxide or other ferromagnetic materials) \cite{Pankhurst2003,Gijs2004,Espy2006,Jiang2006,Smits2016,Aranson2017,Rogowski2021}, the dipole moment can be as large as $m \!\simeq\! 10^{-13}\,\mathrm{A}\cdot\mathrm{m}^2$ or even higher, enabling access to much larger  values of $\chi$ than noted above. In the theoretical modeling of  prolate spheroids in Ref. \cite{GolestanianFocusing}, effective particle radius $R_{\mathrm{eff}} \!=\! 10 \,\mu \mathrm{m}$, channel width $H/R_{\mathrm{eff}}\!=\!20$ and shear rate of  $\dot\gamma \!=\! 100\, \mathrm{s}^{-1}$ have been used at fixed   ratio $ m B/(\eta R_{\mathrm{eff}}^3 \dot\gamma) \!=\!60$. In water at room temperature (shear viscosity $\eta=10^{-3}\, \mathrm{Pa}\cdot \mathrm{s}$), these translate into  $\tilde H\!=\!20$,  $\mathrm{Pe}_\mathrm{f} \!\simeq\! 6.1\times 10^5$ and $\chi \!\simeq\! 1.46 \times 10^6$. 

%%%%%%%%%%%%%%%%%%%%%%%%%%%%%%%%%%%%%
\section{Reduced (1D) probabilistic theory}
\label{app:U_p}

To obtain the reduced probabilistic theory for the noise-induced lateral broadening of  spheroidal distribution   across the channel width, we  focus on the strong-field regime ($\chi\!>\!\chi_\ast$) and adopt the assumptions of Section \ref{subsec:analytic_schemes} according to which the reduced theory will be applicable strictly in  the regimes of centered and off-centered focusing where near-wall effects (Section \ref{sec:NearWallFocusing}) are negligible. We also take advantage of the fact that the PDF, $\tilde \Psi(\tilde y, \theta)$, is exclusively accumulated on and around the deterministic pinning curve $\theta \!=\! \theta_\mathrm{p}(\tilde y)$, see  Figs. \ref{fig:Fig4} and \ref{fig:Fig6}. These enable the following ansatz for the solutions of the full (2D) Smoluchowski equation \eqref{eq:smoluchowski}: 
\begin{equation}
\tilde \Psi(\tilde y, \theta) = \tilde \Phi(\tilde y, \theta)\, G\!\left(\theta-\theta_\mathrm{p}(\tilde y)\right),  
\label{eq:ansatz}
\end{equation}
where $G\!\left(\theta-\theta_\mathrm{p}(\tilde y)\right)$ is a positive definite function with a compact support $\sigma_\theta$ over the $\theta$-axis. Since the angular broadening around the pinning curve is assumed to be small, we have  $\sigma_\theta\!\ll\! 2\pi$. On the leading order (strict pinning), we can thus set this parameter equal to zero to obtain $G\!=\! \delta\!\left(\theta\!-\!\theta_\mathrm{p}(\tilde y)\right)$. The form of the multiplicative function $\tilde \Phi(\tilde y, \theta) $ in Eq. \eqref{eq:ansatz}  is thus expected to be governed only by the translational diffusivity of spheroids and the wall-induced hydrodynamic lift. To establish this property, we plug the ansatz \eqref{eq:ansatz} into Eq. \eqref{eq:smoluchowski} and integrate the resulting equation with respect to $\theta$ over the entire range $[{\theta_B\!-\!\pi},{\theta_B\!+\!\pi})$. 

The above steps lead straightforwardly to the desired {\em reduced} (1D) Smoluchowski equation
\begin{equation}
\frac{\mathrm{d}}{\mathrm{d} \tilde y} \left(\tilde u^{(\mathrm{im})}_\mathrm{p}(\tilde y) \tilde \phi_\mathrm{p}(\tilde y)\right) = \frac{\mathrm{d}^2}{\mathrm{d} \tilde y^2}\Big(\Delta_\mathrm{p}(\tilde y) \tilde \phi_\mathrm{p}(\tilde y)\Big), 
\label{eq:1DFP}
\end{equation}
where $\tilde \phi_\mathrm{p}(\tilde y)\!\equiv\! \tilde \Phi\left(\tilde y, \theta_\mathrm{p}(\tilde y)\right)$ is the reduced PDF or, according to Eq. \eqref{eq:dens_def}, the {\em reduced (lateral) density profile}  of spheroids along the pinning curve, and 
\begin{equation}
\tilde u^{(\mathrm{im})}_\mathrm{p}(\tilde y) \equiv \tilde u^{(\mathrm{im})}_y\!\left(\tilde y, \theta_\mathrm{p}(\tilde y)\right), 
\label{eq:u_Delta_p1}
\end{equation}
\vskip-6mm
\begin{equation}
\Delta_\mathrm{p}(\tilde y) \equiv  \frac{4}{3}\left[\Delta_+(\alpha) - \Delta_-(\alpha) \cos 2\theta_\mathrm{p}(\tilde y) \right],   
\label{eq:u_Delta_p2}
\end{equation}
are the hydrodynamic lift  and  the effective  translational  diffusivity along the pinning curve, respectively. In the regimes of centered and off-centered focusing,  $\tilde \phi_\mathrm{p}(\tilde y)$ and $\mathrm{d}\tilde \phi_\mathrm{p}/\mathrm{d} \tilde y$ diminish on channel walls. These   can be used to obtain the solution of Eq. \eqref{eq:1DFP}  as 
\begin{equation}
\tilde \phi_\mathrm{p}(\tilde y) = \frac{N}{\Delta_\mathrm{p}(\tilde y)} \mathrm{e}^{-\tilde U_\mathrm{p}(\tilde y)},
\label{eq:P1D}
\end{equation}
with the normalization $\int_0^{\tilde H}\!\tilde \phi_\mathrm{p}(\tilde y)\,{\mathrm{d}} \tilde y\!=\!1$ and the {\em virtual lift potential} being defined as
\begin{equation} 
\tilde U_\mathrm{p} (\tilde y) = - \int^{\tilde y}_{0} \frac{\tilde u^{(\mathrm{im})}_\mathrm{p} (\tilde y_1)}{\Delta_\mathrm{p}(\tilde y_1)} \, \mathrm{d} \tilde y_1.   
\label{eq:Up2}
\end{equation}
Despite its resemblance to the Boltzmann weight, Eq. \eqref{eq:P1D} must be viewed as  the PDF of a nonequilibrium steady state. The virtual lift potential is generated by the hydrodynamic lift  along the pinning curve which itself follows from Eqs. \eqref{eq:u_im} and \eqref{eq:pinning} as (Section 3.3, SM)
\begin{widetext}
\begin{eqnarray}
&&\tilde u^{(\mathrm{im})}_\mathrm{p}(\tilde y) =
 - \frac{15}{64} \mathrm{Pe}_\mathrm{f}\, \zeta(\alpha) \left(\frac{1}{\tilde y^2}-\frac{1}{(\tilde H-\tilde y)^2}\right)\!\left(1- \frac{2\tilde y}{\tilde H}\right)  \sin 2\theta_\mathrm{p}(\tilde y) \bigg[\!\left(-X^\mathrm{M}+2Y^\mathrm{M}-Z^\mathrm{M}\right)\cos^2\theta_\mathrm{p}(\tilde y)  \label{eq:lift_on_pinning}\\
 &&\hskip11cm+ \left(2X^\mathrm{M}-2Y^\mathrm{M}\right)\sin^2\theta_\mathrm{p}(\tilde y)  
 -\frac{6}{5}Y^\mathrm{H}\bigg]. \nonumber  
\end{eqnarray}
\end{widetext}

Equations \eqref{eq:P1D} and \eqref{eq:Up2} can be simplified by noting that $\Delta_+\!\gg\! |\Delta_-|$ for a wide range of  aspect ratios (including $1/5\!\leq\! \alpha\!\leq\! 5$ as used in our plots) according to the definition of spheroidal shape functions in Section \ref{sec:prob} and Section 1 of the SM. The effective diffusivity on the pinning curve can thus be approximated as $\Delta_\mathrm{p}(\tilde y)\!\simeq\! \Delta_+ $ and the virtual lift potential as 
\begin{equation} 
\tilde U_\mathrm{p} (\tilde y) \simeq - \frac{1}{\Delta_+}  \int^{\tilde y}_{0}  \tilde u^{(\mathrm{im})}_\mathrm{p} (\tilde y_1) \, \mathrm{d} \tilde y_1. 
\label{eq:Up2b}
\end{equation}

%%%%%%%%%%%%%%%%%%%%%%%%%%%%%%%%%%%%%
\subsection{Deterministic limit from reduced probabilistic theory}
\label{app:deterministic_RT}
  
Before proceeding further, it is useful to note that the {\em relevant} fixed points of the deterministic dynamical system \eqref{eq:dtm_dyn_eqs1} and \eqref{eq:dtm_dyn_eqs2}, i.e., the fixed points located on the pinning curve (Section 3.3, SM) and their characteristic properties (e.g., latitude within the channel, stability class, etc.) are directly embodied into the reduced theory. These are brought in through  the virtual potential  $\tilde U_\mathrm{p}$. This can be seen by a detailed analysis of $\tilde U_\mathrm{p}$ as we discuss in Appendix  \ref{app:centered_off_RT}, or  by inspecting the reduced theory in the full noise-free limit $\Delta_\pm \!\rightarrow\! 0$. In this limit, a steepest descent analysis verifies that the expression \eqref{eq:P1D} tends to $\delta$-function peak(s) at  latitude(s) $\tilde y_\ast$ satisfying      
\begin{equation}
\frac{\mathrm{d}\tilde U_\mathrm{p}}{\mathrm{d} \tilde y}\bigg|_{\tilde y_\ast}=\tilde u^{(\mathrm{im})}_\mathrm{p}(\tilde y_\ast)=0, 
\end{equation} 
or, equivalently, $\tilde u^{(\mathrm{im})}_y\!\left(\tilde y_\ast,\theta_\mathrm{p}(\tilde y_\ast)\right)\!=\!0$. This is nothing but the equation for relevant fixed points  $(\tilde y_\ast,\theta_\ast)\!=\!\left(\tilde y_\ast,\theta_\mathrm{p}(\tilde y_\ast)\right)$. Hence, the  reduced  theory based on Eq. \eqref{eq:P1D} not only consistently yields the desired equation for deterministic fixed points but it also reproduces the expected deterministic $\delta$-peaked density profiles and PDFs,  $\tilde \phi_\mathrm{p}\!\sim\! \delta(\tilde y\! - \!\tilde y_\ast)$ and $\tilde \Psi\!\sim\! \delta(\tilde y \!-\! \tilde y_\ast)\,\delta\!\left(\theta\!-\!\theta_\mathrm{p}(\tilde y_\ast)\right)$, which  we alluded to in Sections \ref{subsubsec:centered_vs_point}, \ref{subsec:defocusing} and \ref{subsec:optimal_vs_shoulder}.
  
%%%%%%%%%%%%%%%%%%%%%%%%%%%%%%%%%%%%%
\subsection{Semiquantitative validity and other considerations}
\label{app:applicability_RT}

Our emphasis on semiquantitative validity of the reduced probabilistic theory can be clarified from the foregoing discussions. That is, in bridging the deterministic approach based on Eqs. \eqref{eq:dtm_dyn_eqs1} and \eqref{eq:dtm_dyn_eqs2} and the full probabilistic approach based on Eq.  \eqref{eq:smoluchowski}, the reduced theory incorporates the deterministic fixed points in a quantitatively accurate manner, while it approximates the Brownian noise effects through the ansatz \eqref{eq:ansatz} (specifically, by neglecting the rotational noise on the leading order). The reduced (1D) expression \eqref{eq:P1D} systematically overestimates the peak amplitude of density profiles relative to their corresponding full (2D) profiles in the main text (see the discussion of Fig. \ref{fig:Up_RT} in Appendix \ref{app:centered_off_RT}) and, hence, a quantitative comparison between them is not attempted. 

%%%%%%%%%%%%%%%%%%%%%%%%%%%%%%%%%%%%%
\subsection{Centered \& off-centered focusing within reduced theory}
\label{app:centered_off_RT}

\begin{figure}[t!]
\begin{center}
\includegraphics[width=0.85\linewidth]{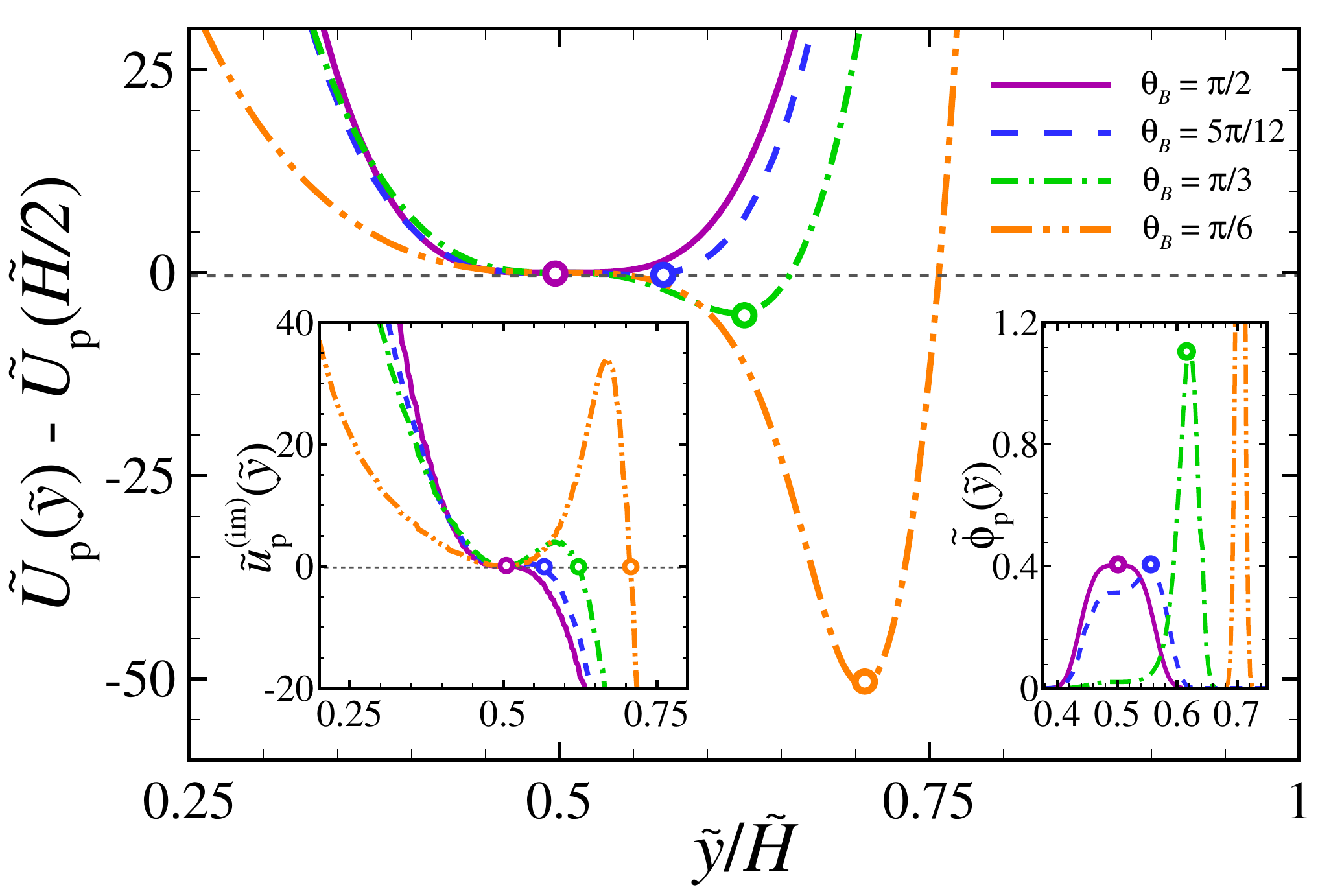}
	\vskip-2mm	
	\caption{Main set: Virtual lift potential $\tilde U_\mathrm{p} (\tilde y)$ from the reduced probabilistic theory as obtained over the pinning curve for $\alpha\!=\!3$, $\tilde H \!=\! 20$, $\mathrm{Pe}_\mathrm{f} \!=\! \chi \!=\! 10^4$, and  different tilt angles $\theta_B \!=\! \pi/2,\,5\pi/12,\,\pi/3,\,\pi/6$. Left and right insets: The corresponding hydrodynamic lift  $\tilde u^\mathrm{(im)}_\mathrm{p} (\tilde y)$ along the pinning curve and the reduced density profile $\tilde \phi_\mathrm{p}(\tilde y)$; see Appendix \ref{app:U_p}. 
		}
\label{fig:Up_RT}
\end{center}
	\vskip-4mm	
\end{figure}

For a given set of parameter values, Eq. \eqref{eq:pinning} can be used to obtain the pinning curve and Eqs. \eqref{eq:lift_on_pinning} and \eqref{eq:Up2b} to obtain $\tilde u^{(\mathrm{im})}_\mathrm{p}(\tilde y)$ and $\tilde U_\mathrm{p}(\tilde y)$ as functions of the latitude $\tilde y$ within the channel. The typical forms of these quantities are shown in Fig. \ref{fig:Up_RT} (main set and left inset) for  $\alpha\!=\!3$, $\tilde H \!=\! 20$, $\mathrm{Pe}_\mathrm{f} \!=\! \chi\! =\!10^4$, $\theta_B\! =\! \pi/2,5\pi/12, \pi/3$ and $\pi/6$. The corresponding reduced density profiles from Eq. \eqref{eq:P1D} are also shown (right inset). For the sake of illustration, in the main set, we have shifted the reference point of the virtual lift potential to channel centerline $\tilde y\!=\!\tilde H/2$ ($ \tilde y\!=\!10$ in the figure) and plotted $\tilde U_\mathrm{p}(\tilde y)\!-\!\tilde U_\mathrm{p}(\tilde H/2)$. This has no effect on the depicted reduced density profiles. In what follows, we discuss this figure and, for  inclusiveness, reproduce some of the key results from the SM (Section 3) where more details can be found.  

{\em Centered focusing.---}In Fig. \ref{fig:Up_RT} and its insets, this case is represented by solid purple curves; i.e., when the field is transverse, $\theta_B \!=\! \pi/2$. The purple curve in the left inset visibly reflects the higher-order nature of the central fixed point (open purple circle) discussed in Section \ref{subsubsec:higherFP_subG}; i.e., the fixed point is a stationary inflection point with $(\partial \tilde u^{(\mathrm{im})}_\mathrm{p}/\partial\tilde y)_{\tilde H/2}\!=\!(\partial^2 \tilde u^{(\mathrm{im})}_\mathrm{p}/\partial\tilde y^2)_{\tilde H/2}\!=\!0$ or, equivalently, a nonhyperbolic, neutrally stable one with $\lambda_y \!=\! 0$ in the linearization as follows from  Eq. \eqref{eq:Jacobian}.  These hold similarly for centered focusing of oblate spheroids (i.e., when the tilt angle is $\theta_B \!=\! 0$; not shown). To the leading order, we find  $\tilde u^{(\mathrm{im})}_\mathrm{p}(\tilde y)\!\simeq\! \frac{1}{3!} (\tilde y\! -\! \tilde H/2)^3 (\partial^3 \tilde u^{(\mathrm{im})}_\mathrm{p}/\partial\tilde y^3)_{\tilde H/2}$ where $(\partial^3 \tilde u^{(\mathrm{im})}_\mathrm{p}/\partial\tilde y^3)_{\tilde H/2}\!<\!0$ as can be inferred from the left inset. We then have the quartic form $\tilde U_\mathrm{p} (\tilde y) \!\simeq\! k (\tilde y \!-\! \tilde H/2)^4$ (main set) where $k\!\sim\! |\partial^3 \tilde u^{(\mathrm{im})}_\mathrm{p}/\partial\tilde y^3|_{\tilde H/2}$ inversely scales  with the translational noise strength; see Eq. \eqref{eq:Up2b}.  This leads to the subGaussian (flat-top) reduced density \eqref{eq:P1D} $\tilde \phi_\mathrm{p}\!\sim\! \exp [- k(\tilde y \!-\! \tilde H/2)^4]$ around the centerline (purple curve, right inset) as we discussed in Section \ref{subsubsec:higherFP_subG}. 

The aforementioned negativity of the third derivative for the lift and, thus, the higher-order stability of central fixed point can be proven by explicit calculation, yielding 
\begin{equation}
\left(\!\frac{\partial^3 \tilde u^{(\mathrm{im})}_\mathrm{p}}{\partial\tilde y^3}\!\right)_{\!\!\tilde H/2}\! = C_\mathrm{R}(\alpha)\, \mathrm{Pe}_\mathrm{f}\,\frac{ 2 \zeta(\alpha)}{  \tilde H^4} \left(\!\frac{\partial \theta_\mathrm{p}}{\partial \tilde y}\!\right)_{\!\!\tilde H/2}, 
\label{eq:3rd_derivative}
\end{equation}
where $C_\mathrm{R}(\alpha)$ is expressed in terms of spheroidal resistance functions. It turns out to be {\em negative} for both prolate and oblate spheroids (Section 3.5, SM).  The quantity $({\partial \theta_\mathrm{p}}/{\partial \tilde y})_{\tilde H/2}$ gives the slope of the  pinning curve at the channel center.  In Figs. \ref{fig:Fig4} and \ref{fig:Fig6} and Fig. 1 of the SM, the pinning curve is visualized by its inverse function $\tilde y = \tilde y_\mathrm{p}(\theta)$. As seen from these figures, the  slope (or its inverse) at the center is {\em positive}. The third derivative \eqref{eq:3rd_derivative} is thus negative for both types of spheroids. The above results also establish the relation $k \!\sim\!|\partial\theta_\mathrm{p}/\partial \tilde y|_{\tilde H/2}$ that we noted  in relation to defocusing in Section \ref{subsec:defocusing}. 

%%%%%%%%%%%%%%%%

{\em Off-centered focusing.---}In Fig. \ref{fig:Up_RT} and its insets, this case is represented by the field tilt angles $\theta_B\! = \!5\pi/12$ (dashed blue), $\pi/3$ (dot-dashed green) and $\pi/6$ (double-dot-dashed orange curves). In all cases, the lift velocity (left inset) displays a nonmonotonic profile featuring two separate fixed points, one at the center $\tilde y\!=\!\tilde H/2$ and the other one at the off-centered latitude $\tilde y_\mathrm{F}\!>\!\tilde H/2$, Eq. \eqref{eq:yF} in the top channel half (open same-color circles). 

The central fixed point is visibly half-stable and, thus, again of higher order. It is a local minimum for the lift velocity, $(\partial \tilde u^{(\mathrm{im})}_\mathrm{p}/\partial\tilde y)_{\tilde H/2}\!=\!0$ and $(\partial^2 \tilde u^{(\mathrm{im})}_\mathrm{p}/\partial\tilde y^2)_{\tilde H/2}\!>\!0$ (left inset). It can thus enable a noise-driven translational flux from the bottom to the top channel half (Section \ref{subsec:optimal_vs_shoulder}). The stability of the off-centered fixed point can also be inferred visually from the negative first derivatives that the lift velocity curves show at its locus (open blue, green and orange circles, left inset). These property are manifested more intuitively by the virtual lift potential (main set) where the half-stable central fixed point appears as a stationary inflection point and the stable off-centered fixed point as a global minimum. The lift potential for oblate spheroids (not shown) has a similar shape except that $\tilde y_\mathrm{F}$ is located in the bottom channel half; also, the half-stable central fixed point emerges as a local maximum of the lift velocity $\tilde u^{(\mathrm{im})}_\mathrm{p}(\tilde y)$. 

For both prolate and oblate spheroids, the foregoing observations based on the lift potential are corroborated by the stability analysis of the fixed points in Section 3.6 of the SM where explicit forms of the mentioned second derivatives, the linearization eigenvalues for both fixed points, and the conditions for existence of the stable off-centered fixed point at $\tilde y_\mathrm{F}$ are detailed and used in our discussions in Sections \ref{subsec:fixedpoints_off} and \ref{subsec:boundaries_off}. 

Finally, as seen in Fig. \ref{fig:Up_RT}, right inset, the reduced density profiles obtained from the lift potential clearly demonstrate both optimal and shouldered subregimes that we discussed using our full (2D) probabilistic results in Sections \ref{subsubsec:optimal_subregime} and \ref{subsubsec:shouldered_subregime}; i.e., the off-centered density peaks can emerge either as sharp localized peaks  or as skewed peaks with a left shoulder stretching toward the channel center. The shoulder can be so pronounced as to mar the off-centered peak (dashed blue curve). 

In line with our discussion in Appendix \ref{app:applicability_RT}, the reduced profiles in Fig. \ref{fig:Up_RT} closely mimic the full probabilistic profiles in Fig. \ref{fig:Fig7}. This is expected as the focusing peaks  and shoulders are linked with the same underlying fixed points. However,  the peak amplitude is clearly overestimated by the reduced profiles (compare the curves in the  right inset with dot-dashed green curves in the three panels of Fig. \ref{fig:Fig7} where we also have $\alpha\!=\!3$; the same effect applies to the centered peak of the purple curve which  can be compared with the dashed blue curve in Fig. \ref{fig:Fig5plus2}b). For this reason, the corresponding shoulder amplitudes are underestimated by the reduced density profiles. Despite these limitations, the reduced probabilistic theory can nevertheless be used to discriminate the two subregimes of optimal and shouldered focusing (Section \ref{subsec:optimal_criterion}).

%%%%%%%%%%%%%%%%%%%%%%%%%%%%%%%%%%%%%
\subsection{Comparison with the quadratic lift potential in Ref. \cite{GolestanianFocusing}}
\label{app:centered_off_comparison}

The existence of an effective lift potential close to the off-centered fixed point was reported using a Langevin equation in Ref. \cite{GolestanianFocusing}, where the potential was shown to be of  quadratic form. This corresponds to the quadratic shape of our lift potential  in the vicinity of its  global off-centered minimum in Fig. \ref{fig:Up_RT}.

 Our derivation in the foregoing appendices shows that a generalized virtual lift potential can be defined systematically along the {\em whole} pinning curve. Also, our elaborations in Sections \ref{subsubsec:higherFP_subG}, \ref{subsec:defocusing}, \ref{subsec:optimal_vs_shoulder} and Appendix  \ref{app:centered_off_RT}  indicate that the knowledge of the lift potential across the channel width and not only around the focusing latitude is important in achieving the full picture; e.g.,  in making a distinction between the subregimes of optimal focusing and the noise-induced shouldered focusing.

\bibliography{HydroMagnetic_v6} 

\end{document}

\title{{\em Supplementary Material for the article:}\\Brownian noise effects on magnetic focusing of prolate and oblate spheroids in channel flow}

\author{Mohammad Reza Shabanniya}
\affiliation{School of Physics, Institute for Research in Fundamental Sciences (IPM), Tehran 19538-33511, Iran}
\author{Ali Naji}
\thanks{Corresponding author. Email: a.naji@ipm.ir}
\affiliation{School of Nano Science, Institute for Research in Fundamental Sciences (IPM), Tehran 19538-33511, Iran}

%\date{\today}

\begin{abstract} 
In this Supplementary Material, we provide the following details on the model system and analytical results that are reported in the main text of the article: (i) Explicit forms of translational and rotational bulk diffusivities of free no-slip spheroids; (ii) hydrodynamic self-mobility tensor of spheroids confined to a planar channel of no-slip walls; (iii) derivation of hydrodynamic stresslet tensor for (ferro)magnetic prolate and oblate spheroids  under a stationary, plane Poiseuille flow and an externally applied static magnetic field; (iv) derivation of wall-induced hydrodynamic translational and angular velocities (including the lift velocity) induced by the low-Reynolds-number hydrodynamic `images' of spheroids in the channel confinement; (v) deterministic (noise-free) pinning of  spheroidal orientation, and (iv) deterministic stability analysis  and classification of the pertinent fixed points for the coupled translational and rotational  dynamics of magnetic spheroids in the aforesaid context.  
\end{abstract}

%\pacs{}
%\keywords{}

\maketitle

%\tableofcontents
%\makeatletter
%\let\toc@pre\relax
%\let\toc@post\relax
%\def\l@subsection#1#2{}
%\def\l@subsubsection#1#2{}
%\makeatother 
%\makeatletter
%\def\l@subsubsection#1#2{}
%\makeatother

%\newpage 

%%%%%%%%%%%%%%%%%%%%%%%%%%%%%%%%%%%%%%%%%%%%%%%
%%%%%%%%%%%%%%%%%%%%%%%%%%%%%%%%%%%%%%%%%%%%%%%
\section{Diffusivities of free spheroids}
\label{sec:diffusivities}

%%%%%%%%%%%%%%%%%%%%%%%%%%%%%%%%%%%%%%%%%%%%%%%
\subsection{Constant-volume spheroids}
\label{subsec:cte_vol}

In the model system described in Section II of the main text, spheroidal particles of different aspect ratio $\alpha$ are considered with equal and constant volume ${\mathcal V}(\alpha)={\mathcal V}_0$ and equal permanent magnetic dipole moment $\mathbf{m}=  m \hat{\mathbf d}$ \cite{GolestanianFocusing,Matsunaga2018,Zhou2017,ZhouPRA2017} which points along their axis of symmetry. This axis is  identified by the unit vector $\hat{\mathbf d}$ and, with no loss of generality, we take $m\geq0$.  For prolate (oblate) spheroids, the axis of symmetry is the major (minor)  body axis and its   half-length is denoted by $a$. The  aspect ratio is defined as $\alpha = a/b$  where $b$ is the half-length of the other two equal spheroidal axes. We shall later require the half-length  of the {\em major} body axis (i.e., $a$ and $b$ for prolate and oblate spheroids, respectively) as a function of $\alpha$. For simplicity, we rename this quantity as $\ell(\alpha)$. For constant-volume spheroids, one has    
\begin{equation}
\ell(\alpha) = \left[\zeta(\alpha)\right]^{1/3} R_{\mathrm{eff}}, 
\label{eq:half-major}
\end{equation}
where $R_{\mathrm{eff}}= \left(3{\mathcal V}_0 /4 \pi\right)^{1/3}$ is the effective particle radius, or the radius of reference sphere with $\alpha = 1$, and we have 
\begin{align}
\zeta(\alpha) =  \left\{\begin{array}{ll}
\!\alpha^2&\,\,:\, \alpha>1 \,\,\,\, \mathrm{(prolate)}, \\ \\
\!\alpha^{-1}&\,\,:\, \alpha<1 \,\,\,\, \mathrm{(oblate)}. 
\end{array}\right.
\label{eq:zeta}
\end{align}

%%%%%%%%%%%%%%%%%%%%%%%%%%%%%%%%%%%%%%%%%%%%%%%
\subsection{Shape functions for translational and rotational diffusivities}
\label{subsec:shape_functions}

Our probabilistic formulation based on Eq. (1) of the main text involves the longitudinal and transverse translational diffusivity of spheroids, ${D}_{\mypara}(\alpha)$ and ${D}_{\myperp}(\alpha)$ respectively, relative to their axis of symmetry as well as their in-plane rotational diffusivity, $D_\mathrm{R}(\alpha)$.  Recall that  spheroidal particles in our model   are  studied within a channel whose wall effects  are  explicitly included through spheroid-wall  steric and hydrodynamic interactions. Therefore,  the said  diffusivities must be treated as bulk or bare diffusivities of spheroids. Adopting the Stokes-Einstein relations for free  no-slip spheroids, these diffusivities can be expressed in terms of well-known spheroidal shape functions  \cite{Perrin,koenig,happelbook}  
\begin{equation}
\Delta_{\mypara,\myperp} (\alpha)={D}_{\mypara,\myperp}(\alpha)/D_{0\mathrm{T}}, \qquad \Delta_\mathrm{R}(\alpha) =D_\mathrm{R}(\alpha)/ D_{0\mathrm{R}},  
\end{equation}
where $D_{0\mathrm{T}}= k_{\mathrm{B}} T / (6\pi\eta R_{\mathrm{eff}})$  and $D_{0\mathrm{R}}= k_{\mathrm{B}} T / (8\pi\eta R_{\mathrm{eff}}^3)$ are the translational and rotational diffusivities  of reference sphere ($\alpha = 1$),  $\eta$ is the fluid shear viscosity and  $k_{\mathrm{B}} T$ is the ambient thermal energy scale. For constant-volume spheroids, the shape functions can be rewritten as \cite{Perrin,koenig,MRSh1} 
\begin{equation}
\Delta_{\mypara,\myperp}(\alpha) =\alpha^{1/3}G_{\mypara,\myperp}^{-1}(\alpha)  \quad \text{and}\quad \Delta_\mathrm{R}(\alpha) =G_\mathrm{R}^{-1}(\alpha),  
\end{equation}
where, for prolate spheroids $(\alpha>1)$, one explicitly has    
\begin{eqnarray}
&&G_\mypara
= 
\frac{8}{3} 
\Bigg[
\frac{-2\alpha}{\alpha^2-1}
+\frac{2\alpha^2-1}{\left(\alpha^2-1\right)^{3/2}} \ln{ \left( \frac{\alpha+\sqrt{\alpha^2-1}}{\alpha-\sqrt{\alpha^2-1}} \right) }
\Bigg]^{-1}\!\!\!\!\!\!,
 \nonumber\\
&&G_\myperp
=
\frac{8}{3} 
\Bigg[
\frac{\alpha}{\alpha^2-1}
+\frac{2\alpha^2-3}{\left(\alpha^2-1\right)^{3/2}} \ln{ \left( {\alpha+\sqrt{\alpha^2-1}}\right) }
\Bigg]^{-1}\!\!\!\!\!\!,
 \nonumber\\
&&G_\mathrm{R} =
\frac{2}{3} \frac{\alpha^4-1}{\alpha}
\Bigg[
\frac{2\alpha^2-1}{\sqrt{\alpha^2-1}} \ln\left({\alpha+\sqrt{\alpha^2-1}}\right)-\alpha
\Bigg]^{-1}\!\!\!\!\!\!,
\label{eq:fullF_prolate}
\end{eqnarray}
and, for oblate spheroids $(\alpha<1)$, 
\begin{eqnarray}
&&G_\mypara
= 
\frac{8}{3} 
\Bigg[
\frac{-2\alpha}{\alpha^2-1}
-\frac{2\left(2\alpha^2-1\right)}{\left(1-\alpha^2\right)^{3/2}} \tan^{-1}{ \left(\frac{\sqrt{1-\alpha^2}}{\alpha} \right) }
\Bigg]^{-1}\!\!\!\!\!\!,
 \nonumber\\
&&G_\myperp
=
\frac{8}{3} 
\Bigg[
\frac{\alpha}{\alpha^2-1}
-\frac{2\alpha^2-3}{\left(1-\alpha^2\right)^{3/2}} \sin^{-1}{\sqrt{1-\alpha^2}   }
\Bigg]^{-1}\!\!\!\!\!\!,
 \nonumber\\
&&G_\mathrm{R} =
\frac{2}{3} \frac{\alpha^4-1}{\alpha}
\Bigg[
\frac{2\alpha^2-1}{\sqrt{1-\alpha^2}} \tan^{-1}{ \left(\frac{\sqrt{1-\alpha^2}}{\alpha} \right) }-\alpha
\Bigg]^{-1}\!\!\!\!\!\!.
\label{eq:fullF_oblate}
\end{eqnarray} 

In the Smoluchowski equation (11) in the main text, $\Delta_{\mypara,\myperp}$ enter through the combinations $\Delta_\pm = (\Delta_{\mypara} \pm \Delta_{\myperp} )/2$ which can be calculated in terms  of $\alpha$ from the above expressions. For a wide range of  aspect ratios (e.g., $1/5\leq \alpha\leq 5$), one can directly verify that $\Delta_+\gg |\Delta_-|$. This  approximation is used in our reduced probabilistic theory (Appendix D, main text), while the exact expressions   \eqref{eq:fullF_prolate} and \eqref{eq:fullF_oblate} are used in our full  numerical  analysis of Eq. (11) in the main text. 

%%%%%%%%%%%%%%%%%%%%%%%%%%%%%%%%%%%%%%%%%%%
%%%%%%%%%%%%%%%%%%%%%%%%%%%%%%%%%%%%%%%%%%%
\section{Hydrodynamic effects due to channel confinement}
\label{sec:hydro_confinement}

%%%%%%%%%%%%%%%%%%%%%%%%%%%%%%%%%%%%%%%%%%%
\subsection{Scope and overview}
\label{subsec:image_scope}

In this section, we detail our derivations for the wall-induced hydrodynamic translational and angular flux velocities that are defined in Section III of the main text; i.e., ${\mathbf u}^{(\mathrm{im})}(\mathbf{r}; {\hat{\mathbf d}})$ and $\boldsymbol{\omega}^{(\mathrm{im})}(\mathbf{r}; {\hat{\mathbf d}})$, respectively. In the context of our model, these quantities represent hydrodynamic {\em self-interactions} of a spheroid at position $\mathbf{r}=(x,y,z)$ and axial orientation  $\hat{\mathbf d}$ within the channel with its  {\em first} hydrodynamic singularity `image' produced at each of the two channel walls.
\vskip2mm

Before we proceed further, it is useful to recall and summarize the hydrodynamic scope of our model:
\begin{itemize}[leftmargin=20pt, rightmargin=13pt]  
\item[$\bullet$]{Our model treats the spheroids as individual or noninteracting particles (as in a dilute suspension) which advect under an imposed, stationary, Poiseuille flow through a long planar channel with rigid no-slip walls. The model is studied under the assumption of low particle and channel Reynolds numbers.}

\item[$\bullet$]{Our forthcoming hydrodynamic calculations are performed under the general  assumptions of far-field theory which involve neglecting finite particle size and near-field hydrodynamic effects. These include lubrication effects, higher-order reflections beyond the first image in the self-mobility tensor  \cite{smart1991,MatsunagaPRE} which is used in the calculation of ${\mathbf u}^{(\mathrm{im})}$ and $\boldsymbol{\omega}^{(\mathrm{im})}$, changes in the shear rate along the spheroids, etc.; see Refs. \cite{kim_microhydrodynamics,Liron1976,ONeill1967,Goldman1967,Goldman1967b,DeMestre1975,lighthill1975,Cichocki1998,Moses2001,Ekiel2008,Cichocki1998,Spagnolie2012,Mathijssen:2016b} and references therein. These assumptions are expected to be valid in channels of  sufficiently large width ($H$) relative to the particle radius ($R_{\mathrm{eff}}$). For the  representative rescaled channel width $\tilde H\!\equiv\!H/R_{\mathrm{eff}}\!=\!20$ used in our analysis in the main text and also in Refs. \cite{GolestanianFocusing,Matsunaga2018}, the validity of far-field hydrodynamic calculations (specifically at spheroid-wall separations beyond a couple particle sizes) has explicitly been established  through boundary element simulations \cite{GolestanianFocusing,Matsunaga2018}.} 
\end{itemize}

It is also useful to recall the scope and advantages of our approach to magnetic focusing  given that it necessitates certain aspects of the hydrodynamic calculations to be generalized relative to  previous works on similar models:
\begin{itemize}[leftmargin=20pt, rightmargin=13pt]  
\item[$\bullet$]{{\em Brownian noise effects.---}The existing literature on lateral migration and focusing of magnetic spheroids under uniform applied fields consider both paramagnetic and ferromagnetic spheroids in simple or pressure-driven microchannel flows \cite{Zhou2017,ZhouPRA2017,Cao2017,GolestanianFocusing, Matsunaga2018,Zhang2018b,Zhang2018,Sobecki2018,Sobecki2020}. Yet, Refs.  \cite{GolestanianFocusing,Zhang2018b} appear to be the only ones that consider {\em noninteracting ferromagnetic} particles (with permanent dipole) in the two-dimensional setting of a {\em plane Poiseuille flow} similar to the present model. These latter studies are conducted on prolate spheroids using deterministic analysis of noise-free particle trajectories and boundary element simulations \cite{GolestanianFocusing} as well as direct finite-element numerical simulations \cite{Zhang2018b}. Except for a brief report of Brownian Dynamics (BD) simulations in Ref. \cite{GolestanianFocusing}, a thorough analysis of noise effects have not been considered before, while it forms the main part of our work in the main text (see detailed discussions of our results in connection with those of Ref. \cite{GolestanianFocusing} in, e.g., Sections IV and V and Appendix D 4 of the main text, and further remarks relating to the aforementioned  BD simulations in Sections V D and VIII therein).} 

\item[$\bullet$]{{\em Pinning constraint (or lack thereof).---}When (ferro)magnetic spheroids are studied subject to a uniform applied field of arbitrary tilt angle relative to the flow, our full probabilistic approach in the main text imposes {\em no} orientational pinning constraints on the spheroids. The field-induced pinning instead follows {\em as an outcome} from our numerical analysis of the governing Smoluchowski equation. A strict pinning constraint on the spheroidal  orientation, as used in the deterministic analysis of Refs. \cite{GolestanianFocusing,Matsunaga2018},  will exclusively be relevant to the strong-field, or the whole-channel pinning, regime that we characterize in detail in Sections IV-VII of the main text.} 

\item[$\bullet$]{{\em Generalized (unconstrained) far-field calculations.---}In the absence of a pinning constraint, our forthcoming far-field expressions for the stresslet tensor, the hydrodynamic lift, etc., generalize those reported in Ref. \cite{GolestanianFocusing} and can be applied to the whole range of weak to strong field strengths. This generalization has not been considered in the literature before.}

\item[$\bullet$]{{\em Other complementary approaches used in the main text.---}The unconstrained far-field expressions as noted above are used not only in our full probabilistic approach but also in our {\em linear} stability analysis of deterministic fixed points describing phase-space dynamics of noise-free spheroids. However, we do use the pinning constraint in our {\em nonlinear} stability analysis of higher-order fixed points by considering the higher-order phase-space flows exclusively over the pinning curve. The results of our linear and nonlinear stability analyses are summarized in the main text and the details are given at length in Section \ref{sec:deterministic} of this Supplementary Material. In the main text (Appendix D), we also present a reduced probabilistic theory which makes use of the pinning constraint on the leading order.}

\item[$\bullet$]{{\em Prolate vs oblate spheroids.---}Our study includes both prolate and oblate spheroids even as the latter have received less attention \cite{Mody2005,Einarsson2015,Michelin2016,Michelin2017} and have not specifically been analyzed  in the context of magnetic focusing under uniform fields.}
\end{itemize}

The rest of this section is organized as follows: In Section \ref{subsec:self-mobility}, we present the explicit form of equal-point Green function, or self-mobility tensor $\mathbb{K}^{(\mathrm{im})}(\mathbf{r},\mathbf{r})$. In Section \ref{subsec:stresslet}, we calculate the second-order stresslet tensor $\mathbb{S}(\mathbf{r}; {\hat{\mathbf d}})$ as customarily used \cite{kim_microhydrodynamics} to model far-field stationary flow of sheared no-slip spheroids at low Reynolds numbers  \cite{kim_microhydrodynamics}. In Section \ref{subsec:u_im}, the term ${\mathbf u}^{(\mathrm{im})}(\mathbf{r}; {\hat{\mathbf d}})$ is calculated by applying the singularity image method of Blake \cite{Blake1971}. In Section \ref{subsec:omega_im}, the term $\boldsymbol{\omega}^{(\mathrm{im})}(\mathbf{r}; {\hat{\mathbf d}})$ is obtained using  ${\mathbf u}^{(\mathrm{im})}(\mathbf{r}; {\hat{\mathbf d}})$  where we also derive the standard forms of the shear and field-induced angular velocities, $\boldsymbol{\omega}_\mathrm{f}(\mathbf{r}; {\hat{\mathbf d}})$ and $\boldsymbol{\omega}_{\mathrm{ext}}({\hat{\mathbf d}})$ respectively, as reproduced in Section III of the main text.

%%%%%%%%%%%%%%%%%%%%%%%%%%%%%%%%%%%%%%%%%%
\subsection{Equal-point Green function (hydrodynamic self-mobility)  tensor}
\label{subsec:self-mobility}

 In a channel with plane-parallel walls, the hydrodynamic self-mobility tensor, $\mathbb{K}^{(\mathrm{im})}(\mathbf{r},\mathbf{r})$, relative to a given  wall of the channel on the first-image level is  given by \cite{smart1991,MatsunagaPRE}
 % \AN{G. K. Batchelor, J. Fluid Mech. 41, 545 (1970). J. R. Smart and D. T. Leighton, Jr., Phys. Fluids A: Fluid Dyn. 3, 21 (1991).} \AN{G. K. Batchelor, J. Fluid Mech. 41, 545 ( 1970). ‘W. R. Schowaiter, C. E. ChaiTey, and H. Brenner, J. Colioid Interface Sci. 26, 152 (1958). “N. A. Frankel and A. Acrivos, J. Fluid Mech. 44,65 ( 1970). ‘*M. Shapira and S. Haber, Jnt. 3. Multiphase Flow 14,483 (1988). IaM. Shapira and S. Haber, int. J. Multiphase Flow 16, 305 ( 1990).‘%, T. Leighton, Jr., Ph.D thesis, Stanford University, 1985. “J, R. Blake, Proc. Cambridge Philos. Sot. 70,303 ( 1971). lclJ. R. Blake and A. T. Chwang, J. Eng. Math, g,23 ( 1974). } 
\begin{equation}
K_{ijk}^{(\mathrm{im})}(\mathbf{r},\mathbf{r})=
\frac{1}{8 h^2}\big(\!
-  5 \delta_{jk} n_i + 3 \delta_{ik} n_j + 3 \delta_{ij} n_k + 3 n_i n_j n_k
\big),  
\label{BlakeFinal}
\end{equation}
where $\delta_{ij}$ is the Kronecker delta and $n_i$ are  components of  the normal-to-wall unit vector, $\hat{\mathbf{n}}$. The subscripts $i, j, k$ indicate the Cartesian coordinates that we may denote interchangeably by either of the label sets $\{x, y, z\}$ or $\{1,2,3\}$. For the bottom and top wall, one sets $\hat{\mathbf{n}}= \pm \hat{\mathbf{y}}$  and uses $h=y$ and $H-y$, respectively,   to indicate the half-distance between the stresslet singularity and its first wall-induced images.  The nonvanishing components of the self-mobility tensor, $ \mathbb{K}^{(\mathrm{im})}\left(\mathbf{r},\mathbf{r}\right)$, follow explicitly from Eq. \eqref{BlakeFinal} by including both the bottom and top wall of the channel (giving the $1/y^2$ and $1/(H-y)^2$ terms below, respectively) as
\begin{equation}
K_{112}^{(\mathrm{im})}\left(\mathbf{r},\mathbf{r}\right) = K_{121}^{(\mathrm{im})}\left(\mathbf{r},\mathbf{r}\right) = -\frac{3}{5}K_{211}^{(\mathrm{im})}\left(\mathbf{r},\mathbf{r}\right) =  -\frac{3}{5}K_{233}^{(\mathrm{im})}\left(\mathbf{r},\mathbf{r}\right)
=  \frac{3}{4} K_{222}^{(\mathrm{im})}\left(\mathbf{r},\mathbf{r}\right)
=  \frac{3}{8} \left( \frac{1}{y^2}- \frac{1}{(H-y)^2}\right).  
\label{eq:GreenComp}
\end{equation} 

%%%%%%%%%%%%%%%%%%%%%%%%%%%%%%%%%%%%%%%%%
\subsection{Stresslet tensor}
\label{subsec:stresslet}

The stresslet tensor, $\mathbb{S}(\mathbf{r}; {\hat{\mathbf d}})$, can be derived using the far-field singularity representation of Ref.  \cite{kim_microhydrodynamics} for a spheroid moving freely in a shear flow at low Reynolds numbers. In the present context, the (ferro)magnetic  spheroids are subjected to an additional torque contribution from the external magnetic field while their overall motion remains  force and torque-free due to the inertialess conditions.  The channel confinement effects are not included in the following stresslet calculations as they will be incorporated through the singularity images in Sections \ref{subsec:u_im} and \ref{subsec:omega_im}. 

%%%%%%%%%%%%%%%%%%%%
\subsubsection{Stresslet singularity representation}

The far-field singularity solution for the stresslet components, $ S_{ij}$, for a spheroid with net  angular velocity $\boldsymbol{\omega}$ is written as \cite{kim_microhydrodynamics} 
\begin{equation}
\label{stresslet_Faxen}
 S_{ij} =
 \frac{20}{3}\pi\eta \big(\ell(\alpha)\big)^3\!
\left(\! X^\mathrm{M} d^{(0)}_{ijkl}\! +\! Y^\mathrm{M} d^{(1)}_{ijkl}\! +\! Z^\mathrm{M} d^{(2)}_{ijkl}\!\right)\! (\mathbb{E}_\mathrm{f})_{kl}  +
4\pi\eta  \big(\ell(\alpha)\big)^3\,  Y^\mathrm{H}   \bigg\{\!\! \left[\left(\boldsymbol{\Omega}_\mathrm{f} -  \boldsymbol{\omega} \right)\times\hat{\mathbf d}\right]\!\hat{\mathbf d} 
+ \hat{\mathbf d}\! \left[\left(\boldsymbol{\Omega}_\mathrm{f} -  \boldsymbol{\omega} \right)\times\hat{\mathbf d}\right]\!\!\bigg\}_{\!ij}\!\!\!, 
\end{equation}
with the explicit form of the half-length $\ell(\alpha)$ for the major body axis to be used from Eq. \eqref{eq:half-major}. The dyads in Eq. \eqref{stresslet_Faxen} are expressed using the components $d_i$ of the axial orientation unit vector $\hat{\mathbf d}$  as \cite{kim_microhydrodynamics}     
\begin{eqnarray}
&&d^{(0)}_{ijkl} =  \frac{3}{2}\big(d_i d_j- \frac{1}{3}\delta_{ij}\big)\big(d_k d_l- \frac{1}{3}\delta_{kl}\big), \quad\,\,
d^{(1)}_{ijkl} = \frac{1}{2}\big(d_i \delta_{jl}d_k + d_j \delta_{il} d_k + d_i \delta_{jk} d_l + d_j \delta_{ik} d_l  - 4 d_i d_j d_k d_l\big), \nonumber
\\
&&d^{(2)}_{ijkl} =  \frac{1}{2} \big( \delta_{ik}\delta_{jl} + \delta_{jk}\delta_{il} - \delta_{ij}
\delta_{kl} + d_i d_j \delta_{kl} + \delta_{ij}d_k d_l  - d_i \delta_{jl} d_k - d_j \delta_{il} d_k - d_i \delta_{jk} d_l - d_j \delta_{ik} d_l 
+ d_i d_j d_k d_l \big). 
\end{eqnarray}
Recall that in the present model $d_3=0$, since $\hat{\mathbf d}$ is constrained to the $x-y$ plane (Section II, main text). Also, in Eq. \eqref{stresslet_Faxen}, $X^\mathrm{M}$, $Y^\mathrm{M}$, $ Z^\mathrm{M}$ and $Y^\mathrm{H}$ denote the spheroidal resistance functions, see Table \ref{Tab:ResistFs} and Ref. \cite{kim_microhydrodynamics}. There is  a particularly useful relationship between  ${Y^\mathrm{H}}$, ${ Y^\mathrm{C}}$ and the Bretherton number $-1<\beta(\alpha)<1$  \cite{Bretherton1962} as 
\begin{equation}
\frac{Y^\mathrm{H}}{ Y^\mathrm{C}} = \frac{\alpha^2 - 1}{\alpha^2+1} =  \beta(\alpha).
\label{eq:YHYC_beta} 
\end{equation}  
Furthermore, in Eq. \eqref{stresslet_Faxen}, $(\mathbb{E}_\mathrm{f})_{kl}$ are components of the rate-of-strain tensor  $\mathbb{E}_\mathrm{f}=\frac{1}{2}[(\nabla \mathbf{u}_\mathrm{f})+(\nabla \mathbf{u}_\mathrm{f})^T]$ and  $\boldsymbol{\Omega}_\mathrm{f} =\frac{1}{2}{\nabla}\times\mathbf{u}_\mathrm{f}$ is the vorticity of the background plane Poiseuille flow  $\mathbf{u}_\mathrm{f}({\mathbf r}) \!=\! u_\mathrm{f}(y) \hat{\mathbf x}$ with $u_\mathrm{f}(y) = \dot{\gamma}\,y\! \left(1- {y}/{H}\right)$ being  the velocity profile across the channel width  and $\dot \gamma={4 U_{\mathrm{max}}}/{H}$ and  $U_{\mathrm{max}}> 0$ being the maximum shear rate and  fluid velocity,  respectively. Hence, 
\begin{eqnarray}
&&\mathbb{E}_\mathrm{f}=
\frac{\dot{\gamma}}{2} \left(1- \frac{2y}{ H}\right)\left(\hat{\mathbf{y}}\hat{\mathbf{x}}+\hat{\mathbf{x}}\hat{\mathbf{y}}\right), 
\label{eq:strain}
\\
&& \boldsymbol{\Omega}_\mathrm{f}= -\frac{\dot\gamma}{2} \left(1 - \frac{2 y}{H}\right)\hat{\mathbf{z}}, 
\label{eq:vorticity}
\end{eqnarray}  
where $\hat{\mathbf z} = \hat{\mathbf x} \times \hat{\mathbf y} $ in the out-of-plane unit vector (see Fig. 1, main text).

%%%%%%%%%%%%%%%%%%
\begin{table*}[t!]
\setlength\extrarowheight{1pt}
\centering
\caption{Spheroidal resistance functions $X^\mathrm{M}$, $Y^\mathrm{M}$, $Z^\mathrm{M}$, $Y^\mathrm{H}$ and $Y^\mathrm{C}$ shown as functions of the spheroidal eccentricity $e$; see Ref.   \cite{kim_microhydrodynamics}.} 
\label{Tab:ResistFs}
\setcellgapes{3pt}\makegapedcells
\resizebox{\textwidth}{!}{
\begin{tabular}{|c|c|c|c|}
\hline
 \multirow{ 2}{*}{Resistance} &
  \multicolumn{1}{c|}{Prolate spheroids ($\alpha>1$)}  &
  Spheres &
  \multicolumn{1}{c|}{Oblate spheroids ($\alpha<1$)}  \\ 
 \multirow{ 1}{*}{function}   &  
 \multicolumn{1}{c|}{$e\equiv\sqrt{1-\alpha^{-2}}$\,;\,\,\, $L(e)\equiv\ln\left(\frac{1+e}{1-e}\right)$} &
($\alpha=1$) &
  \multicolumn{1}{c|}{$e\equiv\sqrt{1-\alpha^{2}}$\,;\,\,\, $L(e)\equiv\cot^{-1} \left(\frac{\sqrt{1-e^2}}{e}\right)$} \\
\hline\hline
$X^\mathrm{M}$ &
$\frac{8}{15}e^5\Big[ (3-e^2) L(e) - 6 e\Big]^{-1}$ &
1 &
$\frac{4}{15}e^5\Big[ (3-2e^2) L(e) - 3 e\sqrt{1-e^2}\Big]^{-1}$\\ 
\hline
\multirow{2}{*}{$Y^\mathrm{M}$} & 
$\frac{4}{5}e^5 \Big[2e(1- 2 e^2)- (1-e^2) L(e)\Big]$ & 
 1 &
$\frac{2}{5}e^5 \Big[e(1+e^2)-\sqrt{1-e^2}L(e)\Big]$ \\ 
&
$\times \Big[\left( 2 e (2 e^2 - 3) + 3(1-e^2) L(e) \right)\left(- 2 e + (1+ e^2)L(e)\right)\Big]^{-1}$ & 
&
$\times \Big[\left(3 e - e^3 - 3\sqrt{1-e^2} L(e)\right)\left(e\sqrt{1-e^2} - (1-2 e^2) L(e)\right)\Big]^{-1}$ \\ 
\hline
$Z^\mathrm{M}$ &
 $\frac{16}{5}e^5(1-e^2)\Big[ 3(1-e^2)^2 L(e) - 2 e (3-5 e^2)\Big]^{-1}$ &
 1 &
 $\frac{8}{5}e^5\Big[ 3 L(e) - (2 e^3 + 3e)\sqrt{1-e^2}\Big]^{-1}$ \\ 
\hline
$Y^\mathrm{H}$ & 
$\frac{4}{3}e^5 \Big[-2 e + (1+e^2) L(e)\Big]^{-1}$ &
 0 &
$-\frac{2}{3}e^5 \Big[e\sqrt{1-e^2} - (1-2 e^2)L(e)\Big]^{-1}$ \\
\hline
$Y^\mathrm{C}$ & 
$\frac{4}{3}e^3 (2-e^2) \Big[-2 e + (1+e^2) L(e)\Big]^{-1}$ &
 1 &
$\frac{2}{3}e^3 (2-e^2) \Big[e\sqrt{1-e^2} - (1-2 e^2)L(e)\Big]^{-1}$ \\
\hline
\end{tabular}
}
\end{table*}
%%%%%%%%%%%%%%%%%%
  
%%%%%%%%%%%%%%%%%%%%
\subsubsection{Field-induced contributions}
 
Equation \eqref{stresslet_Faxen} involves the particle angular velocity, $\boldsymbol{\omega}$, which can itself be determined from the far-field singularity solutions of Ref. \cite{kim_microhydrodynamics} by expressing  the hydrodynamic torque, $\mathbf{T}_\mathrm{h}$, applied by the background  flow on individual spheroids  as 
\begin{equation}
\mathbf{T}_\mathrm{h} =  8 \pi \eta\, \big(\ell(\alpha)\big)^3\! \left[ X^\mathrm{C} \hat{\mathbf d} \hat{\mathbf d} + Y^\mathrm{C} \left( \mathbb{I}-\hat{\mathbf d} \hat{\mathbf d}\right)\right]\cdot\left(\boldsymbol{\Omega}_\mathrm{f} - \boldsymbol{\omega} \right)
- 8 \pi \eta\, \big(\ell(\alpha)\big)^3\, Y^\mathrm{H} \left(\mathbb{E}_\mathrm{f}\cdot \hat{\mathbf d}\right)\times \hat{\mathbf d}.  
\label{torque_Faxen_Or}
\end{equation}

Since the spheroidal rotation is torque-free, the hydrodynamic torque $\mathbf{T}_\mathrm{h}$ is balanced by the magnetic torque ${\mathbf T}_{\mathrm{ext}}$  on the  spheroids due to the externally applied magnetic field which, in our model,  is a uniform field of strength $B\geq 0$ and tilt angle $\theta_B$ relative to the flow direction, or the $x$-axis (Fig. 1, main text). We thus have 
\begin{equation}
{\mathbf T}_{\mathrm{ext}}  = mB \sin\left(\theta_B-\theta\right) \hat{\mathbf z}. 
\label{eq:T_ext}
\end{equation}
Using the torque-free condition $\mathbf{T}_\mathrm{h} = - \mathbf{T}_{\mathrm{ext}}$, the spheroidal angular velocity follows by rearranging Eq. \eqref{torque_Faxen_Or} as  
\begin{equation} 
 \boldsymbol{\omega}  =  \left[ \frac{ Y^\mathrm{H} }{X^\mathrm{C}} \hat{\mathbf d} \hat{\mathbf d} + \frac{Y^\mathrm{H}}{ Y^\mathrm{C}}\left( \mathbb{I}-\hat{\mathbf d} \hat{\mathbf d}\right)\right]\cdot \left[ \left(\mathbb{E}_\mathrm{f}\cdot \hat{\mathbf d}\right) \times \hat{\mathbf d}  \right]+\boldsymbol{\Omega}_\mathrm{f} + \frac{\mathbf{T}_{\mathrm{ext}}}{8 \pi \eta \zeta(\alpha) R_{\mathrm{eff}}^3 Y^\mathrm{C}}.
 \label{omega_Faxen_1}
\end{equation}
Using Eq. \eqref{eq:YHYC_beta} and the identity $\hat{\mathbf d}\cdot[(\mathbb{E}_\mathrm{f}\cdot \hat{\mathbf d})\times \hat{\mathbf d}] = 0$, Eq. \eqref{omega_Faxen_1} can be written as 
\begin{eqnarray}
 \boldsymbol{\omega}  = \beta(\alpha)\left(\mathbb{E}_\mathrm{f}\cdot \hat{\mathbf d}\right)\times \hat{\mathbf d}
 +\boldsymbol{\Omega}_\mathrm{f} + \frac{ \mathbf{T}_{\mathrm{ext}}}{8 \pi \eta \zeta(\alpha) R_{\mathrm{eff}}^3 Y^\mathrm{C}}. 
 \label{omega_Faxen_2}
\end{eqnarray}
Also, using Eqs. \eqref{eq:strain}, \eqref{eq:vorticity}  and \eqref{eq:T_ext},  $\boldsymbol{\omega}$ can be expressed as the sum of the shear and field-induced contributions, defined as  $\boldsymbol{\omega}_\mathrm{f}=\beta(\alpha)\left(\mathbb{E}_\mathrm{f}\cdot \hat{\mathbf d}\right)\times \hat{\mathbf d}
 +\boldsymbol{\Omega}_\mathrm{f}={\omega}_\mathrm{f}( y, \theta)\,\hat{\mathbf{z}}$ and  $ \boldsymbol{\omega}_{\mathrm{ext}}=\mathbf{T}_{\mathrm{ext}}/(8 \pi \eta \zeta(\alpha) R_{\mathrm{eff}}^3 Y^\mathrm{C}) = {\omega}_{\mathrm{ext}}(\theta)\,\hat{\mathbf{z}}$, respectively. Explicitly, one has    
 \begin{eqnarray}
&& {\omega}_\mathrm{f}( y, \theta)  =\,   \frac{ \dot{\gamma}}{2} \left(1-\frac{2 y}{H}\right)\left(\beta(\alpha)\cos2\theta - 1\right),
\label{omega_f_ESI} 
\\
&& {\omega}_{\mathrm{ext}}(\theta) = \, \frac{ mB }{8 \pi \eta \zeta(\alpha) R_{\mathrm{eff}}^3 Y^\mathrm{C}}  \sin\left(\theta_B-\theta\right),  
 \label{omega_ext_ESI}
\end{eqnarray}
which reproduce the standard expressions in Eqs. (5) and (6) of the main text,  provided that one uses the Stokes-Einstein relation 
\begin{equation}
D_\mathrm{R}(\alpha)= \frac{k_\mathrm{B}T}{8 \pi \eta \zeta(\alpha) R_{\mathrm{eff}}^3 Y^\mathrm{C}}. 
\label{eq:YC_DR} 
\end{equation}
In rescaled units (see Section III C, main text, for the rescaling scheme), this latter relation reads 
\begin{equation}
\Delta_\mathrm{R}(\alpha)= \frac{1}{\zeta(\alpha)Y^\mathrm{C} }, 
\label{eq:YC_DR_rescaled} 
\end{equation}
where  $\Delta_\mathrm{R}(\alpha)$ is given in Section \ref{subsec:shape_functions}, $\zeta(\alpha)$  in Eq. \eqref{eq:zeta} and $Y^\mathrm{C}$ in Table \ref{Tab:ResistFs}.  

%%%%%%%%%%%%%%%%%%%%
\subsubsection{Explicit form of stresslet components}

The explicit form of stresslet components, $S_{ij}$, can now be derived by  recasting Eq. \eqref{torque_Faxen_Or} into the form
\begin{equation}
\!\left(\boldsymbol{\Omega}_\mathrm{f} -  \boldsymbol{\omega}\right) \times \hat{\mathbf d}= 
\beta(\alpha)\bigg[\hat{\mathbf d}\left(\hat{\mathbf d}\cdot \mathbb{E}_\mathrm{f}\cdot\hat{\mathbf d}\right)\!-\!\mathbb{E}_\mathrm{f}\cdot\hat{\mathbf d}\bigg]\! -\!  \frac{{\mathbf T}_{\mathrm{ext}}\times \hat{\mathbf d}}{8\pi\eta \zeta(\alpha) R_{\mathrm{eff}}^3Y^\mathrm{C}}, 
\label{torque_Faxen}
\end{equation}
and by using this expression together with Eq. \eqref{stresslet_Faxen} to obtain
\begin{eqnarray}
&&S_{ij} =
\frac{20}{3}\pi\eta \zeta(\alpha) R_{\mathrm{eff}}^3
\left( X^\mathrm{M} d^{(0)}_{ijkl} + Y^\mathrm{M} d^{(1)}_{ijkl} + Z^\mathrm{M} d^{(2)}_{ijkl}\right) (\mathbb{E}_\mathrm{f})_{kl}
\nonumber\\ 
&&\qquad +\,
4\pi\eta \zeta(\alpha) R_{\mathrm{eff}}^3 \beta(\alpha)\bigg\{ Y^\mathrm{H} \bigg[
2\hat{\mathbf d}\hat{\mathbf d}\left(\hat{\mathbf d}\cdot \mathbb{E}_\mathrm{f}\cdot\hat{\mathbf d}\right)-\hat{\mathbf d}\left(\mathbb{E}_\mathrm{f}\cdot\hat{\mathbf d}\right) 
 -\!\left(\mathbb{E}_\mathrm{f}\cdot\hat{\mathbf d}\right)\hat{\mathbf d} \bigg]
\!-\! \hat{\mathbf d}\!\left(\frac{\mathbf{T}_{\mathrm{ext}}\times \hat{\mathbf d}}{8\pi\eta \zeta(\alpha) R_{\mathrm{eff}}^3}\right)
 \!- \!\left(\frac{\mathbf{T}_{\mathrm{ext}}\times \hat{\mathbf d}}{8\pi\eta \zeta(\alpha) R_{\mathrm{eff}}^3}\right)\!\hat{\mathbf d}
\bigg\}_{\!ij}\!\!\!.\nonumber\\
&&
\label{eq:S_ij_final}
\end{eqnarray}
The nonzero components of the stresslet tensor are thus found as 
\begin{widetext}
\begin{equation}
\begin{split}
S_{11} = & 
\frac{5}{3}\pi\eta \zeta(\alpha) R_{\mathrm{eff}}^3 \dot\gamma \left(1-\frac{2 y}{H}\right) \sin 2\theta \bigg[
X^\mathrm{M} \left(3\cos^2\theta- 1\right)
+ 2 Y^\mathrm{M}\left(1-2\cos^2\theta\right)
+ Z^\mathrm{M} \left(\cos^2\theta-1\right)
\bigg] 
\\
&\hskip5.5cm+
4\pi\eta \zeta(\alpha) R_{\mathrm{eff}}^3 \beta(\alpha) \sin 2\theta \left[\frac{1}{2}
 Y^\mathrm{H} \dot\gamma \left(1-\frac{2 y}{H}\right) \cos 2\theta
+ \frac{mB \sin\left(\theta_B-\theta\right)}{8\pi\eta \zeta(\alpha) R_{\mathrm{eff}}^3} 
\right], 
\\
S_{22} = & 
\frac{5}{3}\pi\eta \zeta(\alpha) R_{\mathrm{eff}}^3 \dot\gamma \left(1-\frac{2 y}{H}\right) \sin 2\theta \bigg[
 X^\mathrm{M} \left(3 \sin^2\theta - 1\right)
 + 2 Y^\mathrm{M} \left(1-2\sin^2\theta\right)
+ Z^\mathrm{M} \left(\sin^2\theta-1\right)
\bigg]
\\
&\hskip5.5cm - 4 \pi\eta \zeta(\alpha) R_{\mathrm{eff}}^3 \beta(\alpha) \sin 2\theta \left[
\frac{1}{2} Y^\mathrm{H} \dot\gamma \left(1-\frac{2 y}{H}\right) \cos 2\theta 
+ \frac{mB \sin\left(\theta_B-\theta\right)}{8\pi\eta \zeta(\alpha) R_{\mathrm{eff}}^3}
\right],
\\
S_{12} = &\, S_{21} =
\frac{5}{6}\pi\eta \zeta(\alpha) R_{\mathrm{eff}}^3 \dot\gamma \left(1-\frac{2 y}{H}\right)\! \bigg[
3 X^\mathrm{M} \sin^2 2\theta
 +  4 Y^\mathrm{M} \cos^2 2\theta
+ Z^\mathrm{M} \sin^2 2\theta
\bigg]
\\
&\hskip5.5cm -4\pi\eta \zeta(\alpha) R_{\mathrm{eff}}^3 \beta(\alpha) \cos 2\theta\left[\frac{1}{2}
Y^\mathrm{H} \dot\gamma \left(1-\frac{2 y}{H}\right) \cos 2\theta
+ \frac{mB \sin\left(\theta_B-\theta\right)}{8\pi\eta \zeta(\alpha) R_{\mathrm{eff}}^3}
\right],
\\
S_{33} = & 
\frac{5}{3}\pi\eta \zeta(\alpha) R_{\mathrm{eff}}^3 \dot\gamma \left(1-\frac{2 y}{H}\right) \sin 2\theta \left[
- X^\mathrm{M} 
+ Z^\mathrm{M} 
\right].
\label{eq:stressletsExplicit}
\end{split}
\vspace{1mm}
\end{equation}
\end{widetext} 
 
%%%%%%%%%%%%%%%%%%%%%%%%%%%%%%%%%%%%%%%%%
\subsection{Hydrodynamic lift velocity}
\label{subsec:u_im}

The $i$th component of the   wall-induced hydrodynamic velocity, ${\mathbf u}^{(\mathrm{im})}(\mathbf{r}; {\hat{\mathbf d}})$,  which is produced by the stresslet singularity image relative to a given channel wall follows by  setting $\mathbf{r}'=\mathbf{r}$ in Eq. (3) of the main text; i.e.,  
 \begin{equation}
u_i^{(\mathrm{im})}(\mathbf{r}; {\hat{\mathbf d}}) = - \frac{1}{8\pi\eta} K_{ijk}^{(\mathrm{im})}(\mathbf{r},\mathbf{r}) S_{jk}(\mathbf{r}; {\hat{\mathbf d}}), 
\label{eq:u_im_main}
\end{equation}
where Einstein's summation convention is used.  Using the expressions \eqref{eq:GreenComp} for the self-mobility components with $h=y$ and $H-y$ for the calculation of image interactions relative to the bottom and top wall of the channel, respectively, as well as the expressions \eqref{eq:stressletsExplicit} for the stresslet components, and then by superimposing the contributions from the two walls, one obtains the components of ${\mathbf u}^{(\mathrm{im})}(\mathbf{r}; {\hat{\mathbf d}})$ as 
\begin{eqnarray}
&& u^{(\mathrm{im})}_x( y, \theta) \! = \! - {\zeta(\alpha) R_{\mathrm{eff}}^3} \left(\frac{1}{ y^2}\!-\!\frac{1}{( H- y)^2}\!\right) \!
\Bigg\{\frac{5  \dot\gamma}{64} \!  \left(1\!-\! \frac{2 y}{ H}\right) \!
 \bigg[4Y^\mathrm{M} \!\cos^2 2\theta +\! \left(3X^\mathrm{M}\!+\! Z^\mathrm{M}\right)\sin^2 2\theta \bigg]
 \nonumber\\
&&\hskip6.5cm
 -\frac{3\beta(\alpha)}{8}\cos2\theta\bigg[\frac{\dot\gamma}{2} Y^\mathrm{H}  \left(1\!- \!\frac{2 y}{ H}\right)\!\cos 2\theta  + \frac{mB}{ 8\pi\eta \zeta(\alpha) R_{\mathrm{eff}}^3} \sin\left(\theta_B-\theta\right)\! \bigg] \!\Bigg\},
\label{eq:uxExplicit}
\\
&& u^{(\mathrm{im})}_y( y, \theta) \! = \! 
 - {\zeta(\alpha) R_{\mathrm{eff}}^3} \left(\frac{1}{ y^2}\!-\!\frac{1}{( H- y)^2}\!\right)\!\sin 2\theta 
 \Bigg\{\!\frac{15  \dot\gamma}{64}\!   \left(1\!-\! \frac{2 y}{ H}\right)\!
 \bigg[\!\left(-X^\mathrm{M}\!+2Y^\mathrm{M}\!-\!Z^\mathrm{M}\right)\cos^2\theta 
 \!+\! \left(2X^\mathrm{M}\!-2Y^\mathrm{M}\right)\sin^2\theta \bigg]
 \nonumber\\
&&\hskip7.4cm
 -\frac{9\beta(\alpha)}{16}\bigg[\frac{\dot\gamma}{2} Y^\mathrm{H}  \left(1\!-\! \frac{2 y}{ H}\right)\cos 2\theta  + \frac{mB}{ 8\pi\eta \zeta(\alpha) R_{\mathrm{eff}}^3} \sin\left(\theta_B-\theta\right)\! \bigg]\! \Bigg\}.  
 \label{eq:LiftExplicit}
\end{eqnarray}
As noted in the main text (Section III C), for symmetry reasons, only the transverse component $u^{(\mathrm{im})}_y( y, \theta)$ enters the governing Smoluchowski equation. This component represents the desired wall-induced {\em hydrodynamic lift} experienced by a freely moving spheroid within the channel. In dimensionless form, Eq. \eqref{eq:LiftExplicit} reduces to Eq. (9) in the main text; i.e., 
\begin{eqnarray}
\label{eq:u_im}
&&\hskip-3mm\tilde u^{(\mathrm{im})}_y(\tilde y, \theta)\! =\!
 - \zeta(\alpha)\! \left(\frac{1}{\tilde y^2}\!-\!\frac{1}{(\tilde H-\tilde y)^2}\!\right)\!\sin 2\theta 
 \Bigg\{\!\frac{15  \mathrm{Pe}_\mathrm{f}}{64} \!  \left(\!1\!-\! \frac{2\tilde y}{\tilde H}\!\right)\!
 \bigg[\!\left(-X^\mathrm{M}\!+\!2Y^\mathrm{M}\!-\!Z^\mathrm{M}\right)\!\cos^2\theta 
 \!+\! \left(2X^\mathrm{M}\!-\!2Y^\mathrm{M}\right)\!\sin^2\theta \bigg]
\\
&&\hskip8.5cm
 -\frac{9\beta(\alpha)}{16}\bigg[\frac{1}{2} Y^\mathrm{H} \mathrm{Pe}_\mathrm{f} \! \left(\!1\!- \!\frac{2\tilde y}{\tilde H}\!\right)\!\cos 2\theta \! +\! \frac{\chi}{ \zeta(\alpha)} \sin\left(\theta_B\!-\theta\right)\! \bigg] \!\Bigg\}.   
  \nonumber
\end{eqnarray} 
We recall that the rescaled channel width is defined as $\tilde H=H/{R_{\mathrm{eff}}}$,  the flow P\'eclet number (rescaled shear strength) as $\mathrm{Pe}_\mathrm{f}={\dot\gamma}/{D_{0\mathrm{R}}}$, and the magnetic coupling parameter (rescaled field strength) as $\chi={mB }/{k_\mathrm{B} T}$.

%%%%%%%%%%%%%%%%%%%%%%%%%%%%%%%%%%%%%%%%%%
\subsection{Wall-induced hydrodynamic angular velocity}
\label{subsec:omega_im}

The expression for angular velocity of a torque-free spheroid can be used to determine the wall-induced hydrodynamic angular  velocity, $\boldsymbol{\omega}^{(\mathrm{im})}$, by formally replacing $\mathbf{u}_\mathrm{f}\rightarrow {\mathbf u}^{(\mathrm{im})}$ to calculate $\mathbb{E}_\mathrm{f}$ and $\boldsymbol{\Omega}_\mathrm{f}$ in Eq.  \eqref{torque_Faxen_Or}. This yields  
\begin{equation}
  \boldsymbol{\omega}^{(\mathrm{im})}  = \beta(\alpha) \left(\mathbb{E}_\mathrm{f}^{(\mathrm{im})}\cdot \hat{\mathbf d}\right)\times \hat{\mathbf d}+\boldsymbol{\Omega}_\mathrm{f}^{(\mathrm{im})},
\label{eq:w_im_1}
\end{equation}
where $ \mathbb{E}_\mathrm{f}^{(\mathrm{im})}= \frac{1}{2}[\nabla\mathbf{u}^{(\mathrm{im})}+(\nabla \mathbf{u}^{(\mathrm{im})})^T]$ and $\boldsymbol{\Omega}_\mathrm{f}^{(\mathrm{im})} = \frac{1}{2}\nabla \times\mathbf{u}^{(\mathrm{im})}$.   The above relation is to be understood as an extra additive contribution to the r.h.s. of Eq. \eqref{omega_Faxen_2}.  Using Eqs. \eqref{eq:strain} and \eqref{eq:vorticity}, we find   $\boldsymbol{\omega}^{(\mathrm{im})}={\omega}^{(\mathrm{im})}(y, \theta) \,\hat{\mathbf z}$ where  
\begin{equation} 
{\omega}^{(\mathrm{im})}(y, \theta) = \frac{1}{2}\! \left( \beta(\alpha) \cos 2\theta - 1  \right)\! \left(\!\frac{\partial u_x^{(\mathrm{im})}}{\partial y}\! \right)+ \frac{ 1}{2} \beta(\alpha)\sin 2 \theta\! \left(\!\frac{\partial u_y^{(\mathrm{im})}}{\partial y}\!\right).   
\label{eq:u_omega_in}
\end{equation}
Finally, using Eqs. \eqref{eq:uxExplicit}, \eqref{eq:LiftExplicit} and  \eqref{eq:u_omega_in}, we obtain the explicit form of  ${\omega}^{(\mathrm{im})}(y, \theta)$ as 
\begin{widetext}
\begin{eqnarray}
&&{\omega}^{(\mathrm{im})}( y, \theta)
= - \frac{\dot\gamma}{64}\frac{\zeta(\alpha)R_{\mathrm{eff}}^3}{{H}}\left(\frac{H - y}{y^3}-\frac{y}{\big(H-y\big)^3}\right) \beta(\alpha)  \bigg\{\cos^3 2\theta \left(15X^\mathrm{M}-20Y^\mathrm{M}+5Z^\mathrm{M}-6\beta(\alpha) Y^\mathrm{H}\right) 
\nonumber\\
&& \hskip2cm
+  \frac{\sin 4\theta}{4} \left(45X^\mathrm{M}-30Y^\mathrm{M}-15Z^\mathrm{M}\right)
 - \cos^2 2\theta \left(15X^\mathrm{M} -20Y^\mathrm{M} + 5 Z^\mathrm{M} +12 \beta(\alpha) Y^\mathrm{H}\right) 
 \nonumber\\
&& \hskip2 cm
 - \frac{\sin 2\theta}{2}   \left( 15 X^\mathrm{M} + 30 Y^\mathrm{M} - 45 Z^\mathrm{M} \right)
 -  \cos 2\theta \left(15X^\mathrm{M} + 5 Z^\mathrm{M} - 18 Y^\mathrm{H} \right) + \frac{ 15 X^\mathrm{M}+5 Z^\mathrm{M}}{\beta(\alpha)}
\bigg\}
\nonumber\\
&& \hskip2 cm  
+  \frac{3  }{16} \frac{mB}{8\pi\eta}\left(\frac{1}{{y}^3}+\frac{1}{({H}-{y})^3}\right)
\sin\left(\theta_B-\theta\right)
\left(  \beta^2(\alpha) \cos^2 2\theta  +2 \beta(\alpha) \cos 2\theta - 3 \beta^2(\alpha) \right), 
\label{eq:w_im}
\end{eqnarray}
\end{widetext}
which, upon rescaling (Section III C, main text), reduces to its dimensionless form (Appendix B, main text) as 
\begin{eqnarray}
&&\tilde{{\omega}}^{(\mathrm{im})}(\tilde y, \theta)
= - \frac{\mathrm{Pe}_\mathrm{f}}{64}\frac{\zeta(\alpha)}{\tilde{H}}\left(\!\frac{\tilde{H} - \tilde{y}}{\tilde{y}^3}-\frac{\tilde{y}}{\big(\tilde{H}-\tilde{y}\big)^3}\!\right) \beta(\alpha)  \bigg\{\frac{ 15 X^\mathrm{M}+5 Z^\mathrm{M}}{\beta(\alpha)}+\cos^3 2\theta \left(15X^\mathrm{M}-20Y^\mathrm{M}+5Z^\mathrm{M}-6\beta(\alpha) Y^\mathrm{H}\right) 
\nonumber\\
&& \hskip0cm
 - \cos^2 2\theta \left(15X^\mathrm{M} -20Y^\mathrm{M} + 5 Z^\mathrm{M} +12 \beta(\alpha) Y^\mathrm{H}\right) 
  -  \cos 2\theta \left(15X^\mathrm{M} + 5 Z^\mathrm{M} - 18 Y^\mathrm{H} \right)
 +  \frac{\sin 4\theta}{4} \left(45X^\mathrm{M}-30Y^\mathrm{M}-15Z^\mathrm{M}\right)
\nonumber\\
&& \hskip0cm
  - \frac{\sin 2\theta}{2}   \left( 15 X^\mathrm{M} + 30 Y^\mathrm{M} - 45 Z^\mathrm{M} \right)
\!\!\bigg\}  
+  \frac{3\chi  }{16}\!\left(\frac{1}{\tilde{y}^3}\!+\!\frac{1}{(\tilde{H}\!-\!\tilde{y})^3}\right)\beta(\alpha)
\sin\left(\theta_B\!-\!\theta\right)
\left(  \beta(\alpha) \cos^2 2\theta  +2  \cos 2\theta - 3 \beta(\alpha) \right).
\label{eq:w_im_res}
\end{eqnarray}

%%%%%%%%%%%%%%%%%%%%%%%%%%%%%%%%%%%%%%%%%%
%%%%%%%%%%%%%%%%%%%%%%%%%%%%%%%%%%%%%%%%%%
\section{Deterministic dynamics, orientational pinning and fixed points} 
\label{sec:deterministic}

In this section, we detail our linear and nonlinear stability analysis for the fixed points that emerge within the deterministic dynamics of noninteracting (ferro)magnetic spheroids under  plane Poiseuille flow and uniform applied field as used in our model. The connection between the deterministic fixed points and our full and reduced probabilistic results are thoroughly discussed in the main text (Sections IV-VI and Appendix D therein).  

As noted in the main text and the outline of Section \ref{subsec:image_scope}, our linear stability analysis and determination of eigenvalues associated with the fixed points, to be discussed in  Sections \ref{subsec:centered}-\ref{subsec:near_wall}) below, require the generalized, or  unconstrained, form of the hydrodynamic lift  derived in Section \ref{subsec:u_im} rather than its limiting strict-pinning form  given for prolate spheroids in Ref. \cite{GolestanianFocusing}. To our knowledge, a thorough stability analysis as we detail here 
%by determining the  eigenvalues in the linearization and, where required, nonlinear stability of fixed points, and by associating the fixed points with boundaries of the identified regimes of focusing, etc.,  
has not previously been considered for the current problem. 

While following the forthcoming details, one should bear in mind the scope of our model (Section \ref{subsec:image_scope}) and the approximations involved in our deterministic analysis  (Sections III E, main text). For the sake of clarity and completeness, we reproduce  some of the discussions from the main text in what follows. We also present a short discussion of field-induced pinning within the deterministic orientational dynamics of magnetic spheroids in relation to our model. The pinning effect has been explored elsewhere in the literature on  prolate paramagnetic and ferromagnetic spheroids in simple and pressure-driven flows \cite{Zhou2017,ZhouPRA2017,Cao2017,Zhang2018,Zhang2018b,Sobecki2018,Sobecki2019,Sobecki2020,Kumaran2020,GolestanianFocusing,Matsunaga2018}.

%%%%%%%%%%%%%%%%%%%%%%%%%%%%%%%%%%%%%%%%%%
\subsection{Dynamical equations for noise-free spheroids} 
\label{subsec:det_eqs}

The deterministic equations  governing  the  translational and rotational dynamics of noise-free   spheroids in the  latitude-orientation ($ y-\theta$)  coordinate space are written as  
\begin{equation}
{\dot y}= {v}_y(y, \theta), \quad {\dot\theta}= {\omega}(y, \theta),
\end{equation}   
where $ {v}_y$ and $ {\omega}$ are the net deterministic translational and angular velocity of spheroids (Section III, main text). 

As discussed in the main text, since we are mainly  interested in centered and off-centered focusing of the spheroids away from the channel walls, we shall only consider particle trajectories that do not bring the spheroids into contact with the walls. Hence, the local steric contribution to ${v}_y(y, \theta)$, i.e., the first term in Eq. (8) of the main text, is discarded and ${v}_y(y, \theta)$ is replaced  with the hydrodynamic lift $ u^{(\mathrm{im})}_y( y, \theta)$ whose rescaled form can be found in Eq. \eqref{eq:u_im}. 

Also, to enable analytical progress, we  ignore the rapidly decaying image contribution to the net angular velocity, ${\omega}^{(\mathrm{im})}(y, \theta)$, Eq. \eqref{eq:w_im}. This approximations is in line with the observations made in Appendix B of the main text. We shall thus replace the net angular velocity of spheroids with the sum of shear and field-induced terms, Eqs. \eqref{omega_f_ESI} and \eqref{omega_ext_ESI} as  
 \begin{equation}
 {\omega}( y, \theta) \simeq  {\omega}_\mathrm{f}( y, \theta)+{\omega}_{\mathrm{ext}}(\theta),
 \label{eq:w_approx_det}
 \end{equation} 
or, explicitly and in rescaled units, 
 \begin{equation}
\tilde\omega(\tilde y, \theta) \simeq \frac{\mathrm{Pe}_\mathrm{f}}{2}\left(1- \frac{2\tilde y}{\tilde H}\right) \left(\beta(\alpha) \cos2\theta-1 \right) + \chi\Delta_\mathrm{R}(\alpha)  \sin\left(\theta_B-\theta\right).  
 \label{eq:w_tot_rescaled}
\end{equation} 
With the above considerations, the system of equations to be used for the deterministic dynamics of the spheroids are  
\begin{eqnarray}
&&\tilde{\dot y}= {\tilde v}_y(\tilde y, \theta)\simeq \tilde u^{(\mathrm{im})}_y(\tilde y, \theta),
\label{eq:dtm_dyn_eqs1} \\
&& \tilde{\dot\theta} = {\tilde \omega}(\tilde y, \theta) \simeq \tilde {\omega}_\mathrm{f}(\tilde y, \theta)+\tilde{\omega}_{\mathrm{ext}}(\theta).    
\label{eq:dtm_dyn_eqs2}
\end{eqnarray}

%%%%%%%%%%%%%%%%%%%%%%%%%%%%%%%%%%%%%%%%%%
\subsection{The pinning curve}
\label{subsec:pinning_curve}

The balance between counteracting torques exerted on (ferro)magnetic spheroids by the shear flow and the applied magnetic field can lead  to stabilization or pinning of the spheroidal orientation at a particular angle  $\theta_\mathrm{p}(\tilde y)$ at lateral coordinate $\tilde y$ within the channel. In a deterministic setting, the pinning emerges in a strict sense, i.e., on specific {\em pinning curve(s)} in the $\tilde y-\theta$ coordinate space.  In the probabilistic setting, the pinning is reproduced with a small broadening around the deterministic solution along the $\theta$-axis due to the presence of rotational noise (see Section IV B, Appendix D and also Figs. 3, 4 and 7 in the main text). The deterministic pinning curve  $\theta= \theta_\mathrm{p}(\tilde y)$ can thus be obtained by requiring 
\begin{equation}
\tilde{\dot\theta}=0, \quad \partial \tilde{\dot\theta}/\partial \theta<0, 
 \label{eq:pinning_eqs}
\end{equation}
in which the inequality condition is to ensure orientational stability of the solutions.  Using Eqs. \eqref{eq:w_tot_rescaled} and \eqref{eq:dtm_dyn_eqs2}, the relations in \eqref{eq:pinning_eqs} can be rewritten as
\begin{equation}
\tilde {\omega}\!\left(\tilde y, \theta_\mathrm{p}(\tilde y)\right)\!=0, \quad \frac{\partial \tilde {\omega}(\tilde y, \theta)}{\partial \theta}\bigg|_{\theta=\theta_\mathrm{p}(\tilde y)}\!\!<0. 
 \label{eq:pinning_curve}
\end{equation}
The first relation above immediately reproduces Eq. (15) of the main text; i.e., 
\begin{equation}
 \cos^2 \theta_\mathrm{p}(\tilde y) +\frac{ \chi \Delta_\mathrm{R}(\alpha) / \beta(\alpha)}{\mathrm{Pe}_\mathrm{f}(1- 2\tilde y/\tilde H) }\sin\left(\theta_B - \theta_\mathrm{p}(\tilde y)\right)- \frac{1+ \beta(\alpha)}{2\beta(\alpha)}=0,     
\label{eq:pinning}
\end{equation}
in which  $\tilde y$ is restricted to the domain $\tilde y\in [0, \tilde H]$ and the  tilt angle of the field can conveniently be restricted to the first quadrant  $0\leq\theta_B \leq \pi/2$ due to specific symmetry properties of the model at hand (Section III D, main text).  

It is clear from Eq. \eqref{eq:pinning} that once $\tilde y$ is further rescaled with the channel width  $\tilde y\rightarrow \tilde y/\tilde H$, the pinning solutions will depend only on the particle aspect ratio $\alpha$, the tilt angle $\theta_B$, and the field-to-shear strength ratio $\chi/\mathrm{Pe}_\mathrm{f}$. It also follows immediately from Eq. \eqref{eq:pinning} that there is no solution for $\theta_\mathrm{p}(\tilde y)$  (no orientational pinning) in the absence of an applied field ($\chi=0$). Equation \eqref{eq:pinning} can in general admit both stable and unstable solutions when $\chi\neq 0$, but only the stable or pinning solutions will be relevant to field-induced  focusing of the spheroids. 

For finite $\chi$ and arbitrary $\theta_B$, the pinning curve is obtained numerically by solving Eq. \eqref{eq:pinning}. Some of the examples are shown in Figs. 3 and 4 ($\theta_B=\pi/2$) and Fig. 7 ($\theta_B=\pi/6, \pi/4$, and  $\pi/3$) of the main text. When the applied field is tilted ($0<\theta_B<\pi/2$), the point-reflection symmetry relative to the coordinate-space center, i.e.,  $(\tilde H/2, \theta_B)$, is broken and the pinning curve is shifted to the top and bottom half of the channel for prolate and oblate spheroids, respectively (Fig. 7, main text). Also, despite the close, albeit reverse, similarities in the pinning behavior of prolate and oblate spheroids,  Eq. \eqref{eq:pinning} indicates that there is no exact duality or correspondence between the two cases, i.e., prolate and oblate spheroids with aspect ratios $\alpha$ and $1/\alpha$. 

%%%%%%%%%%%%%%%%%%
\subsubsection{Partial pinning within the channel}
\label{subsubsec:partial_pinning}

In the case of prolate spheroids in a transverse field ($\alpha>1$, $\theta_B=\pi/2$) and oblate spheroids in a longitudinal field ($\alpha<1$, $\theta_B=0$), certain aspects of the deterministic pinning curve can be derived analytically (Section IV, main text). The pinning curves turn  out to be similar in these two cases. It appears as a single-valued  monotonous function with point-reflection symmetry relative to the coordinate-space center in each case; i.e.,   $(\tilde H/2,\pi/2)$ and $(\tilde H/2,0)$, respectively. At a given lateral position $\tilde y$ within the channel, solutions of  Eq. \eqref{eq:pinning} indeed emerge as stable/unstable pairs due to a  saddle-node bifurcation  \cite{Strogatz2000} that sets in when the rescaled field strength is increased beyond a  local bifurcation threshold $\chi_\mathrm{p}(\tilde y)$; see Ref. \cite{MRSh1} for relevant details. 

\begin{figure}[t!]
\begin{center}
\includegraphics[width=0.45\linewidth]{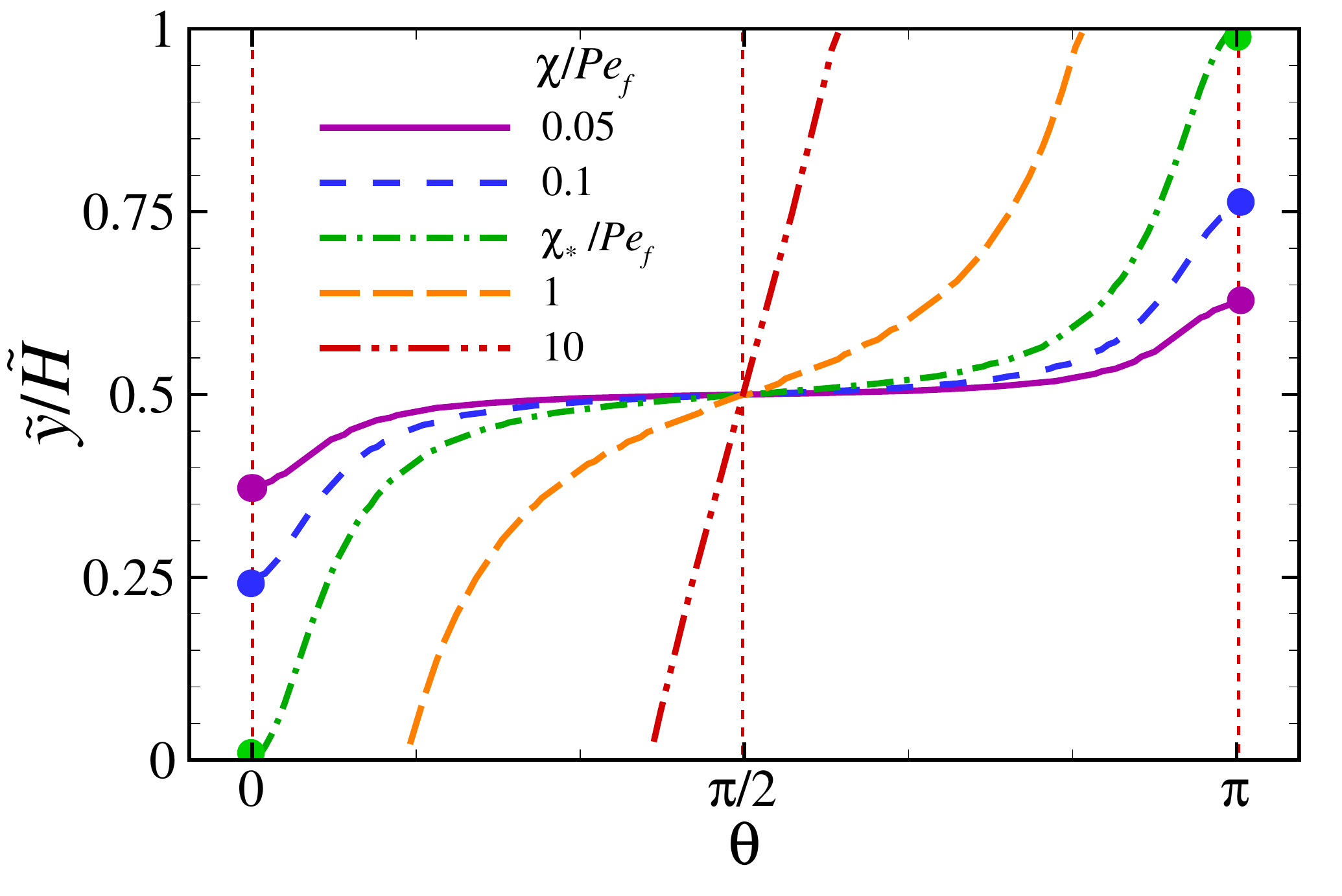}	
\caption{Pinning curves for prolate spheroids of aspect ratio $\alpha=4$ in a transverse magnetic field ($\theta_B = \pi/2$) and different field-to-shear strength ratios $\chi/\mathrm{Pe}_\mathrm{f}$ are shown by their {\em inverse} functions $\tilde y = \tilde y_\mathrm{p}(\theta)$. For the given parameter values, the threshold of the whole-channel pinning is obtained as $\chi_\ast/\mathrm{Pe}_\mathrm{f} \simeq 0.2$. Note the increasing slope of the inverse pinning curve at the centerline as $\chi/\mathrm{Pe}_\mathrm{f}$ increases (Section \ref{subsec:centered}).
}
\label{fig:Fig1ESI}
\end{center}
\vskip-4mm
\end{figure}

Figure  \ref{fig:Fig1ESI} shows typical forms of the pinning curve, depicted by its {\em inverse function} $\tilde y = \tilde y_\mathrm{p}(\theta)$, in the case of prolate spheroids in a transverse field at different values of the ratio $\chi/\mathrm{Pe}_\mathrm{f}$.  For small to intermediate values  of $\chi/\mathrm{Pe}_\mathrm{f}$ (solid purple and dashed blue curves), the pinning solutions are obtained only over a symmetric interval (pinning region) $\tilde y\in [\tilde H/2-{\tilde w_\mathrm{p}}/2, \tilde H/2+{\tilde w_\mathrm{p}}/2]$ around the midchannel. In these cases, the width of pinning region ${\tilde w_\mathrm{p}}$ stays smaller than the channel width, ${\tilde w_\mathrm{p}}<\tilde H$, and the angular coordinates of its endpoints  (shown by bullets in the figure) remain fixed on $\theta=0$ and $\pi$ as $\chi$ is varied. For oblate spheroids in a longitudinal field, the angular coordinates of the endpoints remain fixed on $\mp\pi/2$. Hence, using  Eq. \eqref{eq:pinning},  we find 
\begin{align}
\frac{\tilde w_\mathrm{p}}{\tilde H}  = \frac{\chi}{\mathrm{Pe}_\mathrm{f}} \cdot
\left\{ \begin{array}{ll}
\dfrac{2\Delta_\mathrm{R}(\alpha)}{1-\beta(\alpha)} & \,\,\,\, : \,\,\,  \alpha>1,\, \theta_B=\pi/2, \\ \\
\dfrac{2\Delta_\mathrm{R}(\alpha)}{1+\beta(\alpha)} & \,\,\,\, : \,\,\,  \alpha<1,\, \theta_B=0. 
\end{array}\right.
\label{eq:w_p}
\end{align}

We refer to the regime of parameters where the pinning curve does not extend all the way to the channel walls as the regime of   {\em partial pinning} (Section IV B, main text). This behavior arises when the field-induced angular velocity, $\tilde \omega_{\mathrm{ext}}$, is large enough to  stabilize  Jeffery oscillations of the spheroidal orientation \cite{Jeffery,kim_microhydrodynamics} caused by the shear-induced angular velocity, $\tilde \omega_\mathrm{f}$,  but only around the midchannel and not too close to the walls where $\tilde \omega_\mathrm{f}$ dominates and maintains the (field-modified) Jeffery oscillations (Fig. 3, main text). In fact, at the channel centerline, $\tilde \omega_{\mathrm{ext}}$ always dominates and the spheroidal orientation remains pinned for all nonzero values of the rescaled field strength $\chi$; i.e., the local threshold $\chi_\mathrm{p}(\tilde H/2)=0$. This is because the shear rate from the Poiseuille flow vanishes at the center. 

%-------------------------------

%%%%%%%%%%%%%%%%%%
\subsubsection{Whole-channel pinning}
\label{subsubsec:whole_pinning}

When the pinning curve extends beyond the channel width, $\tilde w_\mathrm{p}>\tilde H$, no Jeffery orbits are permitted anywhere and the spheroidal orientation is  pinned at all lateral positions within the channel. This occurs at the global threshold $\chi_{\ast}$ obtained by setting  $\tilde w_\mathrm{p}=\tilde H$ in Eq. \eqref{eq:w_p} as  
\begin{align}
\frac{\chi_\ast}{\mathrm{Pe}_\mathrm{f}}  = 
\left\{ \begin{array}{ll}
\dfrac{1-\beta(\alpha)}{2\Delta_\mathrm{R}(\alpha)} & \,\,\,\, : \,\,\,  \alpha>1,\, \theta_B=\pi/2, \\ \\
\dfrac{1+\beta(\alpha)}{2\Delta_\mathrm{R}(\alpha)} & \,\,\,\, : \,\,\,  \alpha<1,\, \theta_B=0, 
\end{array}\right.
\label{eq:chi_ast}
\end{align}
where we have $0<\beta(\alpha)<1$ for prolate spheroids and $-1<\beta(\alpha)<0$ for oblate spheroids. As discussed in the main text (Section IV C), ${\chi_\ast}$ is identified  the onset of the {\em whole-channel-pinning regime} or, interchangeably, the {\em strong-field regime}. 

For the parameter values in Fig. \ref{fig:Fig1ESI}, we have $\chi_\ast/\mathrm{Pe}_\mathrm{f} \simeq 0.2$ (dot-dashed green curve) and, thus,  the long-dashed orange and double-dot-dashed red curves correspond to the whole-channel pinning of spheroids. Note that here we have shown the inverse pinning curve $\tilde y = \tilde y_\mathrm{p}(\theta)$ which, as seen, becomes steeper as $\chi/\mathrm{Pe}_\mathrm{f}$ is increased. This inverse curve eventually tends to the vertical line $\theta = \theta_B=\pi/2$ and 0 in the special cases of prolate and oblate, respectively,  as $\chi$ tends to infinity.  

As noted in the main text (Sections IV C and V), the onset $\chi_\ast(\theta_B)$ of whole channel pinning in an applied field with tilt angle $0<\theta_B < \pi/2$ is determined numerically from Eq. \eqref{eq:pinning} and by inspecting the conditions where (by definition) the endpoints of the pinning curve reach the channel walls.

Equation  \eqref{eq:chi_ast} also predicts pinning of spheres at $\chi_\ast/\mathrm{Pe}_\mathrm{f} =[2\Delta_\mathrm{R}(\alpha=1)]^{-1}$. Given that our model involves ferromagnetic particles, this means that ferromagnetic spheres, unlike paramagnetic ones, are also pinned under a strong field \cite{Sobecki2018}, even as neither one can be focused due to a vanishing hydrodynamic lift in their cases (Section III C, main text). 

%%%%%%%%%%%%%%%%%%%%%%%%%%%%%%%%%%%%%%%%%%
%%%%%%%%%%%%%%%%%%%%%%%%%%%%%%%%%%%%%%%%

%%%%%%%%%%%%%%%%%%%%%%%%%%%%%%%%%%%%%%%%%%
\subsection{Relevant fixed points and nullclines}
\label{subsec:FP_on_pinning}

We proceed by  identifying and classifying the relevant {\em fixed points} associated with the deterministic phase-space dynamics of spheroids according to   Eqs. \eqref{eq:dtm_dyn_eqs1} and \eqref{eq:dtm_dyn_eqs2}. The fixed points of interest here emerge in the strong-field regime  where the spheroidal orientation is pinned at all lateral positions across the channel width (Section \ref{subsubsec:whole_pinning}). 

The fixed points correspond to coordinate-space points $(y_\ast, \theta_\ast)$ where the deterministic translational and rotational phase-space flows  concurrently vanish; i.e., 
\begin{equation} 
{\dot y}=0, \quad {\dot\theta}=0, 
  \label{eq:zerolift-pinning_0}
\end{equation}
or, in an equivalent form using Eqs. \eqref{eq:dtm_dyn_eqs1} and \eqref{eq:dtm_dyn_eqs2}, as 
\begin{equation} 
\tilde{\dot y}\simeq \tilde u^{(\mathrm{im})}_y(\tilde y_\ast,\theta_\ast)=0, \quad \tilde{\dot\theta} \simeq \tilde{\omega}(\tilde y_\ast, \theta_\ast)=0, 
  \label{eq:zerolift-pinning}
\end{equation}
which are to be used with  Eqs. \eqref{eq:u_im} and \eqref{eq:w_tot_rescaled}. 

The fixed points can be determined from the intersections between {\em nullclines} of the above dynamical system. The nullclines are defined as the loci of coordinate-space points where {\em either} ${\dot y}$ {\em or} ${\dot\theta}$ vanishes \cite{Rasband1990,Strogatz2000,Nayfeh2008}.  The equation   ${\dot y}=0$ produces  {\em zero-lift nullclines} which include the channel centerline  $\tilde y=\tilde H/2$ and the contours $\theta= n\pi/2$ for integer $n$ (see below). The equation  ${\dot \theta}=0$ produces two sets of nullclines; one with ${\partial {\dot\theta}}/{\partial \theta}<0$ and another with ${\partial {\dot\theta}}/{\partial \theta}>0$. The former nullcline is nothing but the pinning curve $\theta = \theta_\mathrm{p}(\tilde y)$; see Section \ref{subsec:pinning_curve}. The coordinate-space points on the latter nullcline are   orientationally unstable and they turn out to be of no consequence to the magnetic   focusing of spheroids in the strong-field regime. The {\em relevant fixed points} of the system thus result from the intersection of the zero-lift nullclines with the pinning  curve. Since the relevant fixed points will be located on the pinning curve, their coordinates $(y_\ast, \theta_\ast)$ will be related as $\theta_\ast = \theta_\mathrm{p}(\tilde y_\ast)$. These coordinates can thus be found merely by setting the  {\em hydrodynamic lift along the pinning curve}, $\tilde u^{(\mathrm{im})}_\mathrm{p}(\tilde y) \equiv \tilde u^{(\mathrm{im})}_y(\tilde y,\theta_\mathrm{p}(\tilde y))$, equal to zero.   This restricted lift follows by combining Eqs. \eqref{eq:YC_DR_rescaled}, \eqref{eq:u_im} and \eqref{eq:pinning} as 
\begin{eqnarray}
\begin{split}
&\tilde u^{(\mathrm{im})}_\mathrm{p}(\tilde y) =
 - \frac{15}{64} \mathrm{Pe}_\mathrm{f}\, \zeta(\alpha) \left(\frac{1}{\tilde y^2}-\frac{1}{(\tilde H-\tilde y)^2}\right)\!\left(1- \frac{2\tilde y}{\tilde H}\right)  \sin 2\theta_\mathrm{p}(\tilde y)  \\
&
\qquad\qquad\qquad\quad \times \bigg[ \left(-X^\mathrm{M}+2Y^\mathrm{M}-Z^\mathrm{M}\right)\cos^2\theta_\mathrm{p}(\tilde y) + \left(2X^\mathrm{M}-2Y^\mathrm{M}\right)\sin^2\theta_\mathrm{p}(\tilde y)  
 -\frac{6}{5}Y^\mathrm{H}\bigg].   
 \label{eq:lift_on_pinning}
\end{split}
\end{eqnarray}
Requiring $\tilde u^{(\mathrm{im})}_\mathrm{p}(\tilde y_\ast)=\tilde u^{(\mathrm{im})}_y(\tilde y_\ast,\theta_\mathrm{p}(\tilde y_\ast))=0$, we find the following solutions for the relevant fixed points and nullclines:

\begin{itemize}[leftmargin=20pt, rightmargin=13pt] 
\item[(i)]{$1- 2\tilde y_\ast/{\tilde H} = 0$: This produces the intersection of the pinning curve with the zero-lift  nullcline  $\tilde y=\tilde H/2$, leading to the fixed point $(\tilde y_\ast, \theta_\ast) = (\tilde H/2, \theta_B )$ at the  center of coordinate space $\tilde y\in [0, \tilde H]$ and $\theta\in [{\theta_B-\pi},{\theta_B+\pi})$ (Section III D,  main text). This is an omnipresent  {\em central fixed point} in {\em all} regimes of behavior that we identify here and in the main text, given that its corresponding nullcline intersection always exists.}
\item[(ii)]{$\sin 2\theta_\ast = 0$: This gives the intersections of the pinning curve with the zero-lift nullclines $\theta= n\pi/2$ for integer $n$. Since the tilt angle of the applied field can be restricted as $0\leq\theta_B \leq \pi/2$ (Section III D, main text), only the  nullclines with  $n= 0, \pm 1, 2$ will be relevant.  When present, the lateral coordinate of the fixed points $(\tilde y_\ast, n\pi/2)$ in this case can be obtained using the pinning relation  $n\pi/2 = \theta_\mathrm{p}(\tilde y_\ast)$, with the admissible solutions required to fulfill $0\leq \tilde y_\ast\leq \tilde H$.}
\end{itemize}

Before proceeding further, we emphasize that the deterministic solutions for the pinning curve and the fixed points admit a  scaling with the ratio $\chi/\mathrm{Pe}_\mathrm{f}$ rather than depending on $\chi$ and $\mathrm{Pe}_\mathrm{f}$ separately. This can be seen directly from Eqs. \eqref{eq:u_im},  \eqref{eq:w_tot_rescaled} and  \eqref{eq:zerolift-pinning}, and also from our discussion of the pinning solutions in Section \ref{subsec:pinning_curve} as well as the boundary curves which we give later for the different focusing regimes. The $\chi/\mathrm{Pe}_\mathrm{f}$-scaling is in general broken in the probabilistic setting because of the noise terms in the Smoluchowski equation (11), main text. The breakdown is however manifested only when noise-induced effects can produce qualitative changes  (such as field-induced defocusing; Section IV D, main text)  in the system behavior. 

%%%%%%%%%%%%%%%%%%%%%%%%%%%%%%%%%%%%%%%%%%%
\subsection{Linearization and fixed-point stability}

On the linearization level \cite{Rasband1990,Strogatz2000,Nayfeh2008}, small perturbations $\delta\tilde y =  \tilde y - \tilde y_\ast $ and $\delta\theta =\theta - \theta_\ast$  around a given fixed point $( y_\ast, \theta_\ast)$ are governed by the linearized form of the dynamical equations \eqref{eq:dtm_dyn_eqs1} and \eqref{eq:dtm_dyn_eqs2}; i.e., $\dot{\mathbf w}= {\mathbb J}\,{\mathbf w}$ where ${\mathbf w}^T=[\delta\tilde y\,\,\,\, \delta\theta]$  and 
\begin{equation}
{\mathbb J}  = \frac{\partial (\tilde{\dot y}, \tilde{\dot\theta})}{\partial (\tilde y, \theta)} \simeq  
\begin{bmatrix}
 \frac{\partial \tilde u^{(\mathrm{im})}_y}{\partial \tilde y} & \frac{\partial \tilde u^{(\mathrm{im})}_y}{\partial \theta} 
\vspace{2mm} \\
\frac{\partial \tilde\omega}{\partial \tilde y } & \frac{\partial \tilde\omega}{\partial \theta }
\end{bmatrix}_{\tilde y_\ast,\theta_\ast}
 \label{eq:Jacobian}
\end{equation}
is the Jacobian matrix whose eigenvalues $\lambda_y$ and $\lambda_\theta$   determine the {\em stability class} of the fixed point in question. For the most part in our forthcoming analysis, these eigenvalues  turn out to be real-valued. Also, as we deal with fixed points on the pinning curve, one of the eigenvalues, i.e., $\lambda_\theta$ (or its real part where applicable), always turns out to be negative. 

Among the different relevant fixed points, we shall encounter one with $\lambda_y<0$ \& $\lambda_\theta<0$  and   another  with  $\lambda_y>0$ \& $\lambda_\theta<0$. These  amount to  {\em stable} and {\em unstable hyperbolic} fixed points, specifically  a  {\em stable node} and a {\em saddle point}, respectively \cite{Rasband1990,Strogatz2000,Nayfeh2008}.  We shall also encounter a fixed point with  $\lambda_y=0$ \& $\lambda_\theta<0$. This corresponds to a {\em nonhyperbolic, neutrally stable} fixed point on the {\em linearization} level; however, since one of the eigenvalues vanishes, the true nature of this fixed point needs to be determined through an analysis of higher-order contributions (as implied by the so-called Shoshitaishvili theorem   \cite{Nayfeh2008}) as we shall consider in Sections \ref{subsec:centered} and \ref{subsec:offcentered} below. It is useful to note that the stability class of hyperbolic fixed points whose eigenvalues have nonzero real part remains unchanged upon higher-order perturbation according to the Hartman-Grobman theorem  \cite{Nayfeh2008}. There will also be cases of fixed points with complex-valued eigenvalues, including a {\em stable focus}, which we shall return to in Sections \ref{subsec:offcentered} and  \ref{subsec:near_wall}.  

In what follows, we do not provide the explicit expressions for elements of the Jacobian matrix ${\mathbb J}$ and, for the sake of brevity, only give  expressions for the eigenvalues that we obtain  at the fixed points. It is to be emphasized that the elements of ${\mathbb J}$ and, hence, the eigenvalues, are to be determined using our unconstrained or generalized form of the hydrodynamic lift \eqref{eq:LiftExplicit} rather than its strict-pinning form  \eqref{eq:lift_on_pinning}.  

As noted before, the stability analysis as we develop here (i.e., determination of eigenvalues associated with the fixed points in the linearization, nonlinear stability of the fixed points,  association of the fixed points with boundaries of the identified regimes of magnetic focusing, etc.)  has not been discussed in the previous literature on the subject. 

%%%%%%%%%%%%%%%%%%%%%%%%%%%%%%%%%%%%%%%%%%%%%%%
\subsection{Regime of centered focusing (regime {\bf I})}
\label{subsec:centered}

We first consider  the deterministic fixed points in the regime of centered focusing (Section IV, main text). For prolate (oblate) spheroids, this regime of behavior transpires when the external magnetic field is applied transversally (longitudinally) with tilt angle $\theta_B = \pi/2$   ($\theta_B = 0$) and rescaled strength $\chi>\chi_\ast$. Of the nullclines noted under case (ii) in Section \ref{subsec:FP_on_pinning}, only $\theta=\pi/2$ ($\theta=0$) intersects the pinning curve for prolate (oblate)  spheroids. The resulting fixed point coincides with the central fixed point   $(\tilde y_\ast, \theta_\ast) = (\tilde H/2, \theta_B )$ that follows from case (i) in Section \ref{subsec:FP_on_pinning}. In other words, the intersecting nullclines, i.e., the pinning curve and  the zero-lift nullclines $\tilde y=\tilde H/2$ and  $\theta=\pi/2$ ($0$) for prolate  (oblate) spheroids, meet jointly at a {\em single} fixed point. 

Using Eq. \eqref{eq:LiftExplicit}, the first-row elements of the Jacobian matrix \eqref{eq:Jacobian} are found to vanish at this fixed point, ${\mathbb J}_{11}={\mathbb J}_{12}=0$. This yields $\lambda_y = 0$ and $\lambda_\theta = {\mathbb J}_{22}$ where $\lambda_\theta<0$ since ${\mathbb J}_{22}<0$ according to Eq.  \eqref{eq:pinning_curve} (second relation). Then, using Eq. \eqref{eq:w_tot_rescaled}, we  find   
 \begin{equation}
 \lambda_y = 0 \,\,\,\, \& \,\,\,\, \lambda_\theta= - \chi\Delta_\mathrm{R}(\alpha) <0.  
 \label{eq:eigenvalues_case_i}
\end{equation}
On the linearization level, the solitary central fixed point found in this regime is then  nonhyperbolic, neutrally stable. 

To clarify the higher-order nature of the fixed point in $\tilde y$ direction,  we  take advantage of the strong pinning of spheroidal orientation and consider nonlinear perturbations only along the pinning curve. This is done through the quantity $\tilde u^{(\mathrm{im})}_\mathrm{p}(\tilde y)$, Eq. \eqref{eq:lift_on_pinning}, in the proximity of $\tilde y_\ast = \tilde H/2$, bearing in mind that we necessarily have $\tilde H/2 = \tilde y_\mathrm{p}(\theta_B)$. On the  leading order,  we find   
\begin{equation}
\left(\!\frac{\partial \tilde u^{(\mathrm{im})}_\mathrm{p}}{\partial\tilde y}\!\right)_{\!\!\tilde H/2}=\left(\!\frac{\partial^2 \tilde u^{(\mathrm{im})}_\mathrm{p}}{\partial\tilde y^2}\!\right)_{\!\!\tilde H/2}=0.  
\end{equation}
That is, not only the {\em first} $\tilde y$-derivative of the lift velocity  but also its {\em second} $\tilde y$-derivative vanishes at the fixed point; hence,  
\begin{equation}
\tilde u^{(\mathrm{im})}_\mathrm{p}(\tilde y)\simeq   \frac{1}{3!}\left(\!\frac{\partial^3 \tilde u^{(\mathrm{im})}_\mathrm{p}}{\partial\tilde y^3}\!\right)_{\!\!\tilde H/2} (\tilde y - \tilde H/2)^3.  
\label{eq:TaylorSeries_transverse}
\end{equation}
The fixed-point stability is thus determined by the third derivative at the centerline which we obtain  as   
\begin{equation}
\left(\!\frac{\partial^3 \tilde u^{(\mathrm{im})}_\mathrm{p}}{\partial\tilde y^3}\!\right)_{\!\!\tilde H/2}\! =  C_\mathrm{R}(\alpha)\,\mathrm{Pe}_\mathrm{f}\,\frac{ 2 \zeta(\alpha)}{  \tilde H^4} \left(\!\frac{\partial \theta_\mathrm{p}}{\partial \tilde y}\!\right)_{\!\!\tilde H/2}, 
\label{eq:3rd_derivative}
\end{equation}
where $C_\mathrm{R}(\alpha)$ is expressed in terms of spheroidal resistance functions  as 
\begin{eqnarray}
&& C_\mathrm{R}(\alpha) =  
\left\{ \begin{array}{ll}
\!2(15X^\mathrm{M}\!-15Y^\mathrm{M}\! -9 Y^\mathrm{H})  &\,\,: \,\,\alpha>1\,\,\, (\theta_\ast\!=\!\pi/2),\\
\\
\!15X^\mathrm{M}\!-30Y^\mathrm{M} \!+15Z^\mathrm{M} \!+18Y^\mathrm{H} &\,\,: \,\,\alpha<1\,\,\, (\theta_\ast\!=\!0).
\end{array}\right.
\label{eq:3rd_derivative_2}
\end{eqnarray}

The expression \eqref{eq:3rd_derivative} depends on $\mathrm{Pe}_\mathrm{f}$, $\alpha$, $\chi/\mathrm{Pe}_\mathrm{f}$ and $\theta_B$ but its dependences on $\chi/\mathrm{Pe}_\mathrm{f}$ and $\theta_B$ enter only through the  pinning curve  $\theta_\mathrm{p}(\tilde y)$ as seen from from Eq. \eqref{eq:pinning}. Also, the factor $({\partial \theta_\mathrm{p}}/{\partial \tilde y})_{\tilde H/2}$ in  Eq. \eqref{eq:3rd_derivative} is always {\em positive}. It gives the slope of the  pinning curve $\theta = \theta_\mathrm{p}(\tilde y)$ or, equivalently, its inverse slope when the  pinning curve is visualized by its inverse function $\tilde y = \tilde y_\mathrm{p}(\theta)$ as in Fig. \ref{fig:Fig1ESI} (see also Figs. 3, 4 and 7 in the main text). On the other hand, one can show using Table \ref{Tab:ResistFs} that both factors $15X^\mathrm{M}-15Y^\mathrm{M} -9 Y^\mathrm{H}<0$ and $15X^\mathrm{M}-30Y^\mathrm{M} +15Z^\mathrm{M} +18Y^\mathrm{H}<0$; i.e.,  $C_\mathrm{R}(\alpha)<0$ and, hence,  
\begin{equation}
 \left(\!\frac{\partial^3 \tilde u^{(\mathrm{im})}_\mathrm{p}}{\partial\tilde y^3}\!\right)_{\!\!\tilde H/2}\!\!<0 
\label{eq:3rd_derivative_3} 
\end{equation}
 for  {\em both} types of spheroids. This proves that the coordinate-space center $(\tilde H/2, \theta_B)$ is a {\em stable higher-order} fixed point for prolate (oblate) spheroids under  transverse (longitudinal) magnetic field.    
	
The fact that the leading-order variations of $\tilde u^{(\mathrm{im})}_\mathrm{p}$ around the above fixed point are of the third order can also be inferred from  Fig. 11  (purple curve, left inset)  in Appendix D of the main text where the lift velocity and the corresponding virtual lift potential along the pinning curve, Eq. (D8),  are plotted as functions of $\tilde y$. It is this higher-order stability of the solitary central fixed point $(\tilde H/2, \theta_B)$ that gives  the regime of  centered focusing its key features such as the flat-top (subGaussian) spheroidal density profiles (Section IV and Appendix D, main text). 

Equation  \eqref{eq:3rd_derivative} also elucidates how the flatness around the central fixed point is linked with the slope of pinning curve at that point. The said slope decreases (or, equivalently, the slope of the inverse pinning curve in Fig. \ref{fig:Fig1ESI} increases) as the rescaled field strength $\chi$ increases at a given  $\mathrm{Pe}_\mathrm{f}$, leading  to the field-induced defocusing as thoroughly discussed in the main text. This indicates  {\em progressive weakening} of the fixed point; i.e., the nonvanishing third derivative tends to zero as the field is amplified.

%%%%%%%%%%%%%%%%%%%%%%%%%%%%%%%%%%%%%%%%%%
\subsection{Regime of  off-centered focusing (regime {\bf II})}
\label{subsec:offcentered}

The regime of off-centered focusing is introduced and characterized in Section V of the main text. It transpires when the external field is applied with a tilt angle $0<\theta_B < \pi/2$ relative to the flow direction. This regime  spans a relatively wide area of the parameter space as shown in the `phase' diagrams, Figs. 9a and b, of the main text. As we show in this section, the boundaries of off-centered focusing can be determined using the deterministic analysis of fixed points, despite the fact that some its other qualitative aspects (specifically, the distinction between the optimal and shouldered focusing subregimes) are inherently probabilistic and remain beyond the scope of the forthcoming deterministic analysis.  
\vskip2mm 

In contrast to the solitary higher-order fixed point found in the regime of centered focusing, the regime of off-centered focusing turns out to be associated with {\em two} distinct fixed points:
\begin{itemize}[leftmargin=20pt, rightmargin=13pt] 
\item[$\bullet$]{{\em Half-stable central fixed point.---}This fixed point is produced from case (i) in Section \ref{subsec:FP_on_pinning} at the coordinate-space center $(\tilde y_\ast, \theta_\ast) = (\tilde H/2, \theta_B )$ and, similar to centered focusing in Section \ref{subsec:centered}, it again turns out to be a nonhyperbolic, neutrally stable  fixed point on the linearization level with exactly the same eigenvalues $\lambda_{ y}= 0$ \& $\lambda_\theta  = - \chi\Delta_\mathrm{R}(\alpha) < 0$. The higher-order analysis however reveals a different behavior than we found for the higher-order central fixed point in centered focusing. Here, the leading nonzero contribution to small perturbations around the  fixed point is found to be of second order, 
\begin{equation}
\tilde u^{(\mathrm{im})}_\mathrm{p}(\tilde y)\simeq   \frac{1}{2}\left(\!\frac{\partial^2 \tilde u^{(\mathrm{im})}_\mathrm{p}}{\partial\tilde y^2}\!\right)_{\!\!\tilde H/2} (\tilde y - \tilde H/2)^2, 
\label{eq:TaylorSeries_tilted}
\end{equation}
where the second $\tilde y$-derivative of the lift velocity along the pinning curve at the centerline is found as 
\begin{equation}
\left(\!\frac{\partial^2 \tilde u^{(\mathrm{im})}_\mathrm{p}}{\partial\tilde y^2}\!\right)_{\!\!\tilde H/2}\! = - 15  \mathrm{Pe}_\mathrm{f}\, \frac{\zeta(\alpha)}{\tilde H^4} \sin2\theta_B \bigg[ \left(-X^M+2Y^M-Z^M\right)\cos^2\theta_B + \left(2X^M-2Y^M\right)\sin^2\theta_B  
 -\frac{6}{5}Y^H\bigg]. 
\label{eq:2nd_derivative}
\end{equation}
One can directly verify  using Table \ref{Tab:ResistFs} that 
\begin{eqnarray}
&& 
\left\{ \begin{array}{ll}
\left(\!\dfrac{\partial^2 \tilde u^{(\mathrm{im})}_\mathrm{p}}{\partial\tilde y^2}\!\right)_{\!\!\tilde H/2}\!\!>0  &\,\,\,: \,\,\alpha>1,\\
\\
\left(\!\dfrac{\partial^2 \tilde u^{(\mathrm{im})}_\mathrm{p}}{\partial\tilde y^2}\!\right)_{\!\!\tilde H/2}\!\!<0 &\,\,\,: \,\,\alpha<1, 
\end{array}\right.
\label{eq:2nd_derivative_2}
\end{eqnarray}
for the whole span of tilt angles $0<\theta_B < \pi/2$ being considered in this regime. This implies that, in the case of prolate spheroids, the trajectories in the bottom half of the channel ($ 0 \leq \tilde y \leq \tilde H /2 $) are attracted toward the central fixed point and those in the top half of the channel ($ \tilde H /2 < \tilde y \leq \tilde H $) are repelled from it. In the top half of the channel, the trajectories are in fact attracted to another fixed point which will be discussed next. The foregoing properties reflect the {\em half-stable} nature of the central fixed point; i.e., it is attractive from the bottom and repulsive from the top side in $\tilde y$-direction. In the case of oblate spheroids, the central fixed point is similarly half-stable except that the situation w.r.t. the top and bottom  halves of the channel is exactly reversed. 
}
\item[$\bullet$]{{\em Off-centered fixed point.---}This fixed point is produced from case (ii) in Section \ref{subsec:FP_on_pinning}; specifically, by the intersection of the pinning curve and the nullclines  $\theta= \pi/2$ and $0$ for prolate and oblate spheroids, respectively.  Using the angular coordinates $\theta_\ast = \pi/2$ and 0 in Eq. \eqref{eq:pinning}, we find the lateral coordinate of the fixed point as $\tilde y_\ast = \tilde y_\mathrm{F}$  where 
\begin{align}
\frac{\tilde y_\mathrm{F}}{\tilde H}  = 
\left\{ \begin{array}{ll}
\!\dfrac{1}{2} + \dfrac{\chi}{\mathrm{Pe}_\mathrm{f}} \dfrac{\Delta_\mathrm{R}(\alpha)}{1+ \beta(\alpha)} \cos\theta_B &\,\,\,\, : \,\,\, \alpha>1, \\ 
\\
\!\dfrac{1}{2} -  \dfrac{\chi}{\mathrm{Pe}_\mathrm{f}} \dfrac{\Delta_\mathrm{R}(\alpha)}{1- \beta(\alpha) } \sin\theta_B &\,\,\,\, : \,\,\, \alpha<1.  
\end{array}\right.
\label{eq:yF}
\end{align}
This gives the focusing latitude within the channel where magnetic focusing of spheroids takes place. Given that,  by symmetry arguments, the tilt angle of the applied filed is restricted as $0<\theta_B < \pi/2$ (Section III D, main text), the off-centered fixed point, i.e., $(\tilde y_\ast, \theta_\ast) = (\tilde y_\mathrm{F}, \pi/2)$ and $(\tilde y_\mathrm{F}, 0)$  for prolate and oblate spheroids, is located in the {\em top} and {\em bottom} half of the channel, respectively. Note that,  using the definition \eqref{eq:YHYC_beta} of the Bretherton number,  $0<1\pm \beta(\alpha)<2$ for the respective type of spheroids. It is also useful to note that Eq. \eqref{eq:yF} reproduces centered focusing of prolate (oblate) spheroids at  $\tilde y_\mathrm{F} \rightarrow \tilde H/2$ when the limit of transverse (longitudinal) applied field, i.e.,   $\theta_B \rightarrow \pi/2$ ($\theta_B \rightarrow 0$), is taken. 
} 
\item[$\bullet$]{{\em Boundaries of the off-centered focusing regime.---}The {\em upper bound} $\chi_{\ast\ast}(\theta_B)$ on the field strength to achieve off-centered focusing can be defined by the requirement that the off-centered fixed point stays within the channel, $0<\tilde y<\tilde H$. It is in fact sufficient to demand $\tilde y_\mathrm{F} < \tilde H$ ($\tilde y_\mathrm{F} > 0$) given that the fixed point is found within the top (bottom) channel half for prolate (oblate) spheroids.  Using Eq. \eqref{eq:yF}, we find the upper boundary of off-centered focusing as 
\begin{align}
\frac{\chi_{\ast\ast}(\theta_B)}{\mathrm{Pe}_\mathrm{f}}  = 
\left\{ \begin{array}{ll}
\!\dfrac{1+\beta(\alpha)}{2 \Delta_\mathrm{R}(\alpha)} \dfrac{1}{\cos\theta_B} &\,\,\,\, : \,\,\,  \alpha>1, \\ \\
\!\dfrac{ 1-\beta(\alpha)}{2 \Delta_\mathrm{R}(\alpha)} \dfrac{1}{\sin\theta_B} &\,\,\,\, : \,\,\,  \alpha<1. 
\end{array}\right.
\label{eq:chi_2ast}
\end{align}
The field strength $\chi_{\ast\ast}$ is itself excluded from the off-centered focusing regime as it gives $\tilde y_\mathrm{F} = \tilde H$ ($\tilde y_\mathrm{F} = 0$) for prolate (oblate) spheroids. This brings in possible effects from the channel walls and is, thus, included in the near-wall accumulation regime {\bf IV} in Section \ref{subsec:near_wall}. In a strict sense, the aforementioned conditions for the existence of the off-centered focusing fixed point within the channel should be amended as $\tilde y_\mathrm{F} < \tilde H-\tilde y_0(\pi/2)$ and $\tilde y_\mathrm{F} > \tilde y_0(0)$ for prolate and oblate spheroids, respectively, where $\tilde y_0(\pi/2)=\alpha^{2/3}$ and $\tilde y_0(0)=\alpha^{-1/3}$ are the respective closest-approach distances of the spheroidal center to the walls (Appendix A, main text).  These are nothing but the rescaled half-length of the major body axis $\ell(\alpha)/R_{\mathrm{eff}}$ according to Eq. \eqref{eq:half-major} that follow from the  field-aligned configuration of the spheroids in each case (Fig. 1, main text). Since  the said closest-approach distances are typically much smaller than the channel width, the above amendments amount to relatively small changes in the boundaries of the off-centered focusing regime and are thus not taken into account here for the sake of simplicity. 

\hskip10pt{}The {\em lower bound} on the field strength to achieve off-centered focusing can be obtained by demanding that no new nullcline intersection, other than the two specific ones  noted above, transpires in this regime. This will be clarified further in Section \ref{subsec:near_wall} where we discuss the possibility of a third nullcline intersection when $\chi$ is lowered below a certain threshold $\chi_{\ast\ast}^\prime(\theta_B)$. This latter threshold indeed yields the lower boundary of off-centered focusing as 
\begin{align}
\frac{\chi_{\ast\ast}^\prime(\theta_B)}{\mathrm{Pe}_\mathrm{f}}  = 
\left\{ \begin{array}{ll}
\!\dfrac{1-\beta(\alpha)}{2\Delta_\mathrm{R}(\alpha)}\dfrac{1}{\sin\theta_B} &\,\,\,\, : \,\,\,  \alpha>1, \\ \\
\!\dfrac{1+\beta(\alpha)}{2\Delta_\mathrm{R}(\alpha)}\dfrac{1}{\cos\theta_B} &\,\,\,\, : \,\,\,  \alpha<1. 
\end{array}\right.
\label{eq:chi_2astPrime}
\end{align}
The regime of off-centered focusing is thus bracketed as   $\chi_{\ast\ast}^\prime(\theta_B) < \chi < \chi_{\ast\ast}(\theta_B)$.  Using Eqs. \eqref{eq:chi_2ast} and \eqref{eq:chi_2astPrime},  the self-consistency condition $\chi_{\ast\ast}^\prime(\theta_B)<\chi_{\ast\ast}(\theta_B)$ can be written as a condition on the tilt angle of the applied field as  $\theta_B^{\ast} <\theta_B < \pi/2$  and $0<\theta_B < \theta_B^{\ast} $ for prolate and oblate spheroids, respectively, where 
\begin{equation}
\theta_B^{\ast}= \tan^{-1} \left(\frac{1-\beta}{1+\beta}\right). 
 \label{eq:theta_B_ast}
 \end{equation}
This is in fact the tilt angle where the upper ($\chi_{\ast\ast}$) and lower ($\chi_{\ast\ast}^\prime$) boundaries meet in the $\theta_B-\chi$ plane. The off-centered focusing regime is thus bracketed  as (Section V C, main text)
\begin{equation}
\chi_{\ast\ast}^\prime(\theta_B)<\chi<\chi_{\ast\ast}(\theta_B)\quad \&\quad 
 \left\{ \begin{array}{ll}
\!\theta_B^{\ast} <\theta_B < \pi/2 &\,\,\,\, : \,\,\,  \alpha>1, \\  \\
\!0<\theta_B < \theta_B^{\ast}  &\,\,\,\, : \,\,\,  \alpha<1. 
\end{array}\right.
 \label{eq:offcentered_condition}
 \end{equation}
}
\item[$\bullet$]{{\em Stability of the off-centered fixed point.---}To determine the eigenvalues and the stability class of the off-centered  fixed point, it is useful to note that the first element of the Jacobian matrix, ${\mathbb J}_{11}=\partial \tilde u^{(\mathrm{im})}_y/{\partial \tilde y}$, vanishes identically over the nullclines $n\pi/2$ according to Eq. \eqref{eq:LiftExplicit}. This implies  that  the sum of eigenvalues (trace of $ {\mathbb J}$) $\lambda_y+\lambda_\theta = {\mathbb J}_{22}<0$ according to Eqs.  \eqref{eq:pinning_curve} (second relation) and \eqref{eq:Jacobian}. Using Eq. \eqref{eq:w_tot_rescaled},  we thus have  
 \begin{equation}
\lambda_y+\lambda_\theta  =   - \chi\Delta_\mathrm{R}(\alpha)  \cos\left(\theta_B - \theta_\ast \right) <0. 
 \label{eq:eigenvalues_case_ii}
\end{equation}
The eigenvalues can be calculated straightforwardly from the explicit form of the Jacobian elements as 
\begin{eqnarray}
&&\lambda_{y, \theta} = - \frac{\chi \Delta_\mathrm{R}}{2}  \sin \theta_B \pm \Bigg\{\!\!
\left(\frac{\chi \Delta_\mathrm{R}}{2}  \sin \theta_B\! \right)^2\! + \frac{ 3\zeta(\alpha) \mathrm{Pe}_\mathrm{f}}{4\tilde H} \left(\frac{1}{\tilde y_\mathrm{F}^2}-\frac{1}{(\tilde H-\tilde y_\mathrm{F})^2}\right) \nonumber\\
&&\hskip8cm \times \left(-5X^\mathrm{M}+5Y^\mathrm{M} + 3Y^\mathrm{H} \right) \chi\Delta_\mathrm{R}\cos\theta_B
\!\Bigg\}^{1/2}
\label{eq:EigenFP1_prolate}
\end{eqnarray}
for prolate  spheroids at their corresponding fixed-point angular coordinate $\theta_\ast = \pi/2$, and 
\begin{eqnarray}
&&\lambda_{y, \theta} = - \frac{\chi \Delta_\mathrm{R}}{2}  \cos \theta_B  \pm \Bigg\{\!\!
\left(\frac{\chi \Delta_\mathrm{R}}{2}  \cos \theta_B \!\right)^2\! + \frac{ 3\zeta(\alpha) \mathrm{Pe}_\mathrm{f}}{8\tilde H} \left(\frac{1}{\tilde y_\mathrm{F}^2}-\frac{1}{(\tilde H-\tilde y_\mathrm{F})^2}\right) \nonumber\\
&&\hskip7cm \times \left(5X^\mathrm{M} -10 Y^\mathrm{M} + 5 Z^\mathrm{M} + 6 Y^\mathrm{H}\right)   \chi\Delta_\mathrm{R}\sin\theta_B
\!\Bigg\}^{1/2}
\label{eq:EigenFP1_oblate}
\end{eqnarray}
for  oblate  spheroids at $\theta_\ast = 0$. The second term inside the curly brackets in Eqs. \eqref{eq:EigenFP1_prolate} and \eqref{eq:EigenFP1_oblate} is always negative as one can verify directly using Eq. \eqref{eq:yF} and Table \ref{Tab:ResistFs}. Hence, the eigenvalues are either both negative or have negative real parts. Therefore,  the corresponding prolate and oblate fixed point $(\tilde y_\mathrm{F} , \pi/2)$ and  $(\tilde y_\mathrm{F} , 0)$, respectively, is always {\em stable} and its basin of attraction includes the entire channel half in which it is located; i.e., $ \tilde H /2 < \tilde y \leq \tilde H $ (top half) and $ 0 \leq \tilde y < \tilde H/2 $ (bottom half), respectively. When the eigenvalues are real-valued, the fixed point is a  {\em stable node} and when they are complex-valued, it is a   {\em stable focus} \cite{Rasband1990,Nayfeh2008}. Our numerical inspections show that the latter case occurs only in narrow margins of the off-centered focusing regime close to $\theta_B^\ast$ for both prolate and oblate spheroids. With no discernible impact found on the shape of the probability distributions, we do not delve further into exploring the distinctions between these two subtypes of stable off-centered  fixed points. 
}
\end{itemize}

The first relation in Eq. \eqref{eq:yF} and also the presence of a finite negative first derivative for the lift velocity (as a stability condition for focusing of prolate spheroids at $\tilde y_\mathrm{F}$) were previously reported in  Ref. \cite{GolestanianFocusing}. The eigenvalue analysis, the association of off-centered focusing with two distinct fixed points, the implications of fixed-point stability for the boundaries of off-centered focusing and the connection with spheroidal density profiles (Section V D, main text)  have not however been addressed before. 
 
%%%%%%%%%%%%%%%%%%%%%%%%%%%%%%%%%%%%%%%%%%%
\subsection{Beyond off-centered focusing: Near-wall accumulation (regimes {\bf III} and {\bf IV})}
\label{subsec:near_wall}

The focusing `phase'  diagrams in Section VII of the main text include two other  regimes of behavior than discussed above. These are designated as regimes {\bf III} and {\bf IV}. They are also identified based on the analysis of deterministic fixed points. 

In these regimes, there are two main changes in the deterministic phase-space dynamics of spheroids relative to what we found in the regime of off-centered focusing (regime {\bf II})  in Section \ref{subsec:offcentered}: Either the stable off-centered fixed point present in the latter regime ceases to exist  or a new (third) fixed point emerges as a result of an intersection between the pinning curve and the nullcline $\theta = 0$ ($\theta =\pi/2$) for prolate (oblate) spheroids.

In both regimes {\bf III} and {\bf IV}, one deals with situations where the spheroids are partially or entirely pushed toward either or both of the channel walls by the lift velocity along the pinning curve. Due to their near-wall accumulation, the density profile of spheroids in these regimes is expected to be dependent on specific modeling details of the channel walls such as the precise form of the spheroid-wall steric  interaction potential,  any specific  boundary  features, near-field hydrodynamic effects, etc. Analyzing such potentially model-dependent aspects go beyond the scope of our study in the main text where we focus on the more robust field and shear-induced features of magnetic focusing away from the walls. For the sake of completeness, however, we include a brief discussion of the aforesaid regimes  in the main text (Section VI) and show them in the `phase' diagrams of Figs. 9a and b therein. In what follows, we provide further details, specifically on the finer structure of regimes  {\bf III} and {\bf IV}. 

\begin{itemize}[leftmargin=20pt, rightmargin=13pt] 
\item[$\bullet$]{{\em Emergence of a saddle point (regime {\bf III}).---}We begin the discussion of this regime by overviewing its outer boundaries from the aforementioned `phase' diagrams (Fig. 9a and b, main text). As seen from these diagrams,  regime {\bf III} falls below the lower boundary $\chi_{\ast\ast}^\prime(\theta_B)$, Eq. \eqref{eq:chi_2astPrime}, of the off-centered focusing regime ({\bf II}). It also lies above $\chi_\ast(\theta_B)$ as it is still part of the strong-field or whole-channel-pinning regime (Section \ref{subsubsec:whole_pinning}). Regime {\bf III} is thus bracketed as 
\begin{equation}
\chi_\ast(\theta_B) < \chi \leq \chi_{\ast\ast}^\prime(\theta_B)\quad \&\quad 
 \left\{ \begin{array}{ll}
\!0\leq\theta_B \leq \theta_B^{\ast\ast} &\,\,\,\, : \,\,\,  \alpha>1, \\  \\
\!\theta_B^{\ast\ast} \leq \theta_B \leq \pi/2  &\,\,\,\, : \,\,\,  \alpha<1,  
\end{array}\right.
 \label{eq:regime_III_condition}
 \end{equation}
where $\theta_B^{\ast\ast}$ denotes the tilt angle where $\chi_\ast(\theta_B)$ and $\chi_{\ast\ast}^\prime(\theta_B)$ meet. It is approximately obtained as
\begin{equation}
\theta_B^{\ast\ast} \simeq  \left\{ \begin{array}{ll}
\!\sin^{-1}\! \left[ \dfrac{\mathrm{Pe}_\mathrm{f}}{\chi}\dfrac{(1-\beta)}{2  \Delta_\mathrm{R}} \right] &\,\,\,\, : \,\,\,  \alpha>1, \\  \\
\!\cos^{-1}\! \left[ \dfrac{\mathrm{Pe}_\mathrm{f}}{\chi} \dfrac{(1+\beta)}{2 \Delta_\mathrm{R}} \right]   &\,\,\,\, : \,\,\,  \alpha<1.  
\end{array}\right.
\label{eq:thetaB_astast}
\end{equation}
 Our numerical inspections of the deterministic dynamical equations indicate that the fixed points for this regime differ in two subdivisions of it. These can be identified as subregime {\bf III(a)} for $\chi_\ast(\theta_B) < \chi \leq \chi_{\ast\ast}^{\prime\prime}(\theta_B)$  and subregime {\bf III(b)} for $\chi_{\ast\ast}^{\prime\prime}(\theta_B) < \chi \leq \chi_{\ast\ast}^\prime(\theta_B)$. Here,  $\chi_{\ast\ast}^{\prime\prime}(\theta_B)$ is the extrapolation $\chi_{\ast\ast}(\theta_B)$ into regime {\bf III} (Fig. 9a and b, main text) and is thus given by the same explicit form as in Eq. \eqref{eq:chi_2ast}. For the sake of brevity and clarity, this separating boundary between the two subregimes {\bf III(a)} and {\bf III(b)} is not explicitly indicated in the `phase' diagrams in question.  

\hskip10pt{}$\triangleright$\,\,{\em Subregime {\em\bf III(a)} with three fixed points (accumulation on a single channel wall).---}For $\chi_\ast(\theta_B) < \chi \leq \chi_{\ast\ast}^{\prime\prime}(\theta_B)$, three different fixed points coexist. First, the fixed point from case (i) in Section \ref{subsec:FP_on_pinning} continues to exist and, similar to the regime of off-centered focusing in Section \ref{subsec:offcentered}, this central fixed point turns out to be nonhyperbolic, neutrally stable in the linearization and, upon nonlinear perturbation, indeed a half-stable one on the higher order. Likewise, the stable `side' of this fixed point faces the bottom (top) half of the channel and, hence, it enables `funneling' of the spheroids from the bottom to top (top to bottom) channel half   for prolate (oblate) spheroids. Second, the off-centered fixed point at the latitude $\tilde y_\mathrm{F}$, Eq. \eqref{eq:yF}, from case (ii) in Section \ref{subsec:FP_on_pinning} also continues to exist in this subregime but only as a {\em stable focus}. Recall from Section \ref{subsec:offcentered} that this latter fixed point results from the intersection of the pinning curve and the zero-lift nullcline  $\theta= \pi/2$ ($\theta=0$)  for prolate (oblate) spheroids. Third, and as noted above, an additional fixed point arises from an intersection that now transpires  between the pinning curve and the zero-lift  nullcline  $\theta = 0$ ($\theta=\pi/2$) for prolate (oblate) spheroids. The lateral coordinate for this third fixed point is found as  
\begin{align}
\frac{\tilde y_\mathrm{S}}{\tilde H}  = 
\left\{ \begin{array}{ll}
\!\dfrac{1}{2} - \dfrac{\chi}{\mathrm{Pe}_\mathrm{f}}\dfrac{\Delta_\mathrm{R}}{1-\beta}\sin\theta_B  &\,\,\,\, : \,\,\, \alpha>1, \\ 
 \\
\!\dfrac{1}{2} + \dfrac{\chi}{\mathrm{Pe}_\mathrm{f}}\dfrac{\Delta_\mathrm{R}}{1+\beta} \cos\theta_B  &\,\,\,\, : \,\,\, \alpha<1. 
\end{array}\right.
\label{eq:yS}
\end{align} 
 The eigenvalues of the Jacobian matrix corresponding to this latter fixed point, which is thus located at $(\tilde y_\ast, \theta_\ast)=(\tilde y_\mathrm{S} ,  0)$ and $(\tilde y_\mathrm{S} ,\pi/2)$ for prolate and oblate spheroids, respectively, are found as 
\begin{eqnarray}
\begin{split}
&\lambda_{y, \theta} = -\frac{1}{2} \chi \Delta_\mathrm{R} \cos\theta_B 
\pm 
\left[ 
\left(\frac{1}{2} \chi \Delta_\mathrm{R} \cos\theta_B \right)^2 + \frac{2 \mathrm{Pe}_\mathrm{f}}{\tilde H} (1-\beta) \left(\!\frac{\partial \tilde u^{(\mathrm{im})}_y}{\partial \theta}\!\right)_{\!\!(\tilde y_\mathrm{S}, 0)}
\right]^{1/2} &:\,\,  \alpha>1, \\
&
\lambda_{y, \theta} = -\frac{1}{2} \chi \Delta_\mathrm{R} \sin\theta_B 
\pm 
\left[ 
\left(\frac{1}{2} \chi \Delta_\mathrm{R} \sin\theta_B \right)^2 + \frac{2 \mathrm{Pe}_\mathrm{f}}{\tilde H} (1+\beta) \left(\!\frac{\partial \tilde u^{(\mathrm{im})}_y}{\partial \theta}\!\right)_{\!\!(\tilde y_\mathrm{S}, \pi/2)}
\right]^{1/2} &:\,\,  \alpha<1,
\end{split}
\label{eq:EigenFP3}
\end{eqnarray}
where the derivative  inside the brackets is given at the fixed point by 
 \begin{eqnarray}
\begin{split}
&\!\!\!\!\!\!\left(\!\frac{\partial \tilde u^{(\mathrm{im})}_y}{\partial \theta}\!\right)_{\!\!(\tilde y_\mathrm{S}, 0)} \!\! =\!
 - \frac{  \chi   \zeta(\alpha)}{8} \frac{\Delta_\mathrm{R}}{1-\beta}\sin\theta_B \left(\frac{1}{\tilde y_\mathrm{S}^2}\!-\!\frac{1}{(\tilde H-\tilde y_\mathrm{S})^2}\right)
\! \left[\frac{15  }{2}   
\left(-X^\mathrm{M}+2Y^\mathrm{M}-Z^\mathrm{M}\right)
 -9 Y^\mathrm{H} \right] &:\,\,  \alpha>1, \\
&\!\!\!\!\!\!\left(\!\frac{\partial \tilde u^{(\mathrm{im})}_y}{\partial \theta}\!\right)_{\!\!(\tilde y_\mathrm{S}, \pi/2)} \!\! =\!
 - \frac{  \chi   \zeta(\alpha)}{8} \frac{\Delta_\mathrm{R}}{1 + \beta}\cos\theta_B \left(\frac{1}{\tilde y_\mathrm{S}^2}\!-\!\frac{1}{(\tilde H-\tilde y_\mathrm{S})^2}\right)
\! \left[15  \left(X^\mathrm{M}-Y^\mathrm{M}\right) -9 Y^\mathrm{H} \right]&:\,\,  \alpha<1. 
\end{split}
\label{eq:puptheta_FP3}
\end{eqnarray}
These derivatives turn out to be positive at all tilt angles for both types of spheroids. We thus have $\lambda_y > 0 $ \& $\lambda_\theta <0$, indicating that the fixed point is unstable and, specifically,  a {\em saddle point}. 

\hskip10pt{}It is clear from Eqs. \eqref{eq:yF} and \eqref{eq:yS} that, while the latitude of stable focus $\tilde y_\mathrm{F}$ falls within the top (bottom) channel half, the latitude of saddle point $\tilde y_\mathrm{S}$ falls within the {\em opposite} half, i.e., the bottom (top) half of the channel, for prolate (oblate)  spheroids. Since the saddle point is repulsive in the $\tilde y$-direction ($\lambda_y > 0 $), the spheroids are pushed away from its vicinity toward its nearest channel wall, causing partial accumulation of the spheroids on that wall; i.e., the bottom (top) wall for prolate (oblate)  spheroids. There will be no wall accumulation on the other wall, opposite to the saddle point, as the stable focus attracts all particle trajectories in its respective channel half, including those that are funneled through the half-stable central fixed point, along the pinning curve, from the bottom to the top (top to bottom) channel half for prolate (oblate)  spheroids.  The wall accumulation is itself effected by the assumed impermeability of the walls; i.e., the translational particle  flux into the walls is counterbalanced by an outward steric flux (due to the repulsive wall potential) within our model. As noted before and  discussed also in the main text (Section VI), the spheroidal PDF is this subregime will be dependent on the specific near-wall features of the model. 

\hskip10pt{}$\triangleright$\,\,{\em Subregime {\em\bf III(b)} with two fixed points (accumulation on both channel walls).---}For $\chi_{\ast\ast}^{\prime\prime}(\theta_B) < \chi \leq \chi_{\ast\ast}^\prime(\theta_B)$, the stable focus disappears and one  remains with only the half-stable central fixed point and the saddle point. As a result, particle trajectories can also be pushed toward the wall opposite to the saddle point upon funneling through the half-stable central fixed point; hence, creating partial accumulation of spheroids on {\em both} channel walls in this subregime.  

\hskip10pt{}$\triangleright$\,\,{\em Boundary curve $\chi_{\ast\ast}^\prime(\theta_B)$.---}Equation \eqref{eq:yS} can used to establish $\chi_{\ast\ast}^\prime(\theta_B)$ as the upper boundary of regime {\bf III} or, equivalently, the lower boundary of regime {\bf II}. Recall that $\chi_{\ast\ast}^\prime(\theta_B)$ was introduced (but not derived) in Eq. \eqref{eq:chi_2astPrime}. This task is straightforwardly done by requiring $\tilde y_\mathrm{S}$ to stay within the channel  $\tilde y_\mathrm{S} \geq 0$ ($\tilde y_\mathrm{S} \leq \tilde H$) for prolate (oblate) spheroids. 
}
\item[$\bullet$]{{\em Solitary half-stable central fixed point (regime {\bf IV}).---}As seen in the `phase' diagrams (Figs. 9a and b, main text), regime {\bf IV} dominates on the upper edge of  the `phase' diagrams and can be visited by increasing $\chi$ at fixed $\theta_B$ in two different ways. First, by increasing $\chi$ above $\chi_{\ast\ast}(\theta_B)$, i.e., $\chi \geq \chi_{\ast\ast}(\theta_B)$, when $\theta_B^{\ast}<\theta_B <\pi/2$ ($0< \theta_B<\theta_B^{\ast} $) for prolate (oblate) spheroids; see Eq. \eqref{eq:chi_2ast} for the definition of $\chi_{\ast\ast}(\theta_B)$. Second, by increasing $\chi$ above $\chi_{\ast\ast}^\prime(\theta_B)$, i.e., $\chi > \chi^\prime_{\ast\ast}(\theta_B)$,  when  $0\leq \theta_B \leq \theta_B^{\ast}$ ($\theta_B^{\ast}\leq \theta_B \leq\pi/2 $) for prolate (oblate) spheroids. In either case, all fixed points disappear except for the central one $(\tilde y_\ast, \theta_\ast) = (\tilde H/2, \theta_B )$ which remains half-stable. Thus, particle trajectories are pushed toward only {\em one} of the channel walls; i.e., the one facing the repulsive side of the half-stable fixed point.  
}
\end{itemize}

%%%%%%%%%%%%%%%%%%%%%%%%%%%%%%%%%%%%%%%%%%%%%%%%  
%%%%%%%%%%%%%%%%%%%%%%%%%%%%%%%%%%%%%%%%%%%%%%%%
\bibliography{HydroMagnetic_v6}